\documentclass[conference]{IEEEtran}
\IEEEoverridecommandlockouts
\usepackage{graphicx}
\usepackage{subfigure}
\usepackage{algorithm}
\usepackage{algpseudocode}
\usepackage{amsmath}
\usepackage{bbm}
\usepackage{amsthm}
\usepackage{amsfonts}
\usepackage{stfloats}
\usepackage{url}
\usepackage{xspace}
\usepackage{multirow}
\usepackage{booktabs}
\usepackage{mathtools}
\usepackage{caption}
\usepackage{hyperref} 
\usepackage{color}
\let\labelindent\relax
\usepackage{enumitem}
\usepackage{multirow}
\usepackage{fancyhdr}
\usepackage{colortbl}
\usepackage{textcomp}
\usepackage{comment}
\usepackage{enumerate}
\usepackage{enumitem}
\usepackage{tabularx}
\usepackage{etoolbox}
\setlist[itemize]{leftmargin=*}

\usepackage{xcolor}
\hypersetup{
  colorlinks,
  linkcolor={blue!70!green},
  citecolor={green!70!blue},
  urlcolor={orange!70!red}
}

\usepackage{hyperref}
\makeatletter
\def\UrlAlphabet{%
      \do\a\do\b\do\c\do\d\do\e\do\f\do\g\do\h\do\i\do\j%
      \do\k\do\l\do\m\do\n\do\o\do\p\do\q\do\r\do\s\do\t%
      \do\u\do\v\do\w\do\x\do\y\do\z\do\A\do\B\do\C\do\D%
      \do\E\do\F\do\G\do\H\do\I\do\J\do\K\do\L\do\M\do\N%
      \do\O\do\P\do\Q\do\R\do\S\do\T\do\U\do\V\do\W\do\X%
      \do\Y\do\Z}
\def\UrlDigits{\do\1\do\2\do\3\do\4\do\5\do\6\do\7\do\8\do\9\do\0}
\g@addto@macro{\UrlBreaks}{\UrlOrds}
\g@addto@macro{\UrlBreaks}{\UrlAlphabet}
\g@addto@macro{\UrlBreaks}{\UrlDigits}
\makeatother

\definecolor{mygray}{gray}{.9}

\newcommand{\etal}{\textit{et al.}\xspace}
\newcommand{\ie}{\textit{i.e.}\xspace}
\newcommand{\eg}{\textit{e.g.}\xspace}

\newcommand{\aka}{\textit{a.k.a.}\xspace}

\newcommand{\mypara}[1]{\smallskip\noindent\textbf{#1.} \xspace}

\newcommand{\sysname}{\textsc{{ORL-Auditor}}\xspace}

\makeatletter
\patchcmd{\hyper@makecurrent}{%
    \ifx\Hy@param\Hy@chapterstring
        \let\Hy@param\Hy@chapapp
    \fi
}{
    \iftoggle{inappendix}{
        \@checkappendixparam{chapter}%
        \@checkappendixparam{section}%
        \@checkappendixparam{subsection}%
        \@checkappendixparam{subsubsection}%
        \@checkappendixparam{paragraph}%
        \@checkappendixparam{subparagraph}%
    }{}%
}{}{\errmessage{failed to patch}}

\newcommand*{\@checkappendixparam}[1]{%
    \def\@checkappendixparamtmp{#1}%
    \ifx\Hy@param\@checkappendixparamtmp
        \let\Hy@param\Hy@appendixstring
    \fi
}
\makeatletter
\newtoggle{inappendix}
\togglefalse{inappendix}
\apptocmd{\appendix}{\toggletrue{inappendix}}{}{\errmessage{failed to patch}}

\begin{document}

\title{\sysname: Dataset Auditing in Offline Deep \\ Reinforcement Learning}

\author{\IEEEauthorblockN{
Linkang Du\IEEEauthorrefmark{2}\IEEEauthorrefmark{1},
Min Chen\IEEEauthorrefmark{3}\IEEEauthorrefmark{1}\thanks{\IEEEauthorrefmark{1}The first two authors made equal contribution.}, 
Mingyang Sun\IEEEauthorrefmark{2},
Shouling Ji\IEEEauthorrefmark{2}, 
Peng Cheng\IEEEauthorrefmark{2}, 
Jiming Chen\IEEEauthorrefmark{2}, 
Zhikun Zhang\IEEEauthorrefmark{2}\IEEEauthorrefmark{3}\IEEEauthorrefmark{4}\IEEEauthorrefmark{5}\thanks{\IEEEauthorrefmark{5}Zhikun Zhang is the corresponding author.}
}

\IEEEauthorblockA{\IEEEauthorrefmark{2} Zhejiang University, Hangzhou 310017, China\\ Email: linkangd@gmail.com, sji@zju.edu.cn,  saodiseng@gmail.com, cjm@zju.edu.cn}
\IEEEauthorblockA{\IEEEauthorrefmark{3}CISPA Helmholtz Center for Information Security, 
Saarbr\"ucken 66123, Germany\\
Email: min.chen@cispa.de}
\IEEEauthorblockA{\IEEEauthorrefmark{4}Stanford University, Stanford, California 94305, USA\\
Email: zhikun@stanford.edu}
}

\IEEEoverridecommandlockouts
\makeatletter\def\@IEEEpubidpullup{6.5\baselineskip}\makeatother
\IEEEpubid{\parbox{\columnwidth}{
    Network and Distributed System Security (NDSS) Symposium 2024\\
    26 February - 1 March 2024, San Diego, CA, USA\\
    ISBN 1-891562-93-2\\
    https://dx.doi.org/10.14722/ndss.2024.23184\\
    www.ndss-symposium.org
}
\hspace{\columnsep}\makebox[\columnwidth]{}}

\maketitle

\begin{abstract}
Data is a critical asset in AI, as high-quality datasets can significantly improve the performance of machine learning models. 
In safety-critical domains such as autonomous vehicles, offline deep reinforcement learning (offline DRL) is frequently used to train models on pre-collected datasets, as opposed to training these models by interacting with the real-world environment as the online DRL. 
To support the development of these models, many institutions make datasets publicly available with open-source licenses, but these datasets are at risk of potential misuse or infringement. 
Injecting watermarks to the dataset may protect the intellectual property of the data, but it cannot handle datasets that have already been published and is infeasible to be altered afterward. 
Other existing solutions, such as dataset inference and membership inference, do not work well in the offline DRL scenario due to the diverse model behavior characteristics and offline setting constraints. 

In this paper, we advocate a new paradigm by leveraging the fact that cumulative rewards can act as a unique identifier that distinguishes DRL models trained on a specific dataset. 
To this end, we propose \sysname, which is the first trajectory-level dataset auditing mechanism for offline RL scenarios. 
Our experiments on multiple offline DRL models and tasks reveal the efficacy of \sysname, 
with auditing accuracy over 95\% and false positive rates less than 2.88\%. 
We also provide valuable insights into the practical implementation of \sysname by studying various parameter settings. 
Furthermore, we demonstrate the auditing capability of \sysname on open-source datasets from Google and DeepMind, highlighting its effectiveness in auditing published datasets. 
\sysname is open-sourced at {\url{https://github.com/link-zju/ORL-Auditor}}. 
\end{abstract}

\maketitle
\section{Introduction}
\label{sec:intro}
\textit{Deep reinforcement learning} (DRL) has been successfully applied to many complex decision-making tasks, such as autopilot~\cite{FHOL18}, robot control~\cite{A17, pu2022security}, power systems~\cite{zeng2022physics}, intrusions detection~\cite{lopez2020application, yang2022detecting}. 
However, for safety-critical domains, such as robot control, directly interacting with the environment is unsafe since the partially trained policy may risk damage to robot hardware or surrounding objects~\cite{DBLP:journals/nn/RupprechtW22}. 
To address this issue, researchers propose the \textit{offline deep reinforcement learning} (Offline DRL)~\cite{DBLP:journals/corr/abs-2005-01643} paradigm, also known as full batch DRL~\cite{DBLP:books/sp/12/LangeGR12}. 
The general idea is learning from pre-collected data generated by the expert, handcrafted controller, or even random strategy respecting the system's constraints.

To facilitate the research of offline DRL, several high-quality datasets are published by third parties such as DeepMind~\cite{DBLP:conf/nips/Gulcehre0NPCZAM20, DBLP:journals/corr/BeattieLTWWKLGV16}, Berkeley Artificial Intelligence Research (BAIR)~\cite{DBLP:journals/corr/abs-2004-07219}, 
Polixir Technologies~\cite{DBLP:journals/corr/abs-2102-00714}, TensorFlow~\cite{tensorflow2015-whitepaper}, and Max Planck Institute~\cite{guertler2023benchmarking}.  
These datasets are published with strict open-source licenses, such as GNU General Public License~\cite{DBLP:journals/corr/BeattieLTWWKLGV16}, Apache License~\cite{DBLP:conf/nips/Gulcehre0NPCZAM20, DBLP:journals/corr/abs-2004-07219, tensorflow2015-whitepaper, DBLP:journals/corr/abs-2102-00714}, and BSD 3-Clause License~\cite{guertler2023benchmarking}, to protect the intellectual property (IP) of the data owner. 
The licenses typically encompass two essential terms. 
1) Attribution requires you (the users) to appropriately acknowledge the source, provide a link to the license, and indicate any modifications made. 
2) ShareAlike stipulates that if you remix, transform, or build upon the material, you must distribute your contributions under the same license as the original. 
Furthermore, some datasets are accompanied by additional patent grants aimed at safeguarding the rights of data publishers, \eg StarData~\cite{Lin2017STARDATAAS}. 
Additionally, closed-form datasets have the potential to face misuse from insider attacks or intellectual property infringement (\eg, ex-employees stealing data). 
Biscom's 2021 survey finds that 25\% of respondents admitted to taking the valuable data when leaving their job, with 95\% citing a lack of policies or technologies to prevent data theft~\cite{BiscomReport}. 
Tessian reports that 40\% of US employees take their generated data or trained models when leaving their job~\cite{TessianReport}.
The defense against the above threats comes to the question of \textit{how a data owner can prove that a suspect model was derived from its dataset. } 

\begin{figure}[!t]
\centering
\includegraphics[width=0.7\hsize]{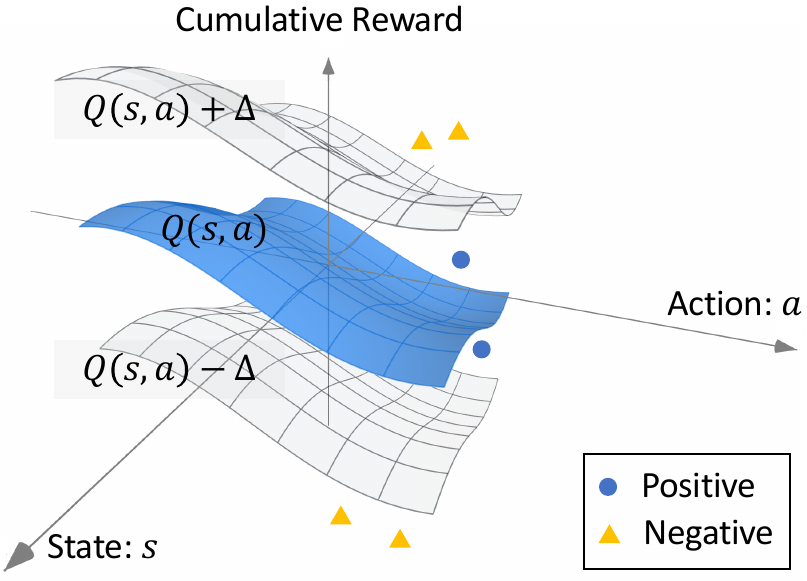}
\caption{Intuitive explanation of \sysname. 
The middle surface is the cumulative rewards of the state-action pairs from a dataset. 
The auditor outputs a positive result if the cumulative rewards of a suspect model's state-action pairs are between the two outer surfaces. }
\vspace{-0.6cm}
\label{fig:intuition}
\end{figure}

\mypara{Existing Solutions}
Recent mainstream solutions for dataset copyright protection can be classified into three categories: Watermarking, dataset inference, and membership inference.
The \textit{watermarking} approach aims to inject samples from a specific distribution prior to publishing the dataset~\cite{DBLP:journals/corr/abs-2010-05821, DBLP:journals/corr/abs-2210-00875}.
However, the auditor needs a post-event mechanism for open-source data since they are already published in the real world. 
In contrast to watermarking techniques, \emph{dataset inference} strategies~\cite{DBLP:conf/iclr/MainiYP21, DBLP:journals/corr/abs-2209-09024} do not require the injection of explicit watermarks~\cite{DBLP:journals/fdata/Boenisch21} into the datasets or trained models.
Maini \etal~\cite{DBLP:conf/iclr/MainiYP21} and Dziedzic \etal~\cite{DBLP:journals/corr/abs-2209-09024} have separately proposed dataset inference methods for supervised learning and self-supervised learning models, enabling the model owner to provide a convincing statistical argument that a particular model is trained on their private data. 
However, the dataset inference with labels~\cite{DBLP:conf/iclr/MainiYP21} needs distances between data and decision boundaries, which is not possible to obtain in RL with continuous outputs. 
The dataset inference without labels~\cite{DBLP:journals/corr/abs-2209-09024} uses the similarity of model behaviors to detect unauthorized dataset usage. 
It requires a public dataset to generate some surrogate models, and forms the auditing basis by comparing the behavioral difference between the surrogate models and the models trained on their private data. 
In offline RL scenes, since the distributions of the collected datasets depend on both environment and operator~\cite{DBLP:journals/corr/abs-2004-07219}, 
it is difficult to determine a suitable public dataset to train the surrogate model, 
making the audit basis hard to establish. 
The third category adopts the notion of \emph{membership inference}~\cite{PWZLYS19, 2020PrivAttackAM, DBLP:journals/corr/abs-2109-03975}.
By collecting the RL models' behaviors on the trained examples (members) and the untrained examples (non-members), a classifier is constructed to determine whether a data sample is used in the model's learning process. 
However, unlike online scenarios in ~\cite{PWZLYS19, 2020PrivAttackAM, DBLP:journals/corr/abs-2109-03975}, 
the auditor cannot collect additional data from the environment as the non-member examples in offline cases, where the auditor does not have access to the environment. 

\mypara{Our Proposal}
In this paper, we propose the first practical dataset auditing paradigm for the offline RL model (\sysname). 
Concretely, we are inspired by the fact that the \textit{cumulative reward}, \ie, the sum of all rewards received over a period of time starting from a given state-action pair, guiding the RL model to learn the behavior policy. 
Thus, the cumulative reward is an intrinsic feature of the datasets, making it suitable as an audit basis. 
\autoref{fig:intuition} provides a schematic diagram of \sysname, where the state, the action, and the cumulative reward compose a three-dimensional space. 
The middle surface illustrates the exact cumulative reward of the dataset, and the other two surfaces show possible offsets of the exact cumulative reward learned by the offline DRL models due to the randomness in the initialization and the learning processes. 
For a \textit{suspect model}, the auditor outputs a positive result, \ie, the data is used to train this model, if the cumulative reward from its state-action pair falls between the two surfaces; otherwise, a negative outcome. 

To implement the auditing, we first train a critic model to predict the cumulative rewards of the state-action pairs in the dataset to be audited, \ie, the target dataset. 
A straightforward strategy to derive the auditing result is to compare the cumulative reward of the state-action pairs from the suspect model to that of the target dataset through a preset judgment threshold of the similarity.
However, designing the threshold value is challenging, as it depends on the distributions of pre-collected datasets, which can vary due to different task settings, collection procedures, and data post-processing methods.
To address this issue, we recognize that the cumulative rewards embedded in the state-action pairs of the models are the estimated cumulative rewards of the target dataset, as the offline DRL models fit the cumulative reward of the dataset during training.
Thus, we train multiple models on the target dataset with varying initializations and optimization, \ie, the shadow models, and collect the cumulative rewards of their state-action pairs. 
Finally, by comparing the cumulative rewards from the suspect model and the shadow models, we make the audit decision through hypothesis testing. 

\mypara{Evaluation}
The experimental results show that the auditing accuracy of \sysname exceeds $95\%$ with false positive rates less than 2.88\% across multiple DRL models and tasks. 
By visualizing the cumulative rewards from the shadow models trained on different datasets, we demonstrate that the cumulative reward is a distinguishable feature for the dataset audit. 
We further evaluate three influential factors for the practical adoption of \sysname, \ie, the number of shadow models, the significance level in hypothesis testing, and the trajectory size. 
First, more shadow models improve the audit accuracy, and \sysname demonstrates exceptional performance with an audit accuracy exceeding 90\%, utilizing a mere 9 shadow models as illustrated in Table \ref{tab:overall audit accuracy based on Grubbs (9 Shadow Models)}. 
Second, the minimum significance level $\alpha$ of \sysname is about 0.001, meaning that the auditor outputs a single result with $99.9\%$ confidence. 
Third, \sysname tends to obtain higher accuracy with a larger trajectory size, yet we also notice that a small trajectory size achieves better results under some tasks~\cite{DBLP:journals/corr/abs-2007-09055}. 
We further implement \sysname to audit the open-source datasets from Google~\cite{DBLP:journals/corr/abs-2004-07219} and DeepMind~\cite{DBLP:conf/nips/Gulcehre0NPCZAM20}, and the experimental results again demonstrate the effectiveness of \sysname in practice. 

\mypara{Robustness}
To evaluate the robustness of \sysname, we have implemented two defense strategies to prevent the auditing. 
The first strategy involves using state-of-the-art membership inference defense techniques, such as the ensemble architecture proposed by Tang \etal~\cite{DBLP:conf/uss/TangMSSNHM22} and Jarin \etal~\cite{DBLP:journals/popets/JarinE23}. 
Despite these defense mechanisms, the audit accuracy of \sysname is still over 85\%. 
In addition to the ensemble architecture, the suspect models may distort actions to hide their training dataset. 
The offline DRL models for real-world decision-making tasks (\ie, self-driving cars) often use Gaussian noise to model natural distortions~\cite{ALHINAI20201}. 
Thus, adding Gaussian noise to the actions is stealthy to avoid the auditor's detection, and Gaussian noise is convenient for mathematical manipulation. 
To simulate strong and weak action distortion, we normalize all dimensions of the action space to $[-1, 1]$ and use Gaussian noise with $(\mu=0, \sigma=0.1)$ and $(\mu=0, \sigma=0.01)$, respectively. 
Our experiments show that \sysname is only slightly affected by Gaussian noise with $(\mu=0, \sigma=0.01)$. 
For $\sigma=0.1$, the TPR values of \sysname decline, yet the strong distortion also impacts the performance of the suspect model, especially in complex tasks. 

\mypara{Contributions}
Our contributions are three-fold:
\begin{itemize}
    \item To our knowledge, \sysname is the first dataset auditing method for the offline DRL models, using the cumulative reward as an intrinsic and stable fingerprint of the dataset. 
    \item We demonstrate the effectiveness of \sysname on four offline DRL models and three tasks. 
    We also systematically analyze various experimental factors, \ie, the hyperparameter settings and the robustness of \sysname, and summarize some important guidelines for adopting \sysname in practice.
    \item By implementing \sysname on the open-source datasets from DeepMind~\cite{DBLP:conf/nips/Gulcehre0NPCZAM20} and Google~\cite{DBLP:journals/corr/abs-2004-07219}, we show that \sysname can serve as a potent audit solution in real-world offline DRL scenarios. 
\end{itemize}

\section{Background}
\label{sec:background}
\subsection{Offline RL Problem}
\label{sec:Offline Reinforcement Learning Problem}
The offline reinforcement learning (offline RL) model aims to learn an optimal (or nearly optimal) policy from a pre-collected dataset $D$ \emph{without} an interactive environment.
We use $\mathbb{S}$ and $\mathbb{A}$ to represent the RL models' input and output space, formally called \textit{state} and \textit{action} in RL scenes. 
$r_t \in \mathbb{R}$ is the temporal reward for each time step, where $\mathbb{R}$ is the real number set. 
A unit in a pre-collected dataset called \emph{transition} is a four-element set: $\{s_t, a_t, r_t, s_{t+1}\}$, where $s_t \in \mathbb{S}$, $a_t \in \mathbb{A}$, and $s_{t+1} \in \mathbb{S}$ is the successive state of $s_t$. 
And a set of transitions in chronological order forms a \emph{trajectory} in dataset $D$. 
Based on the transitions, the offline RL model learns the Markov Decision Process underneath the datasets and forms a policy $\pi_\theta(a\mid{s})$ to maximize  $J(\pi)$. 
\begin{equation}
J(\pi)=\mathbb{E}_{{s_t} \sim d_{\beta}({s,~a}),~{a_t} \sim \pi_\theta({a} \mid {s})}\left[\sum_{t=0}^{{H}} \gamma^{t} r_t\right], \nonumber
\end{equation}
where we use $d_{\beta}$ to denote the distribution over states and actions in dataset $D$, and the actions are sampled according to the behavior policy ${a_t}\sim \pi_\theta(a\mid{s})$.
The discount factor $\gamma$ is applied to discount future rewards in the accumulated reward. 
$H$ is the terminal time step of one trajectory. 

\begin{figure}[!t]
\includegraphics[width=\hsize]{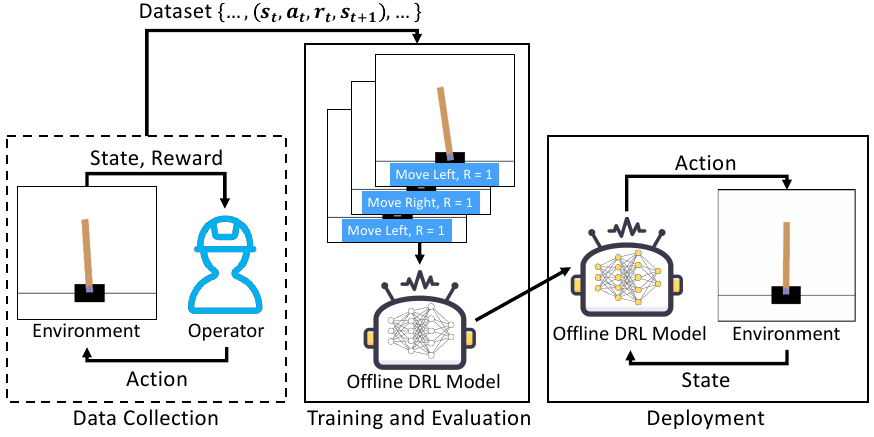}
\caption{A running example of the offline DRL models. }
\vspace{-0.4cm}
\label{fig:running example}
\end{figure}

\mypara{Example}
\autoref{fig:running example} shows an example based on the ``CartPole'' task.\footnote{\url{https://www.gymlibrary.dev/environments/classic_control/cart_pole/}}
In the data collection process, the dataset is generated from the operation logs between the operator and the environment, which contains the position and velocity of the cart and the pole (\ie, state), the operator's force direction (\ie, action), and the corresponding rewards. 
Then, in the training and evaluation process, the offline RL model learns how to play the ``Cartpole'' task from only the pre-collected dataset generated through the data collection process. 
Finally, we deploy the well-trained offline RL model in the environment to perform the task. 

\subsection{Offline RL Models}
\label{sec:offline reinforcement learning model}
In this section, we first introduce two offline RL algorithms~\cite{DBLP:conf/icml/FujimotoMP19, DBLP:journals/corr/abs-1910-01708, DBLP:conf/iclr/KostrikovNL22} separately representing two basic ideas of the offline RL models, \ie, the policy constraints strategy and the value function regularization strategy~\cite{DBLP:journals/corr/abs-2203-01387}. 
Many state-of-the-art model-free offline RL methods~\cite{DBLP:conf/nips/YuKRRLF21, DBLP:conf/nips/KidambiRNJ20, DBLP:conf/nips/FujimotoG21, DBLP:conf/iclr/KostrikovNL22} have been modified from these two approaches. 
We further present a state-of-the-art algorithm~\cite{DBLP:conf/nips/FujimotoG21} which is minimalistic with light computation and hyperparameter setting overhead. 
In addition, we briefly describe the behavior clone method (BC)~\cite{DBLP:conf/nips/Pomerleau88}, which learns the state-action distribution over the dataset via a supervised learning approach. 
Though BC is not a typical reinforcement learning method, it can solve the offline RL problem and usually serves as the baseline method in the offline RL evaluation. 

\mypara{Behavior Clone (BC)~\cite{DBLP:conf/nips/Pomerleau88}}
BC separately takes the pairwise state $s$ and action $a$ in the datasets as input and label, then it optimizes the policy through the following function. 
\begin{equation}
\theta^*=\arg \min _\theta \mathbb{E}_{(s, a) \sim D}\left[\mathcal{L}\left(\pi_\theta(s), a\right)\right], \nonumber
\end{equation}
where $D$ is the pre-collected dataset and $\mathcal{L}$ is the loss function. 
Since BC only imitates action distributions, the performance is close to the mean of the dataset, even though BC works better than online RL algorithms in most cases. 

\mypara{Batch-Constrained Q-learning (BCQ)~\cite{DBLP:conf/icml/FujimotoMP19, DBLP:journals/corr/abs-1910-01708}}
BCQ is the first practical data-driven offline RL algorithm.
The key idea of BCQ is to integrate a generative model to achieve the notion of batch-constrained, \ie, minimizing the deviation between the candidate actions with the action records of the dataset. 
To maintain the diversity of action, BCQ builds a perturbation model to perturb each selected action. 
Then it chooses the highest-valued action through a $Q$-network, that learns to estimate the expected cumulative reward of a given state and action pair. 
Thus, the objective function of BCQ can be defined as the following. 
\begin{equation}
\begin{aligned}
\pi(s)= & \underset{a_i+\xi_\phi\left(s, a_i, \Phi\right)}{\operatorname{argmax}} Q_\theta\left(s, a_i+\xi_\phi\left(s, a_i, \Phi\right)\right) \\
& \left\{a_i \sim G_\omega(s)\right\}_{i=1}^n,  \nonumber
\end{aligned}
\end{equation}
where $G_\omega(s)$ is a conditional variational auto-encoder (VAE)-based~\cite{DBLP:journals/corr/KingmaW13} generative model that can be used to generate candidate actions. 
The value function $Q_\theta$ is used to score the $n$ candidate actions and finds the action with the highest value. 
$\xi_\phi\left(s, a_i, \Phi\right)$ is the perturbation model, which outputs an adjustment to an action $a$ in the range $[-\Phi, \Phi]$. 
Then, the perturbation model can be optimized by the deterministic policy gradient algorithm~\cite{DBLP:conf/icml/SilverLHDWR14} as follows.
\begin{equation}
\phi \leftarrow \underset{\phi}{\operatorname{argmax}} \sum_{(s, a) \in \mathcal{B}} Q_\theta\left(s, a+\xi_\phi(s, a, \Phi)\right),  \nonumber
\end{equation}
where $\mathcal{B}$ represents a mini-batch state-action pair in the dataset. 
To penalize rare states, BCQ takes a convex combination of the values from two $Q$-networks and sets a new target value $y$ to update both $Q$-networks. 
\begin{equation}
y = r+\gamma \max _{a_i}\left[\lambda \min _{j=1,2} Q_{\theta_j^{\prime}}\left(s^{\prime}, a_i\right)+(1-\lambda) \max _{j=1,2} Q_{\theta_j^{\prime}}\left(s^{\prime}, a_i\right)\right], \nonumber
\end{equation}
where $a_i$ corresponds to the perturbed actions, sampled from the generative model $G_\omega(s)$. 

\mypara{Implicit Q-Learning (IQL)~\cite{DBLP:conf/iclr/KostrikovNL22}}
Compared to the batch-constrained idea of BCQ~\cite{DBLP:conf/icml/FujimotoMP19, DBLP:journals/corr/abs-1910-01708}, 
IQL strictly avoids querying values of the actions, which are not in the pre-collected dataset. 
IQL first constructs a model to evaluate the expected returns of state-action pairs. 
The objective function is defined as shown in \autoref{eq:sarsa}. 
\begin{equation}
    L(\theta)=\mathbb{E}_{D}\left[L_2^\tau\left(r(s, a)+\gamma Q_{\hat{\theta}}\left(s^{\prime}, a^{\prime}\right)-Q_\theta(s, a)\right)\right], 
    \label{eq:sarsa}
\end{equation}
where $L_2^\tau(u)=|\tau-\mathbbm{1}(u<0)| u^2$, and $s^{\prime}$ and $a^{\prime}$ represent the successor state and action of $s$ and $a$. 
Both $Q_\theta(s, a)$ and $Q_{\hat{\theta}}$ are used to assess the expected returns of state-action pairs. 
The parameters of $Q_\theta(s, a)$ are adjusted in each optimization round, 
while the parameters of $Q_{\hat{\theta}}$ are updated periodically based on $Q_\theta(s, a)$ to reduce parameter fluctuations during model updates.
\autoref{eq:sarsa} involves the dynamics of the environment, where the environment state $s$ transitions to the next environment state $s^{\prime}$, potentially introducing interference in the evaluation of expected returns for state-action pairs. 
IQL addresses this issue by introducing a new state value model, splitting \autoref{eq:sarsa} into two objective functions. 
\autoref{eq:sarsa-state-value-evaluation} shows the objective function of the state value model $V_\psi$. 
\begin{equation}
    L_V(\psi)=\mathbb{E}_{D}\left[L_2^\tau\left(Q_{\hat{\theta}}(s, a)-V_\psi(s)\right)\right]. 
    \label{eq:sarsa-state-value-evaluation}
\end{equation}

Then, IQL utilizes $V_\psi(s)$ to construct \autoref{eq:sarsa-state-action-value-evaluation} for updating the parameters of the state-action value model $Q_\theta$.
\begin{equation}
    L_Q(\theta)=\mathbb{E}_{D}\left[\left(r(s, a)+\gamma V_\psi\left(s^{\prime}\right)-Q_\theta(s, a)\right)^2\right]. 
    \label{eq:sarsa-state-action-value-evaluation}
\end{equation}

Finally, IQL considers using the state-action value model to construct a behavior policy for deployment. 
This behavior policy also needs to avoid actions that are outside the dataset distribution. 
Thus, IQL employs advantage-weighted regression to update the policy model. 
\begin{equation}
    L_\pi(\phi)=\mathbb{E}_{D}\left[\exp \left(\beta\left(Q_{\theta}(s, a)-V_\psi(s)\right)\right) \log \pi_\phi(a\mid s)\right],
\end{equation}
where $\beta \in [0, \infty)$ represents the inverse temperature. 
For smaller values of $\beta$, IQL is similar to behavior clone, tending to mimic the data collection policy. 
For larger values of $\beta$, IQL is more inclined to select actions corresponding to the highest expected returns according to the state-action value model.
Throughout the entire training process, IQL alternates between optimizing the parameters $\theta$ and $\psi$, and then updates $\phi$ while keeping $\theta$ and $\psi$ fixed. 

\mypara{TD3PlusBC~\cite{DBLP:conf/nips/FujimotoG21}}
The former methods~\cite{DBLP:conf/icml/FujimotoMP19, DBLP:journals/corr/abs-1910-01708, DBLP:conf/iclr/KostrikovNL22} limit or regularize action selection such that the learned policy is easier to evaluate with the given dataset. 
However, they introduce new hyperparameters and often leverage secondary components, such as generative models, while adjusting the underlying RL algorithm. 
TD3PlusBC is a minimalist and highly effective offline RL algorithm based Twin Delayed Deep Deterministic Policy Gradient (TD3)~\cite{DBLP:conf/icml/FujimotoHM18} with BC regularization term, which pushes the policy towards favoring actions contained in the dataset $D$: 
\begin{equation}
\pi=\underset{\pi}{\operatorname{argmax}} \mathbb{E}_{(s, a) \sim \mathcal{D}}\left[\lambda Q(s, \pi(s))-(\pi(s)-a)^2\right], \nonumber
\end{equation}
where $\lambda=\frac{\alpha}{\frac{1}{N} \sum_{\left(s, a\right)}\left|Q\left(s, a\right)\right|}$ for the dataset of $N$ transitions $\left(s, a\right)$. 
To facilitate the policy training, TD3PlusBC normalizes each state in the given dataset by $s_i=\frac{s_i-\mu}{\sigma+\epsilon}$, where $\mu$ and $\sigma$ are the mean and standard deviation respectively. 

The model architectures vary significantly regarding objective function and basic model structure. 
1) Objective Function: BCQ~\cite{DBLP:conf/icml/FujimotoMP19, DBLP:journals/corr/abs-1910-01708} and TD3PlusBC~\cite{DBLP:conf/nips/FujimotoG21} use a policy constraints strategy to maintain the learned policy similar to the one used for collecting the dataset. In contrast, IQL~\cite{DBLP:conf/iclr/KostrikovNL22} adopts a regularization strategy to improve the stochasticity of the learned policy or obtain more accurate Q-value estimations. 
2) Basic Model Structures: BCQ~\cite{DBLP:conf/icml/FujimotoMP19, DBLP:journals/corr/abs-1910-01708} and IQL~\cite{DBLP:conf/iclr/KostrikovNL22} are based on the Q-learning model, while TD3PlusBC~\cite{DBLP:conf/nips/FujimotoG21} builds upon TD3~\cite{DBLP:conf/icml/FujimotoHM18}. 
In \autoref{sec:evaluation}, our experiments are mainly conducted on the above four algorithms. 
However, \sysname can also be applied to any type of offline DRL model as long as the auditor has black-box access to the suspect model. 

\section{Problem Statement and Existing Solutions}
\label{sec:overview}
\subsection{System and Threat Model}
\begin{figure}[!t]
\includegraphics[width=\hsize]{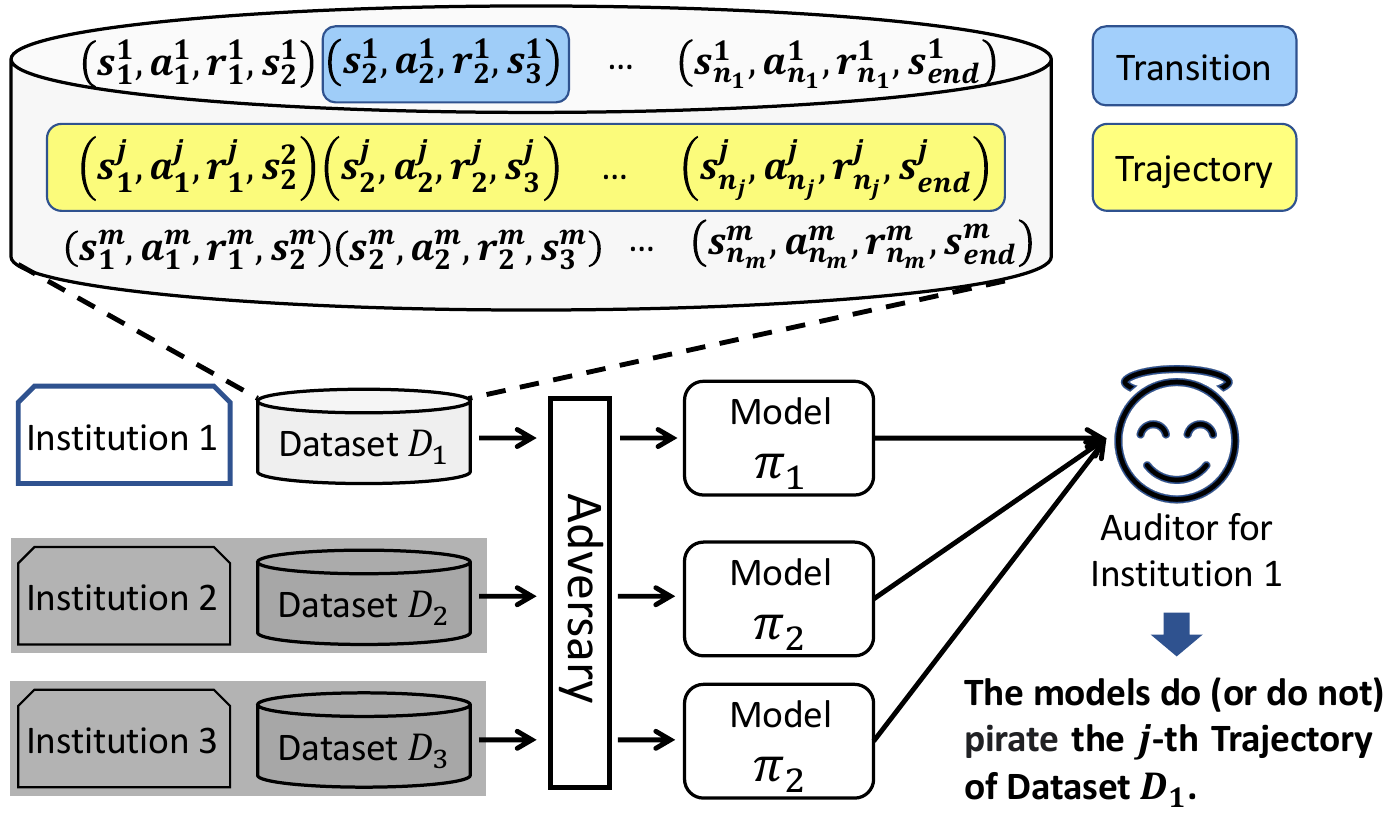}
\caption{An example of the application scenario. 
The auditor can obtain all information about dataset $D_1$ but has no knowledge about the datasets from other institutions. }
\label{fig:application scenario}
\vspace{-0.4cm}
\end{figure}

\mypara{Application Scenarios}
\autoref{fig:application scenario} illustrates a typical application scenario where the data providers collect and then publish or sell the dataset to the customers. 
A malicious customer (adversary) with access to the datasets makes a piracy distribution or illegally builds a Model-as-a-Service (MaaS) platform. 
Institution 1 suspects the models are generated by its dataset, and thus hires an auditor to determine whether the model trainers pirate the trajectories of the dataset $D_1$. 

\mypara{Auditor's Background Knowledge and Capability}
The auditor has full knowledge of the target dataset, such as the number of trajectories and the spaces of state and action. 
In offline RL settings, the auditor is prohibited from interacting with the online environment to collect more data, meaning the entire auditing only depends on the target dataset. 
We consider the auditor has black-box access to the suspect RL model. 
Note that this is the most general and challenging scenario for the auditor. 
A typical application scenario is that an adversary receives the model settings from customers, such as the selected offline RL framework, the model's hyperparameter, and the desired training episodes.
Then, the adversary trains an offline RL model and provides a service interface to the customers. 
The auditor utilizes the states of the dataset (inputs) to query the suspect model and obtain the corresponding actions (outputs). 

\mypara{Discussion}
Compared to the sample-level and dataset-level data in DNN scenes, RL has trajectory-level data, which is the minimum record unit of sequential interactions between the operator and environment. 
Since a single trajectory can guide the model from the initial state to the terminal, the trajectory-level data is regarded as the value unit of the dataset. 
Thus, \sysname is designed to audit the dataset from the trajectory level, where the auditor tries to decide whether the suspect model uses a specific trajectory in the dataset. 
In addition, the auditor can easily extend \sysname to the dataset-level data by setting a piracy alarm threshold. 
If the ratio of misappropriation using trajectories exceeds the preset threshold, the auditor can claim the dataset-level pirate. 

\subsection{Existing Solutions}
\label{sec:existing solutions}
\mypara{Watermarking~\cite{DBLP:journals/corr/abs-2010-05821, DBLP:journals/corr/abs-2210-00875}}
Watermarking-based dataset copyright protection methods inject samples of a specific distribution before publishing the target dataset.
One of its kind is implemented with backdoor attacks against the ML model. 
Li \etal~\cite{DBLP:journals/corr/abs-2010-05821} proposed to modify a dataset by adding a trigger, such as a local patch, to innocent samples in order to make them appear as a pre-defined target class. 
To verify the integrity of the dataset after the attack, they use a hypothesis test approach based on posterior probabilities generated by a third-party model. 
Inspired by this idea, the auditor can employ the backdoor attack against the DRL model~\cite{KWSL20, WJW0XS21, DBLP:journals/tifs/WangSLMJ21} to generate a watermark for the offline RL dataset. 

However, since the open-source datasets are already published, the auditor needs a post-event mechanism that does not require injecting manipulated samples before publishing the dataset. 
Watermarking, on the other hand, is a pre-event mechanism that involves injecting manipulated samples into the dataset before publishing. 
Additionally, it is difficult for the auditor to guarantee that one effective watermarking has a consistent distribution with the original dataset, which inevitably disturbs the model's normal behavior. 

\mypara{Dataset Inferences~\cite{DBLP:conf/iclr/MainiYP21, DBLP:journals/corr/abs-2209-09024}}
The core idea of dataset inference is empowering the model owner to make a compelling statistical argument that a particular model is a copied version of their own model by demonstrating that it is based on their private training data. 
It does not require injecting explicit watermarks~\cite{DBLP:journals/fdata/Boenisch21} to the datasets or the trained models. 
Existing methods~\cite{DBLP:conf/iclr/MainiYP21, DBLP:journals/corr/abs-2209-09024} can be divided into two categories according to whether they have explicit classification labels. 
With the explicit classification labels, \cite{DBLP:conf/iclr/MainiYP21} rely on computing the distances between data points and decision boundaries. 
Without the explicit classification labels, \cite{DBLP:journals/corr/abs-2209-09024} utilizes the similarity of the models' behaviors to detect the unauthorized usage of the dataset, which requires the assumption of an additional public dataset with a similar distribution to form the auditing basis. 

However, the above methods cannot directly be applied to reinforcement learning cases due to two reasons. 
First, the label-based dataset inference~\cite{DBLP:conf/iclr/MainiYP21} cannot be implemented in the RL models since their outputs are usually continuous, and they are guided by the rough reward signals instead of the exact labels. 
Second, the distribution of the offline RL dataset not only depends on the environment but also relies on the strategy of interacting with the environment~\cite{DBLP:journals/corr/abs-2004-07219}. 
Thus, it is challenging to find a proper public dataset in offline RL scenarios. 
As we delve into \autoref{sec:the-behavior-similarity-of-models}, it becomes evident that the behavior similarity of the DRL models varies across different public training data. 
Furthermore, the behavior similarity is also influenced by various offline DRL frameworks. 

\mypara{Membership Inference Attack against RL~\cite{PWZLYS19, 2020PrivAttackAM, DBLP:journals/corr/abs-2109-03975}}
Several membership inference attacks exist against DRL, which seem to address the problem studied in this paper. 
Most of them are targeted at the online RL scenes, assuming that the attacker owns the environment. 
Thus, they can utilize the environment to collect more data and even manipulate some adversarial states to facilitate the inference. 

However, in this paper, we aim at the offline RL cases, which are more challenging since the only thing the auditor can use is the pre-collected dataset. 
That is, in offline RL scenarios, the existing MIA against RL cannot rely on the environment to generate non-member data. 

\section{\sysname}
We instantiate $Q$ of \autoref{fig:intuition} with the cumulative reward, which is an intrinsic feature of the dataset and suitable for auditing. 
$\Delta$ is determined by the shadow models trained on the datasets instead of a preset threshold to adapt the distribution of different datasets. 
Thus, the well-designed $Q$ and $\Delta$ guarantee the adaptiveness and effectiveness of \sysname. 
\subsection{Workflow}
\label{sec:workflow}
\begin{figure*}[!ht]
\includegraphics[width=\hsize]{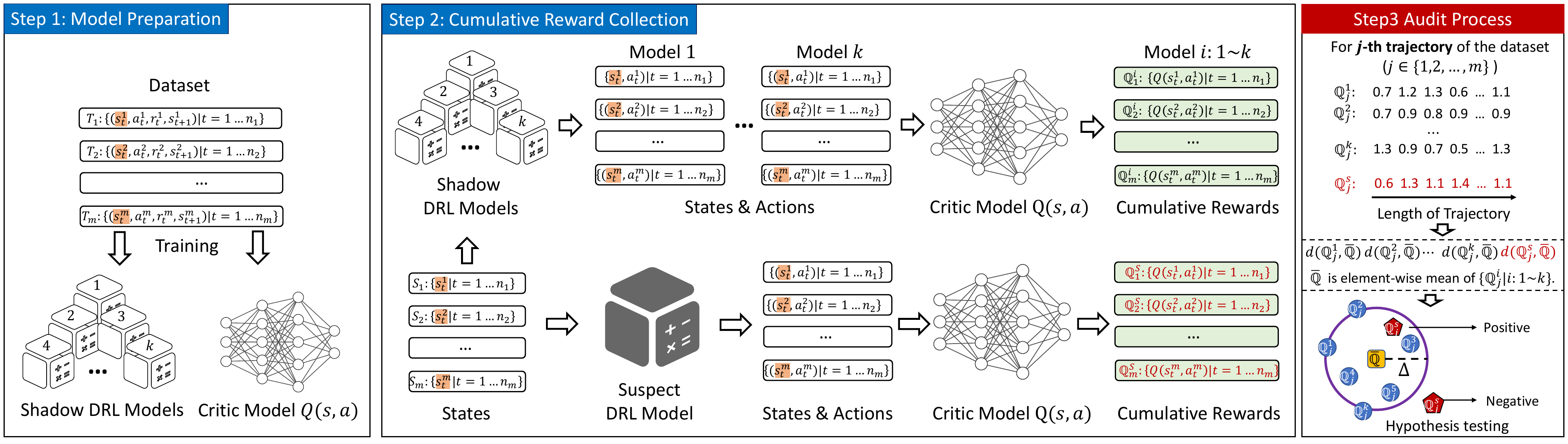}
\caption{The workflow of \sysname contains three steps, \ie, model preparation, cumulative reward collection, and audit process. 
\sysname first trains a set of shadow DRL models and a critic model on the target dataset, then collects the cumulative rewards from the state-action pairs of the shadow models and the suspect model. 
Finally, \sysname audits every trajectory based on hypothesis testing. }
\vspace{-0.4cm}
\label{fig:framework}
\end{figure*}

For ease of understanding, we refer to the \textit{target dataset} as the dataset to be audited and the \textit{actual dataset} as the dataset used by the suspect model. 
If the suspect model is trained on the target dataset, the actual dataset is the same as the target dataset, \ie, positive audit result for the suspect model; otherwise, the suspect model does not use the target dataset, \ie, negative audit result for the suspect model. 
\autoref{fig:framework} illustrates the workflow of \sysname. 

\mypara{Step 1: \underline{M}odel \underline{P}reparation (MP)}
In the left box of \autoref{fig:framework}, the auditor prepares the critic model and the shadow DRL models based on the target dataset, which contains $m$ trajectories $T$ with the length of $n_i$ ($i \in \left\{1,2,\dots, m\right\}$). 
The critic model is optimized to estimate the cumulative reward of each state-action pair. 
For each trajectory in the dataset, a series of predictions for its state-action pairs compose the exclusive feature for auditing. 
There are two ways to optimize the critic model, \ie, the Monte-Carlo-based (MC-based) and the temporal-difference-based (TD-based) strategies. 
We adopt the TD-based learning method and explain the reasons in \autoref{sec:the selection of critic model}. 
In addition, the auditor trains a set of shadow models following the model's objective function introduced in \autoref{sec:background} with different model initializations. 

\mypara{Step 2: \underline{C}umulative \underline{R}eward \underline{C}ollection (CRC)} 
In the middle box, the shadow models observe the states of the dataset and take actions. 
For $i$-th trajectory in the dataset, the auditor records the state $s^{i}_t$ and the action $a^{i}_{t}$ of each shadow model, where $a^{i}_{t}$ represents the shadow model's action at the $t$-th step of trajectory $T_i$. 
After finishing the action collection, the auditor obtains the $k$ sets of state-action pairs from the shadow models, representing the learned policies with different initialization and training processes on the target dataset. 
Using the critic model in Step 1, the auditor calculates the estimations for all state-action records, \ie, the estimated cumulative rewards, which are the samplings of the exact cumulative rewards of the corresponding state-action pairs in the dataset. 
Similarly, the auditor queries the suspect model with state $s^i_t$ and observes the action $a^{i}_t$. 
The state-action pairs are then put into the critic model and to obtain the estimations for the suspect model. 

\mypara{Step 3: \underline{A}udit \underline{P}rocess (AP)}
After the above two steps, the auditor obtains the estimated cumulative rewards from the shadow models and the suspect model and then conducts the audit process. 
For $j$-th ($j \in \{1,2,\dots, m \}$) trajectory of the dataset, the auditor collects $k$ series of the estimated cumulative rewards from the shadow models, \ie, $\{\mathbb{Q}_j^i \mid i \in \{1,2,\dots, k\}\}$, and one from the suspect model, \ie, $\mathbb{Q}_j^s$. 
\sysname conducts hypothesis testing based on the distances of $\mathbb{Q}_j^i$ and $\mathbb{Q}_j^s$ from $\bar{\mathbb{Q}}_j$. 
The auditor can rule out suspicion if $d(\mathbb{Q}_j^s, \bar{\mathbb{Q}}_j)$ is out of the distribution of $\{d(\mathbb{Q}_j^i, \bar{\mathbb{Q}}_j) \mid i \in \{1,2,\dots, k\}\}$. 
Otherwise, the auditor will conclude a positive decision, \ie, the suspect model is trained using this trajectory. 
The auditor repeatedly implements the above processes for other trajectories of the dataset and obtains the final audit report with judgment for all trajectories. 
We discuss more details of the distance metric and the hypothesis testing in \autoref{sec:the details of audit process}. 

\subsection{The Selection of Critic Model}
\label{sec:the selection of critic model}
The auditor can use either Monte Carlo (MC) based or Temporal-Difference (TD) based algorithms to train a critic model from the trajectories of the dataset. 
The main distinction between the two methods lies in their learning targets, which leads to differences in their objective functions. 
In the case of MC-based methods, the learning target $G$ is the empirical cumulative rewards from the dataset.
\begin{equation}
\begin{aligned}
G(s_t, a_t)=r_{t}+\gamma r_{t+1}+\ldots+\gamma^{H-1} r_{H}, \nonumber
\end{aligned}
\end{equation}
where $G(s_t, a_t)$ represents the exact cumulative reward from $(s_t, a_t)$ to the terminal time step $H$ of one trajectory. 
The discount factor $\gamma$ is applied to discount future rewards. 
The critic model is trained by minimizing the following objective. 
\begin{equation}
\label{eq:mc critic}
\begin{aligned}
\mathbb{E}_{\left(s_t, a_t, r_{t+1}, s_{t+1}\right) \sim D}\left[\left(G\left(s_t, a_t\right)-Q_\theta\left(s_t, a_t\right)\right)^2\right]. \nonumber
\end{aligned}
\end{equation}
For TD-based methods, the learning target changes to the expected cumulative reward in a heuristic form, \ie, $r_{t}+\gamma Q\left(s_{t+1}, a_{t+1}\right)$. 
Thus, the critic model is trained by minimizing the following loss function. 
\begin{small}
\begin{equation}
\label{eq:td critic}
\begin{aligned}
\mathbb{E}_{\left(s_t, a_t, r_{t+1}, s_{t+1}\right) \sim D}\left[\left(r_{t+1}+\gamma Q_{\theta^{\prime}}\left(s_{t+1}, a_{t+1}\right)-Q_\theta\left(s_t, a_t\right)\right)^2\right], \nonumber \\
\end{aligned}
\end{equation}
\end{small}
where the critic model starts with arbitrary initialization $\theta$.
Then, it repeatedly evaluates $Q_\theta\left(s_t, a_t\right)$, obtains a reward $r_{t+1}$, and updates the weights. 
The $\theta^\prime$ is a snapshot of $\theta$ and copies from $\theta$ every few updates of $\theta$. 
The MC-based method utilizes the exact cumulative rewards from the dataset to train the critic model, resulting in an unbiased prediction.
It also has strong convergence properties due to the stationary of $G_t$.
However, it cannot be applied to situations where the collected data is truncated, and all trajectories in the dataset must be completed. 
In practice, many sequential decision-making tasks usually have long or infinite time steps. Thus, the dataset provider segments the interaction record into trajectories by a preset maximum length. 
The TD-based method tackles the limitation of the MC-based algorithm and can learn from incomplete sequences. 
Nevertheless, due to the heuristic learning process, the TD-based method has some bias and is more sensitive to model initialization. 
Therefore, we choose the element-wise mean of the shadow models' cumulative rewards $\bar{\mathbb{Q}}$ as the auditing directrix in \autoref{sec:workflow} instead of relying solely on the critic model's predictions to compensate for the shortages of TD-based methods. 

\subsection{The Details of Audit Process}
\label{sec:the details of audit process}
\begin{algorithm}[!t]
\caption{Workflow of \sysname}
\label{alg:Generation Mechanism 2}
\begin{algorithmic}[1]
\Require Dataset $D$, suspect model $\pi_s$, number of shadow models $k$, significance level $\alpha$
\Ensure The trajectory-level audit report
\State $//$ \textbf{Step 1: Model Preparation} 
\State Train shadow models $\{\pi_i \mid i=1,\dots,k\}$ and critic model
\State $//$ \textbf{Step 2: Data Preparation} 
\For {each model $\pi$ in $\{\pi_i \mid i=1,\dots,k\} \cup \{\pi_s\} $}
    \State Query $\pi$ by states $s\in D$ and obtain the actions. 
    \State Evaluate each $(s, a)$ pair based on the critic model $Q$. 
    \State Record the cumulative reward in sequential form $\{\mathbb{Q}_j \mid j=1,\dots,m\}$. 
\EndFor
\State $//$ \textbf{Step 3: Audit Process}
\State audit\_report = []
\For {each trajectory in $\{T_j \mid j=1,\dots,m\}$} 
\State Calculate the element-wise mean $\bar{\mathbb{Q}}_j$ of $\{\mathbb{Q}_j^i \mid i= 1,\dots, k\}$
\State Measure the $d(\mathbb{Q}_j, \bar{\mathbb{Q}}_j)$ of each $\mathbb{Q}_j^i$ and $\mathbb{Q}_j^s$ from $\bar{\mathbb{Q}}_j$. 
\State $//$ Hypothesis testing
\State From $\{d^i \mid i=1,\dots, k\}$ and $d^s$, decide whether the suspect model $M_s$ pirates $T_j$ with significance level $\alpha$. 
\State audit\_report.append($j$-th audit result)
\EndFor
\State Return audit\_report
\end{algorithmic}
\end{algorithm}

In the audit process, the choice of distance metric and the hypothesis testing method play a critical role in \sysname's performance. 
A proper metric is sensitive to the deviations between the estimated cumulative rewards, which can facilitate the hypothesis testing. 
A suitable hypothesis testing method can provide precise results with high confidence. 

\mypara{Distance Metric}
We consider three types of distance metrics, \ie, $\ell_p$ norm, Cosine distance, and Wasserstein distance. 
$\ell_p$ norm is a popular method of measuring the distance between vectors, \ie, the sum of the absolute difference of the components of the vectors. 
In the RL scene, the states and actions are sequential data, meaning the distance metric should measure both the value and the position deviation of the cumulative rewards. 
However, $\ell_p$ norm may fail to reflect the difference from the sequence aspect of the same set of values. 
Cosine distance is a derivative of Cosine similarity, defined as the cosine of the angle between two vectors. 
Cosine distance embodies the difference from both the value and position aspects of the vectors. 
However, Cosine distance normalizes the inner product using the two vectors' norm, which weakens the numerical differences between the cumulative rewards. 
The Wasserstein distance, \aka earth mover's distance (EMD), is a metric of the difference between two probability distributions over a region~\cite{DBLP:conf/iccv/RubnerTG98}. 
It can be defined as follows. 
\begin{equation}
\begin{aligned}
l_1(u, v)=\inf _{\pi \in \Gamma(u, v)} \int_{\mathbb{R} \times \mathbb{R}}|x-y| \mathrm{d} \pi(x, y), \nonumber
\end{aligned}
\end{equation}
where $\Gamma(u, v)$ is the set of distributions on $\mathbb{R} \times \mathbb{R}$ whose marginals are $u$ and $v$ on the first and second factors respectively. 
Wasserstein distance fits well with audit requirements, reflecting numerical and positional deviations of the cumulative rewards. 
Thus, we set Wasserstein distance by default and compare different distance metrics in \autoref{sec:evaluation}. 

\mypara{Hypothesis Testing}
After the selection of the distance metric, the auditor proceeds to hypothesis testing with the distances of $\mathbb{Q}_j^i$ and $\mathbb{Q}_j^s$ from $\bar{\mathbb{Q}}_j$. 
\begin{equation}
\begin{aligned}
&H_0: d(\mathbb{Q}_j^s, \bar{\mathbb{Q}}_j) \text{is not an outlier. }   \nonumber \\
&H_1: d(\mathbb{Q}_j^s, \bar{\mathbb{Q}}_j) \text{is an outlier. }   \nonumber
\end{aligned}
\end{equation}

An intuitive method is to leverage the $3\sigma$ principle, \ie, the normal samples should be distributed within the range of three times the standard deviation $\sigma_{d}$ from the mean $\mu_{d}$. 
The $3\sigma$ principle is an efficient hypothesis testing method, yet the mean $\mu_{d}$ is easily misled by outliers. 
Compared to the $3\sigma$ principle, Grubbs' test~\cite{grubbs1950sample} is a more robust hypothesis testing method for detecting single outliers in univariate datasets.
If the Grubbs' test statistic of $d(\mathbb{Q}_j^s, \bar{\mathbb{Q}}_j)$ exceeds the threshold derived on the significance level, the auditor can claim $d(\mathbb{Q}_j^s, \bar{\mathbb{Q}}_j)$ deviate from the mean value, \ie, reject $H_0$ and output negative audit result. 
For a set of samples $\left\{d_i \mid i=1,2,\dots,n \right\}$, Grubbs' Test locates the outlier by the procedures. 
\begin{itemize}
\item [1)] Calculate the mean $\mu_{d}$ and standard deviation $\sigma_{d}$.
\item [2)] Calculate the Grubbs' test statistic by $G=\frac{|d(\mathbb{Q}_j^s, \bar{\mathbb{Q}}_j)-\mu_{d}|}{\sigma_{d}}$. 
\item [3)] If $G>\frac{n-1}{\sqrt{n}} \sqrt{\frac{t_{\alpha /(n), n-2}^2}{n-2+t_{\alpha /(n), n-2}^2}}$, $H_0$ is invalid, \ie, the suspect model is not trained by this trajectory. 
In the above inequation, $t_{\alpha /(n), n-2}^2$ represents the upper critical value in the $t$-distribution when the degree of freedom is $n-2$, and the significance level is $\frac{\alpha}{n}$. 
\end{itemize}

Both hypothesis testing methods are based on the assumption that the distance values follow Gaussian distribution. 
Thus, \sysname needs to pre-check that the distance values of the shadow models satisfy the Gaussian distribution. 
We adopts Anderson-Darling test~\cite{stephens1974edf} 
since it fits the scenarios where the auditor has a small number of samplings, and the actual distribution is unknown. 
In the evaluation, all the distance values of the shadow models can pass the Anderson-Darling test due to the randomness of the models' initialization and training. 
After that, \sysname conducts the hypothesis testing. 

\section{Evaluation}
\label{sec:evaluation}
We first introduce the tasks and the experimental setup in \autoref{sec:experimental setup}. 
We validate the effectiveness of \sysname on Behavior Clone and three offline DRL models, \ie, Batch-Constrained Q-learning (BCQ)~\cite{DBLP:conf/icml/FujimotoMP19}, Implicit Q-Learning (IQL)~\cite{DBLP:conf/iclr/KostrikovNL22}, and TD3PlusBC~\cite{DBLP:conf/nips/FujimotoG21} in \autoref{sec:overall performance}. 
Then, we visualize the cumulative rewards by t-SNE~\cite{van2008visualizing} to demonstrate that the cumulative rewards are intrinsic and stable features for dataset auditing in \autoref{sec:visualization of cumulative rewards}. 
After that, we further evaluate the impact of three factors on \sysname, \ie, the number of shadow models, the significance level in hypothesis testing, and the trajectory size in \autoref{sec:evalution-of-parameter-setting}. 
Finally, we utilize \sysname to audit the open-source datasets from Google~\cite{DBLP:journals/corr/abs-2004-07219} and DeepMind~\cite{DBLP:conf/nips/Gulcehre0NPCZAM20} in \autoref{sec:real-world application}. 

\subsection{Experimental Setup}
\label{sec:experimental setup}
\mypara{Tasks}
We adopt Lunar Lander, Bipedal Walker, and Ant tasks in Gym~\cite{DBLP:journals/corr/BrockmanCPSSTZ16}, which are widely used in the prior works~\cite{CGZXL21,IUQJAHN20,PTLBC18}. 
The tasks stem from distinct real-world problems, each with numerical vectors containing different physical information, \eg, position, velocity, and acceleration. 
These tasks involve both discrete and continuous variables in observation and action spaces, with the dimension ranging from low (2-dim) to high (111-dim). 
We give an overview in \autoref{tab:overview-of-tasks} and put their details in \autoref{sec:the details of tasks}. 

\begin{table}[t]
\caption{The Overview of Tasks. The ``continuous'' and ``discrete'' illustrate the data type of the state and action with the corresponding number of dimensions in parentheses. }
\label{tab:overview-of-tasks}
    \setlength{\tabcolsep}{0.8em}
    \renewcommand{\arraystretch}{1.3}
    \footnotesize
    \centering
\begin{tabular}{ccc}
\toprule
\textbf{Task Name}                                                 & \textbf{State Shape} & \textbf{Action Shape} \\ \hline
\begin{tabular}[c]{@{}c@{}}Lunar Lander\\ (Continuous)\end{tabular} & \begin{tabular}[c]{@{}c@{}}Continuous(6-dim)\\ Discrete(2-dim)\end{tabular}          & Continuous(2-dim)      \\ \hline
Bipedal Walker                                                     & Continuous(24-dim)           & Continuous(4-dim)           \\ \hline
Ant                                                                & Continuous(111-dim)          & Continuous(8-dim)      \\ \bottomrule
\end{tabular}
\end{table}

\mypara{Dataset Generation and Offline Model Preparation}
To obtain the datasets for tasks in \autoref{tab:overview-of-tasks}, we adopt the same idea as the existing dataset publishers~\cite{DBLP:conf/nips/Gulcehre0NPCZAM20, DBLP:journals/corr/abs-2004-07219, DBLP:journals/corr/abs-2102-00714, tensorflow2015-whitepaper}, \ie, training the online RL models in the interactive environment and recording the interactions as the datasets. 
The datasets consist of numerical vectors. 
In Lunar Lander, each transition includes state, next state (6-dimensional continuous and 2-dimensional discrete variables), action (2-dimensional continuous variables), and reward (scalar). 
Therefore, each transition is a 19-dimensional numerical vector. 
Similarly, the data types of Bipedal Walker and Ant are 53-dimensional and 231-dimensional numerical vectors, respectively. 
The number of transitions for each task is $5\times10^5$ (Lunar Lander), $10^6$ (Bipedal Walker), and $2\times10^6$ (Ant). 

The offline RL models learn from the datasets. 
\autoref{tab:dataset generation and offline model preparation} summarizes the whole process. 
For each task, we use five global random seeds to train five online models separately. 
We collect the datasets from five online models with random seed 0, where every online model only generates one dataset. 
For ease of reading, the datasets share the same name with their online models. 
We train thirty offline DRL models for every dataset with distinct global random seeds in initialization and optimization processes. 
All the online and offline models are implemented by open-source RL libraries~\cite{stable-baselines3, d3rlpy} with default hyperparameter settings. 

\begin{table}[t]
\centering
\caption{The main steps in dataset generation and offline model preparation with the details of the input and output. }
\label{tab:dataset generation and offline model preparation}
\begin{tabular}{c}
\hline
\multicolumn{1}{|c|}{\multirow{2}{*}{\begin{tabular}[c]{@{}c@{}}For each combination of task and \\  offline RL model in the experiment\end{tabular}}}                  \\
\multicolumn{1}{|c|}{}                                                                                                                                                  \\ \hline
$\downarrow$                                                                                                                                                            \\ \hline
\multicolumn{1}{|c|}{Train with 5 random seeds:  \{0, 1, …, 4\}}                                                                                                        \\ \hline
$\downarrow$                                                                                                                                                            \\ \hline
\multicolumn{1}{|c|}{\textbf{5 online RL models} detailed in \autoref{tab:the online models}}                                                                           \\ \hline
$\downarrow$                                                                                                                                                            \\ \hline
\multicolumn{1}{|c|}{Collect with 1 random seed: \{0\}}                                                                                                                 \\ \hline
$\downarrow$                                                                                                                                                            \\ \hline
\multicolumn{1}{|c|}{\textbf{5x1 offline Datasets} detailed in \autoref{tab:the offline datasets}}                                                                      \\ \hline
$\downarrow$                                                                                                                                                            \\ \hline
\multicolumn{1}{|c|}{Train with 30 random seeds:  \{42, 43, …, 72\}}                                                                                                    \\ \hline
$\downarrow$                                                                                                                                                            \\ \hline
\multicolumn{1}{|c|}{\multirow{2}{*}{\begin{tabular}[c]{@{}c@{}}\textbf{5x1x30 offline RL models} detailed in \\ \autoref{tab:BC models}, \autoref{tab:BCQ models}, \autoref{tab:IQL models}, and \autoref{tab:TD3PlusBC models}\end{tabular}}}  
\\ 
\multicolumn{1}{|c|}{}   
\\ \hline
\end{tabular}
\end{table}

\mypara{Critic Model}
We adopt the fully connected neural network as the critic model, which has four hidden layers with 1024 neurons on each layer. 
We optimize the critic model following the TD-based method in \autoref{sec:the selection of critic model} by Adam optimizer with a learning rate of 0.001 and a mini-batch size of 4096. 
The entire training takes 150 epochs, and the learning rate decays to half every 50 epochs.

\mypara{Evaluation Metrics}
Recalling \sysname's application scenario in \autoref{fig:application scenario}, for a single suspect model, the audit accuracy can well characterize the performance of \sysname, \ie, the ratio of the number of correctly auditing trajectory to the total auditing trajectory. 
In our experiment, the positive models (trained on the target dataset) and the negative models (trained on other datasets) are randomly mixed, where the majority may dominate the accuracy. 
Thus, we provide the true positive rate (TPR) and the true negative rate (TNR). 

\mypara{Methods}
We provide the audit performance of 3$\sigma$ principle and Grubbs' test with four distance metrics, \ie, $\ell_1$ norm, $\ell_2$ norm, Cosine distance, and Wasserstein distance. 

\mypara{Competitors}
Recalling \autoref{sec:existing solutions}, existing methods~\cite{PWZLYS19, 2020PrivAttackAM, DBLP:journals/corr/abs-2109-03975} are designed for the online reinforcement learning scenes, assuming that the auditor can continuously interact with the environment to obtain new data as the non-member example. 
Based on the behavioral difference of the model between the member examples and the non-member examples, they build the member inference method to detect whether an example is used to train the suspect model. 
In the offline scenarios, without access to the environment, the auditor only has the pre-collected target dataset. 
Thus, we randomly divide the target dataset into two parts and train offline RL models on the subsets separately. 
Either subset is regarded as the set of non-member examples for the offline RL models trained on the other subset. 
We adopt the same data augmentation, attack classifier architecture, and hyperparameter settings with~\cite{DBLP:journals/corr/abs-2109-03975}. 

\begin{table}
    \centering
    \caption{The performance of existing membership inference attack against offline DRL models.}
    \label{tab:the performance of competitors}
    \begin{tabular}{cccc} 
    \toprule
    \multirow{2}{*}{\begin{tabular}[c]{@{}c@{}}\textbf{Task}\\\textbf{ Name}\end{tabular}} & \multirow{2}{*}{\begin{tabular}[c]{@{}c@{}}\textbf{Offline}\\\textbf{ Model}\end{tabular}} & \multicolumn{2}{c}{\textbf{Accuracy}}  \\ 
    \cline{3-4}
                                                                                           &                                                                                            & \textbf{Training}      & \textbf{Test} \\ 
    \hline
    \multirow{4}{*}{\begin{tabular}[c]{@{}c@{}}Lunar\\ Lander\end{tabular}}                & BC                                                                                         & 50.09$\pm$0.68 & 48.41$\pm$1.87        \\
                                                                                           & BCQ                                                                                        & 49.84$\pm$1.39 & 47.69$\pm$1.45        \\
                                                                                           & IQL                                                                                        & 49.88$\pm$0.76 & 47.34$\pm$1.83        \\
                                                                                           & TD3PlusBC                                                                                  & 50.08$\pm$0.92 & 48.27$\pm$1.81        \\ 
    \cline{2-4}
    \multirow{4}{*}{\begin{tabular}[c]{@{}c@{}}Bipedal\\ Walker\end{tabular}}              & BC                                                                                         & 50.00$\pm$0.63 & 46.27$\pm$2.42        \\
                                                                                           & BCQ                                                                                        & 49.97$\pm$0.69 & 47.38$\pm$2.41        \\
                                                                                           & IQL                                                                                        & 50.17$\pm$0.95 & 47.19$\pm$1.90        \\
                                                                                           & TD3PlusBC                                                                                  & 49.87$\pm$0.94 & 45.48$\pm$1.46        \\ 
    \cline{2-4}
    \multirow{4}{*}{Ant}                                                                   & BC                                                                                         & 50.44$\pm$0.64 & 46.74$\pm$2.37        \\
                                                                                           & BCQ                                                                                        & 50.22$\pm$0.52 & 45.38$\pm$2.16        \\
                                                                                           & IQL                                                                                        & 50.33$\pm$0.35 & 45.89$\pm$1.90        \\
                                                                                           & TD3PlusBC                                                                                  & 50.13$\pm$0.67 & 45.03$\pm$1.55        \\
    \bottomrule
\end{tabular}
\end{table}

\mypara{Implementation}
We use stable-baselines~\cite{stable-baselines3} and d3rlpy~\cite{d3rlpy} to implement online and offline DRL models separately. 
All audit methods are realized with Python 3.8 on a server with 8 NVIDIA GeForce RTX 3090 and 512GB memory. 
\subsection{Overall Audit Performance}
\label{sec:overall performance}
\begin{table*}[t]
\centering
\caption{The TPR and TNR results based on Grubbs' test. 
The mean and standard deviation of TPR and TNR in each row represent the audit results for one combination of task and model by four distance metrics. 
Bold indicates the highest sum of TPR and TNR, \ie, accuracy, in a row. 
Each pair of TPR and TNR is derived from the diagonal and non-diagonal values of the corresponding heatmap in \autoref{fig:audit result on lunarlander}, \autoref{fig:audit result on bipedalwalker} and \autoref{fig:audit result on ant}, which are supplementary to~\cite{DCSJCCZ24}. 
}
\label{tab:overall audit accuracy based on Grubbs}
\setlength{\tabcolsep}{0.8em}
\renewcommand{\arraystretch}{1.1}
\footnotesize
\begin{tabular}{cccccccccc} 
\toprule
\multirow{2}{*}{\begin{tabular}[c]{@{}c@{}}\textbf{Task}\\\textbf{ Name}\end{tabular}} & \multirow{2}{*}{\begin{tabular}[c]{@{}c@{}}\textbf{Offline}\\\textbf{ Model}\end{tabular}} & \multicolumn{2}{c}{\textbf{L1 Norm}}                                                 & \multicolumn{2}{c}{\textbf{L2 Norm}}                               & \multicolumn{2}{c}{\begin{tabular}[c]{@{}c@{}}\textbf{Cosine}\\\textbf{ Distance}\end{tabular}} & \multicolumn{2}{c}{\begin{tabular}[c]{@{}c@{}}\textbf{Wasserstein}\\\textbf{ Distance}\end{tabular}}  \\ 
\cline{3-10}
                                                                                       &                                                                                            & \multicolumn{1}{c}{TPR}                  & \multicolumn{1}{c}{TNR}                   & \multicolumn{1}{c}{TPR}         & \multicolumn{1}{c}{TNR}          & \multicolumn{1}{c}{TPR}                  & \multicolumn{1}{c}{TNR}                              & \multicolumn{1}{c}{TPR}                  & \multicolumn{1}{c}{TNR}                                    \\ 
\hline
\multirow{4}{*}{\begin{tabular}[c]{@{}c@{}}Lunar\\ Lander\end{tabular}}                & BC                                                                                         & \textbf{99.01$\pm$0.46} & \textbf{100.00$\pm$0.00} & 96.96$\pm$0.73 & 100.00$\pm$0.00 & 96.93$\pm$0.77          & 100.00$\pm$0.00                     & 98.40$\pm$0.74          & 99.94$\pm$0.16                            \\
                                                                                       & BCQ                                                                                        & \textbf{98.29$\pm$1.14} & \textbf{100.00$\pm$0.00} & 96.03$\pm$1.15 & 100.00$\pm$0.00 & 95.97$\pm$1.07          & 99.99$\pm$0.04                      & 97.57$\pm$1.17          & 99.91$\pm$0.14                            \\
                                                                                       & IQL                                                                                        & \textbf{98.61$\pm$1.51} & \textbf{99.91$\pm$0.32}  & 97.52$\pm$2.51 & 99.97$\pm$0.12  & 97.49$\pm$2.56          & 99.92$\pm$0.19                      & 98.32$\pm$1.79          & 97.10$\pm$5.66                            \\
                                                                                       & TD3PlusBC                                                                                  & \textbf{98.29$\pm$2.04} & \textbf{99.48$\pm$0.79}  & 96.35$\pm$3.01 & 99.89$\pm$0.22  & 96.27$\pm$3.16          & 99.91$\pm$0.23                      & 98.53$\pm$1.25          & 95.59$\pm$3.77                            \\ 
\cline{2-10}
\multirow{4}{*}{\begin{tabular}[c]{@{}c@{}}Bipedal\\ Walker\end{tabular}}              & BC                                                                                         & 99.20$\pm$1.47          & 100.00$\pm$0.00          & 98.40$\pm$2.70 & 100.00$\pm$0.00 & 98.56$\pm$2.68          & 100.00$\pm$0.00                     & \textbf{99.31$\pm$1.32} & \textbf{100.00$\pm$0.00}                  \\
                                                                                       & BCQ                                                                                        & 99.52$\pm$0.77          & 100.00$\pm$0.00          & 98.16$\pm$2.89 & 100.00$\pm$0.00 & 99.87$\pm$0.15          & 100.00$\pm$0.00                     & \textbf{99.89$\pm$0.13} & \textbf{100.00$\pm$0.00}                  \\
                                                                                       & IQL                                                                                        & 95.10$\pm$7.41          & 100.00$\pm$0.00          & 95.04$\pm$5.45 & 100.00$\pm$0.00 & \textbf{99.84$\pm$0.32} & \textbf{100.00$\pm$0.00}            & 95.01$\pm$6.72          & 100.00$\pm$0.00                           \\
                                                                                       & TD3PlusBC                                                                                  & \textbf{99.36$\pm$1.28} & \textbf{94.77$\pm$19.42} & 97.15$\pm$5.71 & 93.36$\pm$21.46 & 96.96$\pm$5.82          & 91.98$\pm$21.75                     & 98.08$\pm$3.84          & 88.26$\pm$25.34                           \\ 
\cline{2-10}
\multirow{4}{*}{Ant}                                                                   & BC                                                                                         & 97.42$\pm$1.66          & 99.94$\pm$0.11           & 96.48$\pm$1.66 & 99.90$\pm$0.36  & 99.20$\pm$1.08          & 85.66$\pm$28.23                     & \textbf{98.00$\pm$1.19} & \textbf{99.92$\pm$0.14}                   \\
                                                                                       & BCQ                                                                                        & 97.17$\pm$2.96          & 99.80$\pm$0.43           & 95.68$\pm$2.54 & 99.84$\pm$0.43  & 99.66$\pm$0.43          & 86.70$\pm$26.89                     & \textbf{98.67$\pm$1.65} & \textbf{99.79$\pm$0.46}                   \\
                                                                                       & IQL                                                                                        & 97.20$\pm$2.33          & 99.66$\pm$0.73           & 96.61$\pm$2.50 & 99.69$\pm$0.59  & 99.57$\pm$0.79          & 86.25$\pm$27.90                     & \textbf{99.36$\pm$0.42} & \textbf{99.63$\pm$0.78}                   \\
                                                                                       & TD3PlusBC                                                                                  & 98.53$\pm$1.80          & 99.18$\pm$1.72           & 97.17$\pm$1.79 & 99.35$\pm$1.74  & 99.72$\pm$0.40          & 87.79$\pm$26.43                     & \textbf{99.25$\pm$1.24} & \textbf{99.14$\pm$1.81}                   \\
\bottomrule
\end{tabular}
\end{table*}

We assess the effectiveness of \sysname across twelve combinations of three tasks and four models. 
Furthermore, we present an evaluation of the efficacy of the competitors on offline DRL models. 

\mypara{Setup}
From \autoref{tab:dataset generation and offline model preparation}, we train 30 offline RL models for each dataset and obtain 150 offline DRL models for every experimental setting. 
We audit the 5 datasets separately, where the auditor randomly selects 15 models from the target dataset as the shadow models, and the remaining 15 models along with the 120 models from other datasets are the positive and the negative suspect models. 
For the target dataset, we randomly select fifty auditing trajectories to audit. 
Since the unbalanced amount of the positive and the negative models, we report the aggregated mean with a standard deviation of both TPR and TNR for each setting in \autoref{tab:overall audit accuracy based on Grubbs} and provide the audit results between every two datasets in \autoref{fig:audit result on lunarlander} (Lunar Lander), \autoref{fig:audit result on bipedalwalker} (Bipedal Walker), and \autoref{fig:audit result on ant} (Ant). 
Each pair of TPR and TNR  in \autoref{tab:overall audit accuracy based on Grubbs} is derived from the diagonal and non-diagonal values of the corresponding heatmap. 
As a supplementary of~\cite{DCSJCCZ24}, we also show the audit result by 3$\sigma$ principle in \autoref{tab:overall audit accuracy based on 3sigma}. 
The competitors' performance is shown in \autoref{tab:the performance of competitors}, where the values of mean and standard variation are calculated by repeating experiment ten times. 

\mypara{Observations}
We have the following observations from \autoref{tab:overall audit accuracy based on Grubbs}, \autoref{tab:overall audit accuracy based on 3sigma}, and \autoref{tab:the performance of competitors}. 
1) Most TPR and TNR values are higher than 95\%, 
meaning that \sysname is a valid solution to audit the learned dataset of the offline DRL models. 
For instance, all results for \sysname with $\ell_1$ norm are beyond $94\%$ across the experiment settings. 

2) \sysname obtains different audit accuracy over four distance metrics. 
The audit effectiveness with $\ell_1$ norm and Wasserstein distance is better than that of $\ell_2$ norm and Cosine distance. 
In \autoref{tab:overall audit accuracy based on Grubbs} and \autoref{tab:overall audit accuracy based on 3sigma}, \sysname with Wasserstein distance always performs the best or the second place. 
And results of $\ell_2$ norm are usually behind the other three distance metrics. 
Recalling \autoref{sec:the details of audit process}, Wasserstein distance characterizes both the numerical and the positional deviations of the cumulative rewards, which is more sensitive. 
Since the numerical differences between the cumulative rewards are slight, \eg, from 0.01 to 0.1 in our experiment, $\ell_2$ norm may undercut these small but potential differences. 

3) The accuracy of the audit as determined by Grubbs' test outperforms that of the 3$\sigma$ principle.
The 3$\sigma$ principle is an empirical method, which is easily misled by the outlier cumulative rewards of the shadow models. 
Recalling \autoref{sec:the details of audit process}, Grubbs' test first calculates the statistic $G$ and compares $G$ with an adaptive threshold, where the number of samples is also considered in the hypothesis testing. 

4) Without the new data from the environment, the effectiveness of the existing membership inference methods is attenuated. 
From one perspective, the similarity between sub-datasets splited from the same dataset can result in the trained RL models exhibiting undifferentiated behavior, making it difficult to effectively distinguish between members and non-members.
On the other hand, when considering the results presented in \autoref{fig:full behavior similarity}, 
we conclude that the actions of RL models should not be directly utilized as the foundation for membership inference. 

\subsection{Visualization of Cumulative Rewards}
\label{sec:visualization of cumulative rewards}
To further explain the audit results in \autoref{sec:overall performance}, we analyze the cumulative rewards from the shadow models and the suspect models, \ie, $\mathbb{Q}_j^i$ and $\mathbb{Q}_j^s$, by using t-SNE~\cite{van2008visualizing}. 

\begin{figure*}[t]
    \centering
    \includegraphics[width=\hsize]{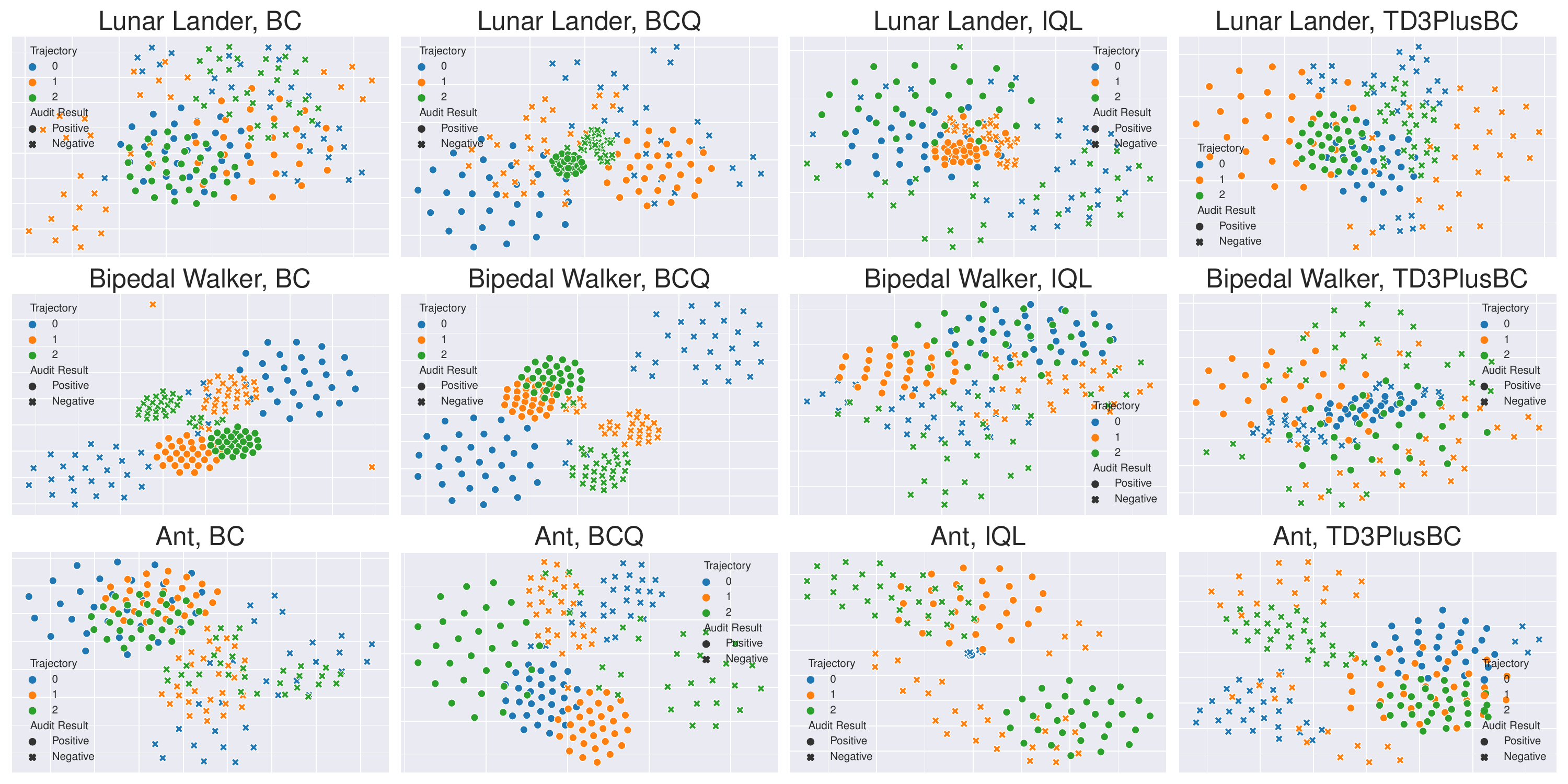}
    \caption{Visualization of cumulative rewards by t-SNE. 
        The caption of each plot demonstrates the offline DRL model's type and task. 
        In a single plot, we randomly select three trajectories from the first dataset for the task, \ie, Lunar Lander dataset 1171, Bipedal Walker dataset 0841, and Ant dataset 2232 in \autoref{tab:the offline datasets}, and then show the cumulative rewards from 30 positive models and 30 negative models for each trajectory. }
    \label{fig:visualization of cumulative rewards by t-SNE}
\end{figure*}

\mypara{Setup}
The caption of each plot in \autoref{fig:visualization of cumulative rewards by t-SNE} indicates the used task and offline DRL model. 
Each point in the plots shows the visualization of a single $\mathbb{Q}_j^i$ (positive) or $\mathbb{Q}_j^s$ (negative). 
In a single plot, we demonstrate the results of three trajectories from each tasks' first datasets. 
For instance, the target dataset of the plot titled ``Lunar Lander, BC'' is dataset ``1171'' in \autoref{tab:the offline datasets}. 
The thirty positive points for each trajectory are collected from the shadow models trained on dataset ``1171'', while the thirty negative points are randomly sampled from the shadow models from the other four datasets. 

\mypara{Observations}
From \autoref{fig:visualization of cumulative rewards by t-SNE}, we have the following observations.  
1) For a trajectory of the target dataset, the cumulative rewards from the shadow models and the suspect models are clearly divided into different groups, meaning that the critic model well reflects the differences in the models' actions. 
Thus, the cumulative reward generated by the critic model is a qualified post-event fingerprint for trajectory-level auditing. 

2) The distribution of points varies on the different trajectories. 
For example, trajectory 1 from the Lunar Lander dataset is harder to cluster than the other two trajectories. 
We speculate that this is because trajectory 1 represents a basic policy, \eg, a local optimum policy to fire the lander's thrusters all the way, and similar trajectories exist in the other four datasets. 
Due to the non-uniqueness of the optimal strategy in RL problems and the impact of randomness in the model training process, the collected trajectories have unique characteristics. 
Thus, other trajectories' cumulative rewards are clearly divided. 

\subsection{Hyperparameter Study}
\label{sec:evalution-of-parameter-setting}
We extend our assessment to scrutinize three pivotal determinants that impact the pragmatic integration of \sysname. 
Specifically, we consider the amount of shadow models, the level of significance in hypothesis testing, and the magnitude of the trajectory size. 
Due to space limitations, we only give brief conclusions in this section. 
Please refer to the specific analysis in \autoref{sec:impact of shadow models' amount}, \autoref{sec:impact of significance level}, and \autoref{sec:impact_of_trajectory_size}. 

\mypara{Impact of Shadow Models' Amount}
\begin{figure*}[ht]
    \centering
    \includegraphics[width=\hsize]{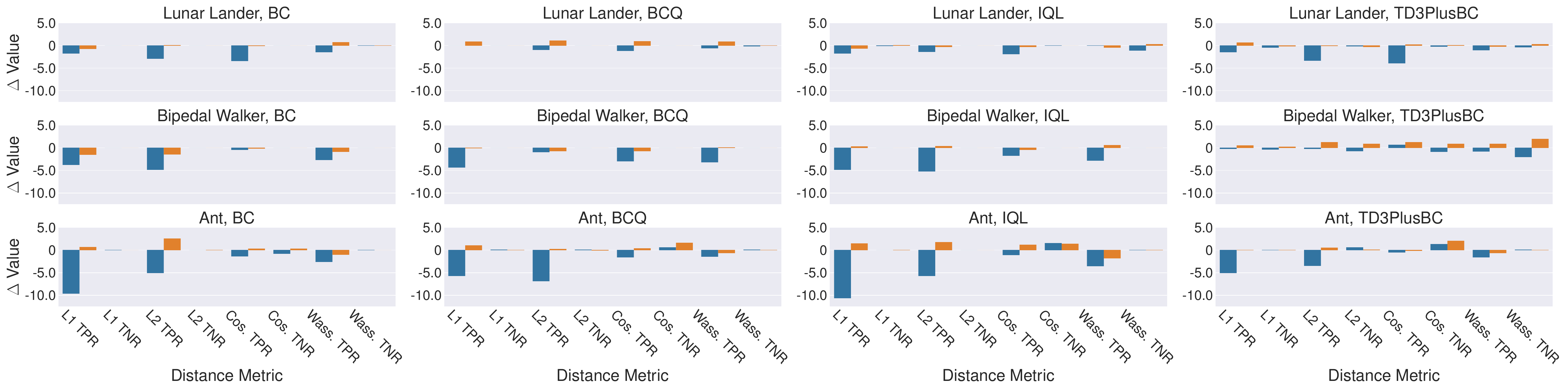}
    \includegraphics[width=0.35\hsize]{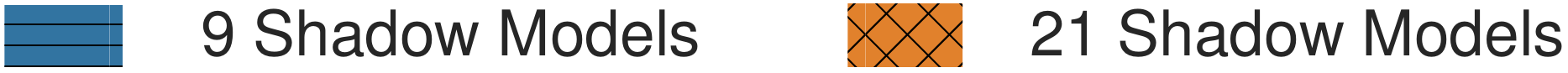}
    \caption{Impact of shadow models’ amount. 
    The change value of TPR and TNR when the number of shadow models varies to 9 and 21 compared to the default 15 shadow models. 
    The caption of each plot demonstrates the offline DRL model's type and task. 
    The x labels display the four distance metrics. 
    The y labels show the absolute fluctuating values of TPR and TNR. } 
    \label{fig:impact of shadow models' amount}
\end{figure*}
We change the shadow models' amount to 9 and 21 with the other settings the same as \autoref{sec:overall performance}. 
\autoref{fig:impact of shadow models' amount} shows the value change of TPR and TNR compared with that of 15 shadow models. 
Each figure's title illustrates the settings of the model and the task, the x-axis indicates the four metrics, and the y-axis is the absolute value change. 
As a supplementary of~\cite{DCSJCCZ24}, we provide the detailed results in \autoref{tab:overall audit accuracy based on Grubbs (9 Shadow Models)} (9 Shadow Models) and \autoref{tab:overall audit accuracy based on Grubbs (21 Shadow Models)} (21 Shadow Models). 

From \autoref{fig:impact of shadow models' amount}, we have the following observations. 
1) The audit accuracy increases with a larger amount of shadow models. 
2) There exists a saturation point for audit accuracy with the expansion of shadow models. 

\mypara{Impact of Significance Level}
\begin{figure*}[ht]
    \centering
    \includegraphics[width=\hsize]{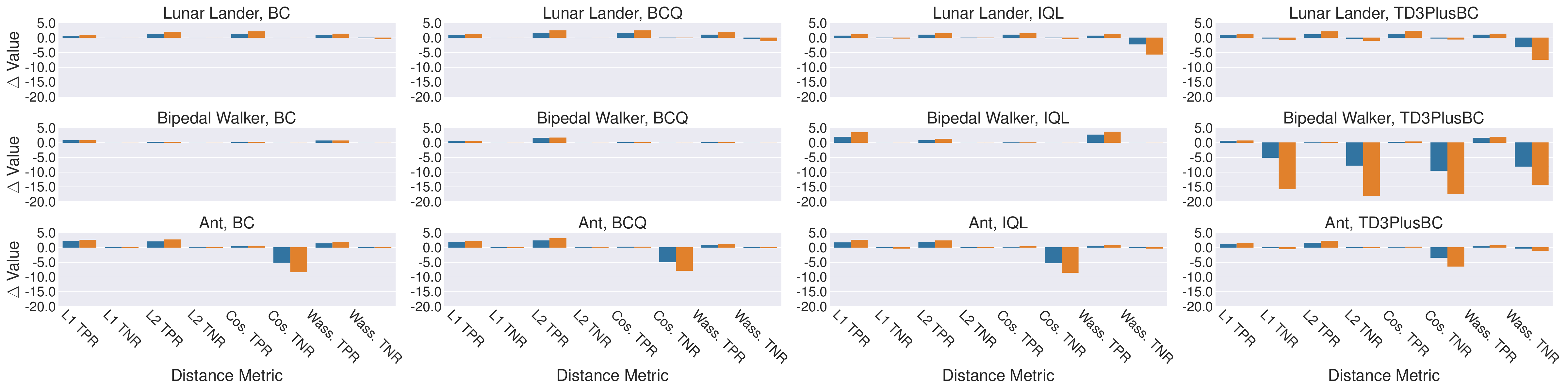}
    \includegraphics[width=0.3\hsize]{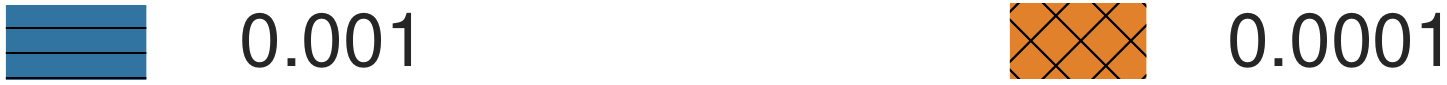}
    \caption{Impact of the significance level. 
    The change value of TPR and TNR when the significance level varies to 0.001 and 0.0001 compared to the default 0.01. 
    The caption of each plot demonstrates the offline DRL model's type and task. 
    The x labels display the four distance metrics. 
    The y labels show the absolute fluctuating values of TPR and TNR. 
    }
    \label{fig:impact of significance level}
\end{figure*}
The significance level represents the auditor's confidence in the auditing results. 
In \autoref{sec:overall performance}, we adopt the significance level $\alpha=0.01$, meaning that the auditor has 99\% confidence in the judgments. 
Generally speaking, the significance level represents the maximum audit capacity of \sysname instead of a hyperparameter setting since it is an audit requirement by the dataset owner. 
We demand the auditor to output a more confident judgment, where the error possibility should be limited to 1\textperthousand~and 0.1\textperthousand, \ie, $\alpha=0.001$ and $\alpha=0.0001$. 
\autoref{fig:impact of significance level} shows the value change of TPR and TNR compared with that when $\alpha=0.01$. 
As a supplementary of~\cite{DCSJCCZ24}, 
the detailed results between every two datasets are in \autoref{tab:overall audit accuracy based on Grubbs (sigma=0.001)} ($\alpha=0.001$) and \autoref{tab:overall audit accuracy based on Grubbs (sigma=0.0001)} ($\alpha=0.0001$). 

From \autoref{fig:impact of significance level}, we have the following observations.  
1) For a complicated task, we recommend the auditor select a large significance level for \sysname. 
2) For the suspect models with low performance, \sysname should adopt a large significance level to guarantee audit accuracy. 
3) In general, $\alpha=0.01$ is a safe bound of \sysname, and a lower $\alpha$ may break through the capability boundary of \sysname, inducing the auditor to misclassify the negative model to the positive set. 

\mypara{Impact of Trajectory Size}
\begin{figure*}[ht]
    \centering
    \includegraphics[width=\hsize]{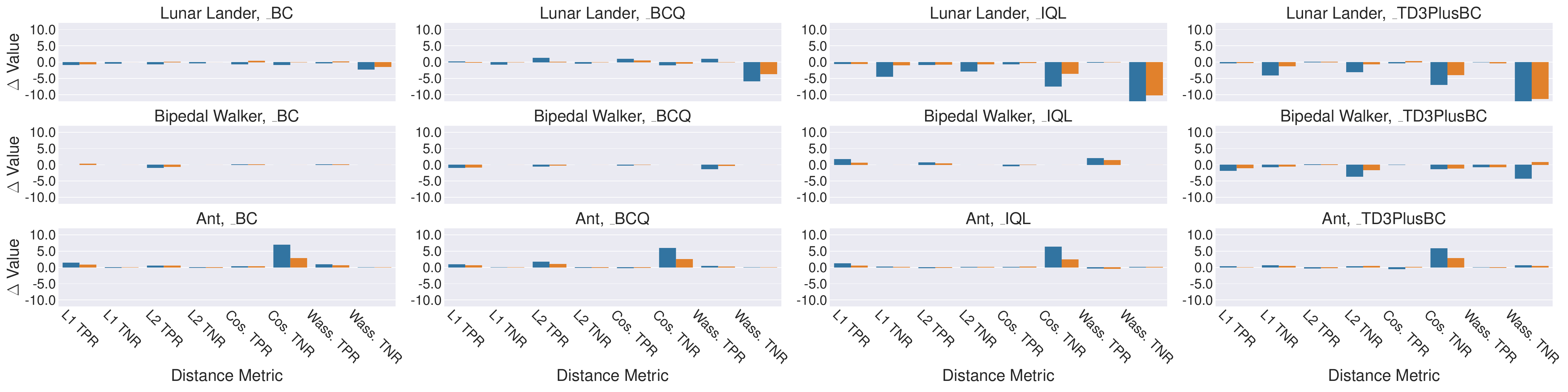}
    \includegraphics[width=0.4\hsize]{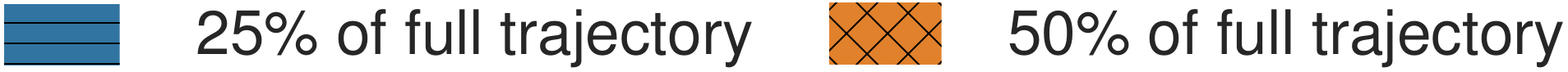}
    \caption{Impact of the trajectory size. 
    The change value of TPR and TNR when the trajectory size varies to 25\% and 50\% compared to the entire trajectories (100\%). 
    The caption of each plot demonstrates the offline DRL model's type and task. 
    The x labels display the four distance metrics. 
    The y labels show the absolute fluctuating values of TPR and TNR. 
    }
    \label{fig:impact_of_trajectory_size}
    \vspace{-0.2cm}
\end{figure*}
We investigate the relationship between the trajectory size and audit accuracy. 
In \autoref{sec:overall performance}, we adopt the full-length trajectory, meaning that the auditor utilizes all states of each trajectory to query the suspect model and obtains the corresponding actions to conduct the dataset auditing. 
We change the trajectory size to 25\% and 50\% of the full length with the other settings the same as \autoref{sec:overall performance}. 
\autoref{fig:impact_of_trajectory_size} shows the value change of TPR and TNR compared with that of the full-length trajectory. 
As a supplementary of~\cite{DCSJCCZ24}, we also provide the detailed results in \autoref{tab:overall_audit_accuracy_based_on_Grubbs_0.25_trajsize} (25\%) and \autoref{tab:overall_audit_accuracy_based_on_Grubbs_0.5_trajsize} (50\%). 

From \autoref{fig:impact_of_trajectory_size}, we have the following observations. 
1) \sysname tends to achieve higher accuracy with a larger trajectory size. 
2) A small trajectory size achieves better results under some tasks since the front states of each trajectory are able to reflect more behavioral information of the model~\cite{DBLP:journals/corr/abs-2007-09055}. 

\subsection{Real-world Application}
\label{sec:real-world application}
In this section, we apply \sysname to audit the open-source datasets from DeepMind~\cite{DBLP:conf/nips/Gulcehre0NPCZAM20} and Google~\cite{DBLP:journals/corr/abs-2004-07219}. 
We choose the ``halfcheetah'' task published by both, where the operator controls a 2-dimensional cheetah robot consisting of 9 links and 8 joints connecting them (including two paws) to make the cheetah run forward (right) as fast as possible. 
The details of the halfcheetah dataset and the offline DRL models are in \autoref{tab:details of the halfcheetah dataset} and \autoref{tab:details of the models trained on the halfcheetah dataset}. 
All experimental settings are consistent with these in \autoref{sec:overall performance}. 

\mypara{Observations}
From \autoref{tab:audit result on halfcheetah}, we have the following observations. 
1) \sysname can be effective in real-world applications. 
The TPR and TNR of \sysname exceed 95\% with $\ell_1$ norm and Wasserstein distance, meaning that \sysname remains valid for the existing open-source datasets. 
2) Wasserstein distance has stable performance on the experimental and the real-world datasets. 
The overall accuracy of \sysname with Wasserstein distance are all higher than the other three metrics. 

\section{Robustness}
\subsection{Ensemble Architecture}
\label{sec:ensemble architecture}
\begin{figure*}[!ht]
    \centering
    \includegraphics[width=\hsize]{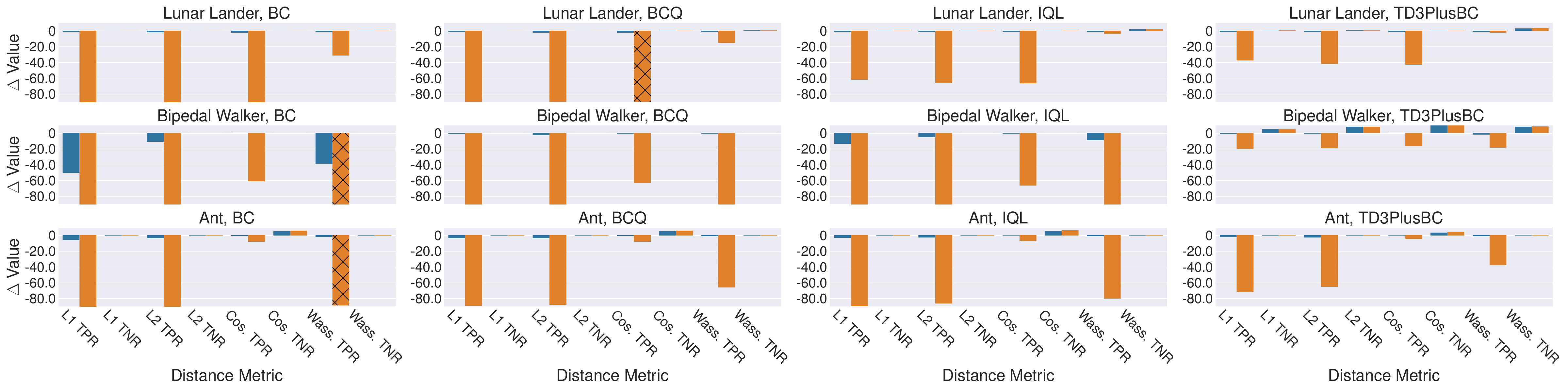}
    \includegraphics[width=0.3\hsize]{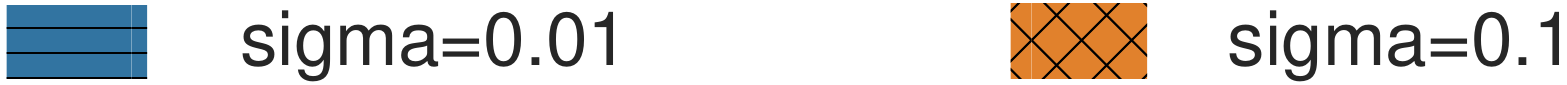}
    \caption{Robustness against action distortion. 
    The change value of TPR and TNR when the suspect model adds Gaussian noise parameterized with $(\mu=0, \sigma=0.01)$ and $(\mu=0, \sigma=0.1)$ to its output. 
    The caption of each plot demonstrates the offline DRL model's type and task. 
    The x labels display the four distance metrics. 
    The y labels show the absolute fluctuating values. }
    \label{fig:robustness}
    \vspace{-0.4cm}
\end{figure*}
To hinder the audit of a dataset, an adversary may utilize state-of-the-art membership inference defense strategies proposed in recent research works~\cite{DBLP:conf/uss/TangMSSNHM22, DBLP:journals/popets/JarinE23}. 
These defense strategies aim to mitigate the influence of a member example on the behavior of a machine learning model.
Based on the idea of model ensemble, in particular, \cite{DBLP:conf/uss/TangMSSNHM22, DBLP:journals/popets/JarinE23, CZWBHZ22} proposed to split the training set into several subsets and train sub-models on each of these subsets. 
Then, when an auditor uses an example from the target dataset to query a suspect model, the adversary aggregates the outputs of the sub-models that have not been trained on this example.

\mypara{Setup}
The number of divided subsets, denoted by $K$, represents a crucial hyperparameter for ensemble-based methods, as discussed in~\cite{DBLP:conf/uss/TangMSSNHM22, DBLP:journals/popets/JarinE23}. 
Considering the analysis conducted in these studies, as well as the size of the offline RL datasets, 
we have established $K=5$ for the present investigation. 
All other experimental settings remain unchanged from those described in Section \ref{sec:overall performance}, 
and the corresponding audit outcomes are presented in \autoref{tab:overall audit accuracy against model ensemble}. 
As a supplementary of~\cite{DCSJCCZ24}, 
the results between every two datasets are in \autoref{fig:model ensemble k5-2 on lunarlander} (Lunar Lander), \autoref{fig:model ensemble k5-2 on bipedalwalker} (Bipedal Walker), \autoref{fig:model ensemble k5-2 on ant} (Ant), and \autoref{fig:model ensemble k5-2 on halfcheetah} (Half Cheetah). 

\begin{table*}
\centering
\caption{The TPR and TNR results of \sysname against model ensemble ($K=5$). 
The mean and standard deviation of TPR and TNR in each row represent the audit results for one combination of task and model by four distance metrics. 
Each pair of TPR and TNR is derived from the diagonal and non-diagonal values of the corresponding heatmap in \autoref{fig:model ensemble k5-2 on lunarlander} (Lunar Lander), \autoref{fig:model ensemble k5-2 on bipedalwalker} (Bipedal Walker), \autoref{fig:model ensemble k5-2 on ant} (Ant), and \autoref{fig:model ensemble k5-2 on halfcheetah} (Half Cheetah), which are supplementary to~\cite{DCSJCCZ24}. 
}

\label{tab:overall audit accuracy against model ensemble}
\setlength{\tabcolsep}{0.6em}
\renewcommand{\arraystretch}{1.2}
\footnotesize
\begin{tabular}{cccccccccc} 
\toprule
\multirow{2}{*}{\begin{tabular}[c]{@{}c@{}}\textbf{Task}\\\textbf{ Name}\end{tabular}} & \multirow{2}{*}{\begin{tabular}[c]{@{}c@{}}\textbf{Offline}\\\textbf{ Model}\end{tabular}} & \multicolumn{2}{c}{\textbf{L1 Norm}}                                                                        & \multicolumn{2}{c}{\textbf{L2 Norm}}                                                                        & \multicolumn{2}{c}{\begin{tabular}[c]{@{}c@{}}\textbf{Cosine}\\\textbf{ Distance}\end{tabular}}             & \multicolumn{2}{c}{\begin{tabular}[c]{@{}c@{}}\textbf{Wasserstein}\\\textbf{ Distance}\end{tabular}}         \\ 
\cline{3-10}
                                                                                       &                                                                                            & TPR                                                  & TNR                                                  & TPR                                                  & TNR                                                  & TPR                                                  & TNR                                                  & TPR                                                  & TNR                                                   \\ 
\hline
\multirow{4}{*}{\begin{tabular}[c]{@{}c@{}}Lunar\\ Lander\end{tabular}}                & BC                                                                                         & 100.00$\pm$0.00                     & 100.00$\pm$0.00                     & 99.20$\pm$0.98                      & 100.00$\pm$0.00                     & 99.20$\pm$0.98                      & 100.00$\pm$0.00                     & 99.60$\pm$0.80                      & 99.90$\pm$0.44                       \\
                                                                                       & BCQ                                                                                        & 99.60$\pm$0.80                      & 100.00$\pm$0.00                     & 98.00$\pm$2.19                      & 100.00$\pm$0.00                     & 98.00$\pm$2.19                      & 100.00$\pm$0.00                     & 99.60$\pm$0.80                      & 100.00$\pm$0.00                      \\
                                                                                       & IQL                                                                                        & 100.00$\pm$0.00                     & 99.90$\pm$0.44                      & 99.20$\pm$0.98                      & 100.00$\pm$0.00                     & 99.60$\pm$0.80                      & 99.90$\pm$0.44                      & 99.60$\pm$0.80                      & 97.60$\pm$4.27                       \\
                                                                                       & TD3PlusBC                                                                                  & 100.00$\pm$0.00                     & 99.30$\pm$0.95                      & 99.60$\pm$0.80                      & 99.90$\pm$0.44                      & 99.60$\pm$0.80                      & 99.80$\pm$0.60                      & 99.60$\pm$0.80                      & 95.80$\pm$3.57                       \\ 
\cline{2-10}
\multirow{4}{*}{\begin{tabular}[c]{@{}c@{}}Bipedal\\ Walker\end{tabular}}              & BC                                                                                         & 100.00$\pm$0.00                     & 100.00$\pm$0.00                     & 100.00$\pm$0.00                     & 100.00$\pm$0.00                     & 100.00$\pm$0.00                     & 100.00$\pm$0.00                     & 100.00$\pm$0.00                     & 100.00$\pm$0.00                      \\
                                                                                       & BCQ                                                                                        & 100.00$\pm$0.00                     & 100.00$\pm$0.00                     & 100.00$\pm$0.00                     & 100.00$\pm$0.00                     & 100.00$\pm$0.00                     & 100.00$\pm$0.00                     & 100.00$\pm$0.00                     & 100.00$\pm$0.00                      \\
                                                                                       & IQL                                                                                        & 100.00$\pm$0.00                     & 100.00$\pm$0.00                     & 100.00$\pm$0.00                     & 100.00$\pm$0.00                     & 100.00$\pm$0.00                     & 100.00$\pm$0.00                     & 100.00$\pm$0.00                     & 100.00$\pm$0.00                      \\
                                                                                       & TD3PlusBC                                                                                  & 100.00$\pm$0.00                     & 94.90$\pm$19.07                     & 100.00$\pm$0.00                     & 93.80$\pm$21.63                     & 100.00$\pm$0.00                     & 92.70$\pm$21.62                     & 100.00$\pm$0.00                     & 89.20$\pm$23.94                      \\ 
\cline{2-10}
\multirow{4}{*}{Ant}                                                                   & BC                                                                                         & 99.60$\pm$0.80                      & 100.00$\pm$0.00                     & 99.60$\pm$0.80                      & 99.90$\pm$0.44                      & 99.60$\pm$0.80                      & 83.20$\pm$31.99                     & 99.20$\pm$1.60                      & 100.00$\pm$0.00                      \\
                                                                                       & BCQ                                                                                        & 100.00$\pm$0.00                     & 99.70$\pm$0.71                      & 99.60$\pm$0.80                      & 99.80$\pm$0.60                      & 100.00$\pm$0.00                     & 85.70$\pm$28.31                     & 100.00$\pm$0.00                     & 99.70$\pm$0.71                       \\
                                                                                       & IQL                                                                                        & 100.00$\pm$0.00                     & 99.80$\pm$0.60                      & 99.20$\pm$0.98                      & 99.70$\pm$0.71                      & 99.20$\pm$0.98                      & 86.80$\pm$28.32                     & 100.00$\pm$0.00                     & 99.80$\pm$0.60                       \\
                                                                                       & TD3PlusBC                                                                                  & 99.60$\pm$0.80                      & 99.30$\pm$1.82                      & 100.00$\pm$0.00                     & 99.40$\pm$2.20                      & 100.00$\pm$0.00                     & 87.80$\pm$25.87                     & 99.60$\pm$0.80                      & 98.50$\pm$3.79                       \\ 
\cline{2-10}
\multirow{4}{*}{\begin{tabular}[c]{@{}c@{}}Half\\Cheetah\end{tabular}}                 & BC                                                                                         & \multicolumn{1}{l}{85.00$\pm$25.98} & \multicolumn{1}{l}{100.00$\pm$0.00} & \multicolumn{1}{l}{84.50$\pm$25.71} & \multicolumn{1}{l}{100.00$\pm$0.00} & \multicolumn{1}{l}{94.00$\pm$10.39} & \multicolumn{1}{l}{67.50$\pm$43.20} & \multicolumn{1}{l}{87.00$\pm$21.38} & \multicolumn{1}{l}{100.00$\pm$0.00}  \\
                                                                                       & BCQ                                                                                        & \multicolumn{1}{l}{91.00$\pm$15.59} & \multicolumn{1}{l}{100.00$\pm$0.00} & \multicolumn{1}{l}{89.00$\pm$16.76} & \multicolumn{1}{l}{100.00$\pm$0.00} & \multicolumn{1}{l}{95.00$\pm$8.66}  & \multicolumn{1}{l}{67.17$\pm$42.30} & \multicolumn{1}{l}{93.00$\pm$12.12} & \multicolumn{1}{l}{100.00$\pm$0.00}  \\
                                                                                       & IQL                                                                                        & \multicolumn{1}{l}{90.00$\pm$12.81} & \multicolumn{1}{l}{100.00$\pm$0.00} & \multicolumn{1}{l}{86.50$\pm$16.70} & \multicolumn{1}{l}{100.00$\pm$0.00} & \multicolumn{1}{l}{94.50$\pm$9.53}  & \multicolumn{1}{l}{71.00$\pm$41.37} & \multicolumn{1}{l}{91.50$\pm$12.52} & \multicolumn{1}{l}{100.00$\pm$0.00}  \\
                                                                                       & TD3PlusBC                                                                                  & \multicolumn{1}{l}{61.50$\pm$20.32} & \multicolumn{1}{l}{100.00$\pm$0.00} & \multicolumn{1}{l}{77.00$\pm$19.42} & \multicolumn{1}{l}{100.00$\pm$0.00} & \multicolumn{1}{l}{95.00$\pm$8.66}  & \multicolumn{1}{l}{65.67$\pm$41.28} & \multicolumn{1}{l}{52.00$\pm$33.26} & \multicolumn{1}{l}{100.00$\pm$0.00}  \\
\bottomrule
\end{tabular}
\end{table*}

\mypara{Observations}
We conclude the following observations based on the above results. 
1) Even when faced with ensemble architecture, \sysname maintains a high level of audit accuracy. 
As shown in \autoref{tab:overall audit accuracy against model ensemble}, both TPR and TNR consistently exceed 80\%. 
As described in \autoref{sec:workflow}, \sysname uses predicted cumulative rewards from the critic model as the basis for auditing. 
During training, the critic model captures the overall features of the dataset distribution, instead of memorizing features from individual samples. Since the ensemble model is trained on the target dataset, its behavior embeds the distribution characteristics of the dataset, which \sysname can detect.

2) The use of ensemble architecture may result in a decrease in model performance for certain tasks. 
Our experimental results, as shown in column ``Model Performance (Model Ensemble)'' of Tables \autoref{tab:BC models}, \autoref{tab:BCQ models}, \autoref{tab:IQL models}, and \autoref{tab:TD3PlusBC models}, demonstrate a decline in the performance of offline RL models when utilizing ensemble architecture. 
For instance, when BCQ models learn from the Ant dataset ``3569'', the mean values of cumulative reward decrease significantly. 
Furthermore, due to the sub-models being trained on subsets of data, they only fit a partial dataset's distribution. 
Consequently, when applying the model ensemble to practical scenarios, the standard deviations of the model's performance are large. 

\subsection{Action Distortion} 
\label{sec:perturbing models output}
The suspect models may perturb the actions, \ie, changing the original models' outputs, to conceal its training dataset in practice. 
The action distortion mechanism should be stealthy and cannot be detected by the auditor easily. 
Considering that the DRL models are usually applied to real-world decision-making tasks, such as self-driving cars and industry automation~\cite{neptune, he2022collaborative}, the natural distortion is often modeled as Gaussian noise. 
For example, thermal noise, which is caused by the random motion of electrons in a conductor, can be modeled as a Gaussian noise with a constant power spectrum~\cite{ALHINAI20201}. 
In addition, Gaussian noise is easy to manipulate mathematically. 
For ease of evaluating the effects of different distortion intensities, all dimensions of the models' action space are normalized into $[-1, 1]$. 
Then, we utilize Gaussian noise with mean $(\mu=0)$ and standard deviation $(\sigma=0.1)$ and $(\sigma=0.01)$ to represent the two levels of distortion. 

\mypara{Setup}
\autoref{fig:robustness} depicts the impact of ``with'' or ``without'' the action distortion. 
The information about the used offline DRL model and task is shown in each figure's title.
The x-axis indicates the four metrics, and the y-axis is the absolute value change. 
As a supplementary of~\cite{DCSJCCZ24}, the detailed results between every two datasets are in \autoref{fig:robustness 0.01 on lunarlander}, \autoref{fig:robustness 0.1 on lunarlander} (Lunar Lander), \autoref{fig:robustness 0.01 on bipedalwalker}, \autoref{fig:robustness 0.1 on bipedalwalker} (Bipedal Walker), \autoref{fig:robustness 0.01 on ant}, and \autoref{fig:robustness 0.1 on ant} (Ant). 

\mypara{Observations}
We conclude the following observations based on the above results. 
1) \sysname is able to resist the potential action distortion from the suspect model, especially with the Cosine metric. 
From \autoref{fig:robustness}, the TPR and TNR vary slightly across most of the settings with weak noise, where the maximum accuracy attenuation is within 3\% for Cosine distance. 
We speculate that Cosine distance has a noise suppression ability when calculating the inner product of two series of cumulative rewards. 
Also, the weak noise may facilitate the dataset auditing since it will move the negative samples farther away from the positive set. 

2) \sysname with a single distance metric faces limitations for heavy distortion. 
The TPR of \sysname suffers an obvious decline with strong noise. 
Since the strong distortion thoroughly changes the distribution of the models' actions, the cumulative rewards of the suspect model trained on the target dataset are different from those of the auditor's shadow models. 
In this case, the auditor cannot identify the positive models from the negative just by a single kind of distance metric. 
From \autoref{fig:robustness 0.1 on ant}, Cosine distance is good at discriminating the positive models (results in the diagonal), and Wasserstein distance is proper for the negative models (results in the non-diagonal). 
Thus, for strong distortion, the combination of multiple distance metrics can enhance the auditing robustness of \sysname. 
In addition, we should note that the models' normal behavior is also destroyed by the strong distortion. 
For example, in \autoref{tab:IQL models}, the noise induces the model performance of IQL to decrease up to 25\%, and the better the model's quality, the more pronounced the performance drop. 

\section{Related Work}
\mypara{Membership and Dataset Inferences}
To infer whether an individual data record was used to train the target model, Shokri \etal~\cite {SSSS17} proposed the first practical membership inference strategy by training a number of shadow classifiers to distinguish the target model’s outputs on members versus non-members of its training dataset.
Since then, researchers have investigated membership inference in various systems, such as machine unlearning~\cite{CZWBHZ21}, facial recognition systems~\cite{CZWBZ23}, and neural architecture search~\cite{HZSBLZ22}.
Liu \etal~\cite{LWHSZBCFZ22} presenting a first-of-its-kind holistic risk assessment of different inference attacks against machine learning models. 
Maini \etal~\cite {DBLP:conf/iclr/MainiYP21} introduced the definition of dataset inference and designed the first mechanism to identify whether a suspect model copy has private knowledge from the dataset. 

Compared with the existing works, 
\sysname is a well-designed solution built for the offline DRL scenes, which overcomes several new challenges. 
First, \sysname is a post-event mechanism that can be directly applied to the existing open-source datasets. 
Second, \sysname does not use any auxiliary datasets. 

\mypara{Knowledge Extraction Against DRL}
The DRL models learn from the interaction with the environment, which can be valuable information in some cases, \eg, indoor robot navigation. 
Pan \etal~\cite{PWZLYS19} demonstrated such knowledge extraction vulnerabilities in DRL under various settings and proposed algorithms to infer floor plans from some trained Grid World navigation DRL models with LiDAR perception. 
For exacting the model functionality, Chen \etal~\cite{CGZXL21} proposed the first method to acquire the approximation model from the victim DRL.
They built a classifier to reveal the targeted black-box DRL model's training algorithm family based only on its predicted actions and then leveraged state-of-the-art imitation learning techniques to replicate the model from the identified algorithm family. 
Ono \etal~\cite{OT20} integrated \textit{differential privacy}~\cite{ZWLHBHCZ21,YZDCCS23,WZWHBCZ23} into the distributed RL algorithm to defend the extraction. 
The local models report noisy gradients designed to satisfy local differential privacy~\cite{DZBLJCC21,DHZFCZG23,WCZSCLLJ21,ZWLHC18}, \ie, keeping the local information from being exploited by adversarial reverse engineering. 
Chen \etal~\cite{chen2022copy} proposed a novel testing framework for deep learning copyright protection, which can be adjusted to detect the knowledge extraction against DRL. 

\section{Disscusion}
\mypara{Highlights of \sysname}
1) \sysname is the first approach to conduct trajectory-level dataset auditing for offline DRL models. 
2) By conducting a comprehensive analysis of \sysname under different experimental settings, such as the shadow model's amount, the significance level in hypothesis testing, the trajectory size, and the robustness against ensemble architecture and action distortion, 
we conclude some useful observations for adopting \sysname. 
3) We apply \sysname to audit the models trained on the open-source datasets from Google and DeepMind. 
All TPR and TNR results are superior than 95\%, demonstrating \sysname is an effective and efficient strategy for the published datasets. 

\mypara{Limitations and Future Work}
Below, we discuss the limitations of \sysname and promising directions for further improvements. 
1) From \autoref{sec:impact of significance level}, the accuracy of \sysname decreases when the significance level downs to 0.001. 
Thus, it is interesting to enhance \sysname to satisfy stricter auditing demands in the future. 
2) \sysname based on a single distance metric may not be sufficiently robust to strong distortion. 
Based on the observations in \autoref{sec:perturbing models output}, integrating more distance metrics in the audit process may be a further promising direction. 

\section{Conclusion}
In this work, we propose a novel trajectory-level dataset auditing method for offline DRL models relying on the insight that cumulative rewards can serve as the dataset's intrinsic fingerprint and exist in all models trained on the target dataset. 
Both the true positive rate and the true negative rate of \sysname exceed 90\% on four offline DRL models and three task combinations. 
We show that \sysname is an effective and efficient solution to protect the IP of the dataset owners through multiple experiments. 
By studying parameter settings about the number of shadow models, the significance level in hypothesis testing, and the trajectory size, we conclude several important observations for adopting \sysname in practice. 
The robustness evaluation demonstrates that \sysname can resist the defenses of the model ensemble and the action distortion of the suspect model. 
Integrating multiple distance metrics to improve the robustness of \sysname against action distortion is a promising direction for future work. 
Finally, we utilize the open-source datasets from Google~\cite{DBLP:journals/corr/abs-2004-07219} and DeepMind~\cite{DBLP:conf/nips/Gulcehre0NPCZAM20} to examine the practicality of \sysname, and show that \sysname behaves excellently on existing published datasets. 

\section*{Acknowledgment}
We would like to thank the anonymous reviewers for their constructive comments. 
We also thank Yanchao Sun for sharing her expertise in reinforcement learning. 
This work was partly supported by the National Key Research and Development Program of China under No. 2022YFB3102100, NSFC under Grants 62088101, 61833015, 62103371, U20A20159, and the Fundamental Research Funds for the Central Universities 226-2022-00107, 226-2023-00111. 
Min Chen was partly sponsored by the Helmholtz Association within the project ``Trustworthy Federated Data Analytics'' (TFDA) (No. ZT-I-OO1 4).
Zhikun Zhang was supported by the CISPA-Stanford Center for Cybersecurity (FKZ:13N1S0762).

\bibliographystyle{abbrv}
\bibliography{main.bib}
\appendix

\subsection{The Behavior Similarity of Models}
\label{sec:the-behavior-similarity-of-models}
\begin{figure*}[!t]
\centering
\includegraphics[width=\hsize]{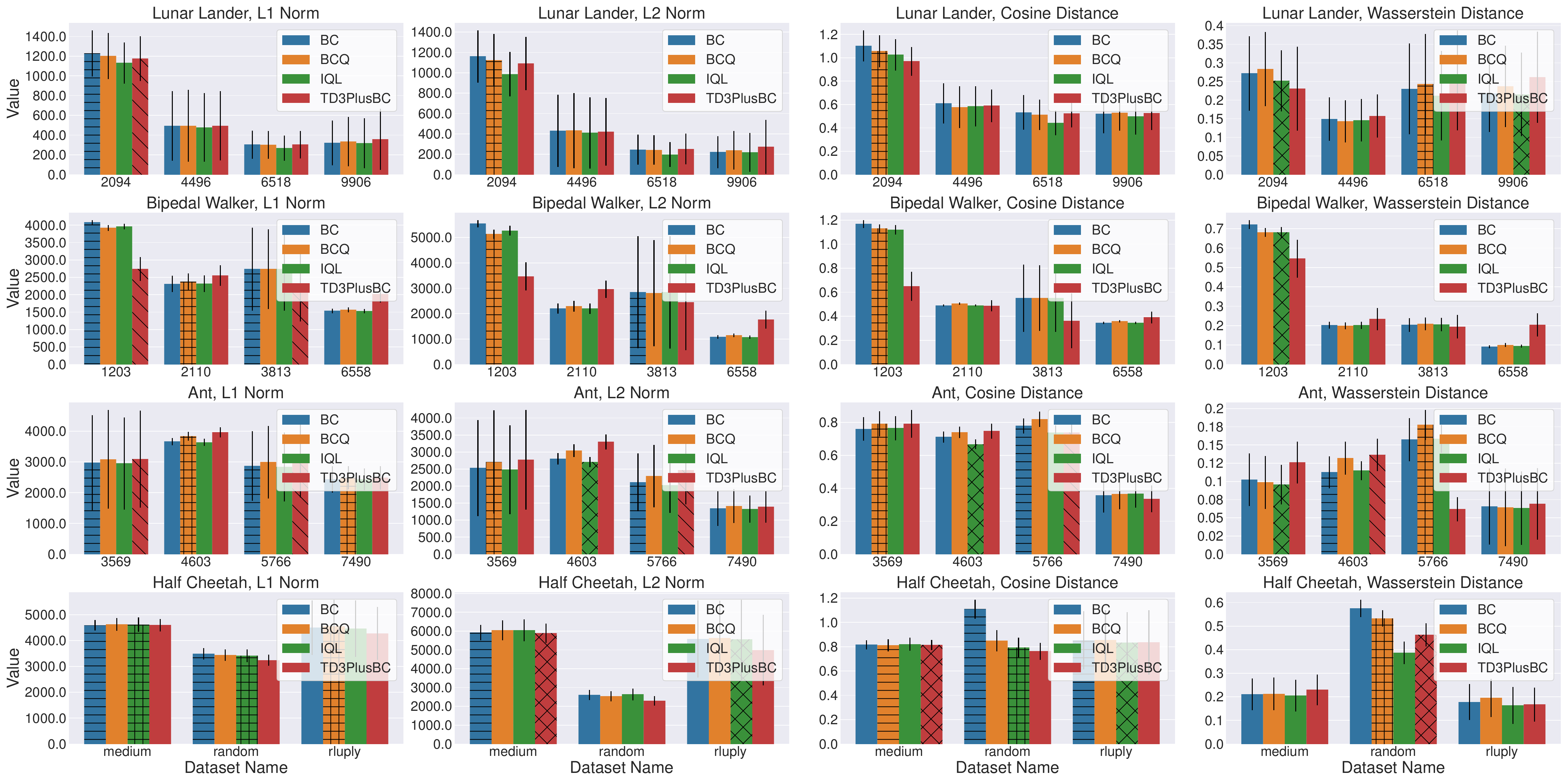}
\caption{Models' behavior similarity measured by $\ell_1$ Norm, $\ell_2$ Norm, Cosine Distance, and Wasserstein Distance. 
From \autoref{tab:the offline datasets}, we use the first dataset of each task as the private training data and the remaining four datasets are the public training data. 
For each plot, the x-axis displays the four public training data, 
and the y-axis shows the absolute fluctuating values of the behavior similarity between the models trained on the private dataset and the public datasets.
BC, BCQ, IQL, and TD3PlusBC are abbreviations for different offline RL frameworks. 
}
\vspace{-0.5cm}
\label{fig:full behavior similarity}
\end{figure*}
In \autoref{fig:full behavior similarity}, we provide the behavior similarity of the offline RL models trained on the datasets in \autoref{tab:the offline datasets}. 
Taking the Bipedal Walker task as an example, the dataset ``0841'' is regarded as the target dataset, and the other four are the public datasets. 
We observe that the behavior similarity of the RL models waves heavily among the different public training data. 
If the auditor adopts the dataset ``1203'' as the public training data, the auditor likely misclassifies the RL models trained on the other three public datasets into the bootleg models. 
In addition, the behavior similarity is also affected by different offline RL frameworks, \ie, BC~\cite{DBLP:conf/nips/Pomerleau88}, BCQ~\cite{DBLP:conf/icml/FujimotoMP19, DBLP:journals/corr/abs-1910-01708}, IQL~\cite{DBLP:conf/iclr/KostrikovNL22}, and TD3PlusBC~\cite{DBLP:conf/nips/FujimotoG21} (detailed in \autoref{sec:offline reinforcement learning model}). 

\subsection{The Details of Tasks}
\label{sec:the details of tasks}
\mypara{Lunar Lander (continuous version)}
The LunarLander task is to smoothly land a spaceship between two flags on the target pad. 
The landing pad is always at coordinates (0,0). 
The ship has three throttles; one throttle points downward (the main engine) and the other two points in the left and right direction (the left and right engines). 
The observation is an 8-dimensional vector: the coordinates of the lander in the x-axis and y-axis, its linear velocities in the x-axis and y-axis, its angle, its angular velocity, and two booleans that represent whether each leg is in contact with the ground or not. 
The action is two real values ranging in $[-1, 1]$. 
The first dimension controls the main engine, 
where the engine is off when the value is in $[-1, 0)$ and increases from 50\% to 100\% throttle when the value rises from 0 to 1. 
The other two points are controlled by the second value, where the spaceship fires the left engine if the value in $[-1.0, -0.5)$, fires the right engine if the value in $[0.5, 1)$, and shuts down both engines if the value in $[-0.5, 0.5]$. 
The reward for moving from the top of the screen to the landing pad and zero speed is about 140 points. 
Landing outside the landing pad is possible. 
Thus, the player loses the terminal reward if the lander moves away from the landing pad. 
The player gets 10 additional points for each leg touching the ground. 
Firing the main engine is -0.3 points in each frame. 
The episode finishes if the lander crashes or lands smoothly, receiving -100 or 100 points. 

\mypara{Bipedal Walker}
The Bipedal Walker task is to operate a 4-joint walker robot to move forward as fast as possible. 
The robot is made of a hull and two legs. 
Each leg has 2 joints at both the hip and knee. 
The observation of the task includes eight continuous physical variables, \ie, hull angle speed, angular velocity, horizontal speed, vertical speed, the position of joints and joints angular speed, legs contact with ground, and 10 lidar rangefinder measurements. 
Actions are motor speed values in the [-1, 1] range for each of the 4 joints at both hips and knees.
The walker starts standing at the left end of the terrain with the hull horizontal, and both legs in the same position with a slight knee angle. 
The reward is given for moving forward, totaling 300+ points up to the far end. If the robot falls, it gets -100. Applying motor torque costs a small amount of points. A more optimal model will get a better score.
The episode will terminate if the hull gets in contact with the ground or the walker exceeds the right end of the terrain length.

\mypara{Ant} 
In this task, the player manipulates a 3D robot (ant), which consists of one torso (free rotational body) with four legs attached to it, with each leg having two links, to move in the forward (right) direction. 
The observation contains positional values of different body parts of the ant, followed by the velocities of those individual parts (their derivatives), with all the positions ordered before all the velocities. 
By default, an observation is a vector with shape (111,) where the elements correspond to the following: position (1-dim), angles (12-dim), velocities(14-dim), and the information about the contact forces (84-dim). 
The player can apply torques on the eight hinges connecting the two links of each leg and the torso (nine parts and eight hinges). 
Thus, the action space is an 8-dim continuous vector representing the torques applied at the hinge joints. 
The reward of the ``Ant'' task consists of four parts: healthy reward, forward reward, control cost, and contact cost. 
The total reward returned is reward = healthy reward + forward reward - control cost - contact cost. 
The task ends when either the ant state is unhealthy, or the episode duration reaches 1000 timesteps. 

\subsection{Impact of Shadow Models' Amount}
\label{sec:impact of shadow models' amount}
We investigate the relationship between the number of shadow models and the audit accuracy. 

\mypara{Setup}
We change the shadow models' amount to 9 and 21 with the other settings the same as \autoref{sec:overall performance}. 
\autoref{fig:impact of shadow models' amount} shows the value change of TPR and TNR compared with that of 15 shadow models. 
Each figure's title illustrates the settings of the model and the task, the x-axis indicates the four metrics, and the y-axis is the absolute value change. 
Also, we provide the detailed results in \autoref{tab:overall audit accuracy based on Grubbs (9 Shadow Models)} (9 Shadow Models) and \autoref{tab:overall audit accuracy based on Grubbs (21 Shadow Models)} (21 Shadow Models). 

\mypara{Observations}
From \autoref{fig:impact of shadow models' amount}, we have the following observations. 
1) The audit accuracy increases with a larger amount of shadow models. 
Since the values of shadow models are the multi-sampling of the true value $Q\left(s, a\right)$ of the dataset, the mean and standard deviation will be more precise with more shadow models. 
For example, \sysname suffers an obvious TPR decline (more than 30\%) with 9 shadow models. 
Since the insufficient knowledge about the diversity of models trained on the target dataset, the auditor easily misclassifies the positive models to the negative group. 

2) There exists a saturation point for audit accuracy with the expansion of shadow models. 
When the shadow models' amount rises from 15 to 21, the TPR usually increases since the auditor observes more possible cumulative rewards originating from the model trained on the target dataset. 
We should note that the value changes slightly in most plots, meaning that similar cumulative rewards appear in the shadow model set, and the diversity does not increase significantly compared to that of 15 shadow models. 
Therefore, excessive shadow models are unnecessary, and the auditor needs to burden more training overhead. 

\subsection{Impact of Significance Level}
\label{sec:impact of significance level}
The significance level represents the auditor's confidence in the audit results. 
In \autoref{sec:overall performance}, we adopt the significance level $\alpha=0.01$, meaning that the auditor has 99\% confidence in the judgments made. 
Generally speaking, the significance level represents the maximum audit capacity of \sysname instead of a hyperparameter setting since it is an audit requirement by the dataset owner. 

\mypara{Setup}
We demand the auditor to output a more confident judgment, where the error possibility should be limited to 1\textperthousand~and 0.1\textperthousand, \ie, $\alpha=0.001$ and $\alpha=0.0001$. 
\autoref{fig:impact of significance level} shows the value change of TPR and TNR compared with that when significance level $\alpha=0.01$. 
The used offline DRL model and task is shown in each figure's title. 
The x-axis indicates the four metrics and the y-axis is the absolute value change. 
The detailed results between every two datasets are in \autoref{tab:overall audit accuracy based on Grubbs (sigma=0.001)} ($\alpha=0.001$) and \autoref{tab:overall audit accuracy based on Grubbs (sigma=0.0001)} ($\alpha=0.0001$). 

\mypara{Observations}
From \autoref{fig:impact of significance level}, we have the following observations.  
1) For a complicated task, we recommend the auditor to select a large significance level for \sysname. 
The task's complexity affects the minimum significance level of \sysname. 
For example, TPR and TNP change a little on the Lunar Lander task when the significance level reduces to 0.001, while they highly shrink on the Ant task. 
From \autoref{tab:overview-of-tasks}, Ant's state and action space are larger than that of Lunar Lander. 
When the auditor leverages the critic model to compress each model's state and action pair into a scalar, the deviation between $Q_j^i$ and $Q_j^s$ (recalling \autoref{fig:framework}) on the Ant task is more imperceptible. 

2) For the suspect models with low performance, \sysname should adopt a large significance level to guarantee audit accuracy. 
For instance, in the figure titled with ``Bipedal Walker, TD3PlusBC'', all TNR results from four distance metrics decrease when $\alpha$ reduces to 0.001 and 0.0001. 
From \autoref{tab:TD3PlusBC models}, most of the TD3PlusBC models' performance on the Bipedal Walker task is around -100, meaning that the TD3PlusBC models do not fully master the knowledge of the dataset. 
Thus, the dataset features reflected in their behavior are ambiguous, which weakens the difference between positive and negative samples. 
Meanwhile, the confidence interval, \ie, $\Delta$ in \autoref{fig:intuition}, expands with a lower significance level. 
For the above two reasons, the TNR results of the TD3PlusBC models on the Bipedal Walker task drop more than 10\% compared with these when $\alpha=0.01$. 

\smallskip

From the above analysis, $\alpha=0.01$ is a safe bound of \sysname, and a lower $\alpha$ may break through the capability boundary of \sysname, inducing the auditor to misclassify the negative model to the positive set. 

\subsection{Impact of Trajectory Size}
\label{sec:impact_of_trajectory_size}
We investigate the relationship between the trajectory size and audit accuracy. 
In \autoref{sec:overall performance}, we adopt the full-length trajectory, meaning that the auditor utilizes all states of each trajectory to query the suspect model and obtains the corresponding actions to conduct the dataset audit. 

\mypara{Setup}
We change the trajectory size to 25\% and 50\% of the full length with the other settings the same as \autoref{sec:overall performance}. 
\autoref{fig:impact_of_trajectory_size} shows the value change of TPR and TNR compared with that of the full-length trajectory. 
Each figure's title illustrates the settings of the model and the task, the x-axis indicates the four metrics, and the y-axis is the absolute value change. 
Also, we provide the detailed results in \autoref{tab:overall_audit_accuracy_based_on_Grubbs_0.25_trajsize} (25\%) and \autoref{tab:overall_audit_accuracy_based_on_Grubbs_0.5_trajsize} (50\%). 

\mypara{Observation}
From \autoref{fig:impact_of_trajectory_size}, we have the following observations. 
1) \sysname tends to achieve higher accuracy with a larger trajectory size. 
Since the predicted cumulative rewards of state-action pairs from the critic model are the audit basis, 
a longer trajectory collects more actions from the suspect model to enhance the significance of hypothesis testing. 
For example, the TNP results decrease at most 13\% when \sysname only leverages 25\% of the trajectory. 

2) It should be noticed that a small trajectory size achieves better results under some tasks. 
For the Ant task, \sysname auditing with 25\% of the full length obtains at most 7\% promotion on the TNR results. 
Based on the analysis of \cite{DBLP:journals/corr/abs-2007-09055}, the front states of each trajectory are able to reflect more behavioral information of the model. 
Thus, in this case, a shorter trajectory truncates the rear state-action pairs, which might be unimportant or even weaken the significance of the hypothesis testing. 
Exploring effective data auditing with shorter trajectory sizes or even using only the first state of each trajectory would be an interesting future direction. 

\subsection{Additional Results}
\label{sec:the outline of figures and tables}
As a supplementary of~\cite{DCSJCCZ24}, we provide additional results about \sysname. 
For ease of reading, we summarize the main figures and tables in \autoref{tab:roadmap_additional_results}. 
\begin{table*}[!ht]
\centering
\caption{The roadmap of the main figures and tables.}
\label{tab:roadmap_additional_results}
\begin{tabularx}{0.85\textwidth}{lllX}
\toprule                                                                                                            
\textbf{Information}                                    & \textbf{Involved Content}                                      & \textbf{Name}                                                                                                                                                                                                                                                                                                                                                & \textbf{Description}                                                                                                                                                                                     \\
Overview of tasks                                       & \autoref{sec:experimental setup}                               & \autoref{tab:overview-of-tasks}                                                                                                                                                                                                                                                                                                                              & The state shape and the action shape of each task.                                                                                                                                                       \\
Online DRL models                                       & \autoref{sec:experimental setup}                               & \autoref{tab:the online models}                                                                                                                                                                                                                                                                                                                              & The performance of the used online models for collecting the offline datasets.                                                                                                                           \\
Offline Datasets                                        & \autoref{sec:experimental setup}                               & \autoref{tab:the offline datasets}                                                                                                                                                                                                                                                                                                                           & The name, the number of trajectories, and the length of   trajectory for each offline dataset.                                                                                                           \\
\multirow{4}{*}{Offline DRL models}                     & \multirow{4}{*}{\autoref{sec:evaluation}}                      & \multirow{4}{*}{\begin{tabular}[c]{@{}c@{}}\autoref{tab:BC models}\\ \autoref{tab:BCQ models}\\ \autoref{tab:IQL models}\\ \autoref{tab:TD3PlusBC models}\end{tabular}}                                                                                                                                                                                      & The offline models' performance with or without defense against \sysname: normal performance (without defense), defended by model ensemble, and defended by perturbing models' output. \\
                                                        &                                                                &                                                                                                                                                                                                                                                                                                                                                              &                                                                                                                                                                                                          \\
                                                        &                                                                &                                                                                                                                                                                                                                                                                                                                                              &                                                                                                                                                                                                          \\
                                                        &                                                                &                                                                                                                                                                                                                                                                                                                                                              &                                                                                                                                                                                                          \\
\multirow{5}{*}{Overall audit performance}              & \multirow{5}{*}{\autoref{sec:overall performance}}             & \multirow{5}{*}{\begin{tabular}[c]{@{}c@{}}\autoref{tab:overall audit accuracy based on Grubbs}\\ \autoref{tab:overall audit accuracy based on 3sigma}\\ \autoref{fig:audit result on lunarlander}\\ \autoref{fig:audit result on bipedalwalker}\\ \autoref{fig:audit result on ant}\end{tabular}}                                                           & The true positive rate (TPR) and true negative rate (TNR)   results based on Grubbs’ test and $3\sigma$ principle.                                                                                       \\
                                                        &                                                                &                                                                                                                                                                                                                                                                                                                                                              &                                                                                                                                                                                                          \\
                                                        &                                                                &                                                                                                                                                                                                                                                                                                                                                              &                                                                                                                                                                                                          \\
                                                        &                                                                &                                                                                                                                                                                                                                                                                                                                                              &                                                                                                                                                                                                          \\
                                                        &                                                                &                                                                                                                                                                                                                                                                                                                                                              &                                                                                                                                                                                                          \\
\multirow{3}{*}{Impact of shadow models’ amount}        & \multirow{3}{*}{\autoref{sec:impact of shadow models' amount}} & \multirow{3}{*}{\begin{tabular}[c]{@{}c@{}}\autoref{fig:impact of shadow models' amount}\\ \autoref{tab:overall audit accuracy based on Grubbs (9 Shadow Models)}\\ \autoref{tab:overall audit accuracy based on Grubbs (21 Shadow Models)}\end{tabular}}                                                                                                    & The change values of TPR and   TNR when the number of shadow models varies to 9 and 21 compared to the default 15 shadow models.                                                                       \\
                                                        &                                                                &                                                                                                                                                                                                                                                                                                                                                              &                                                                                                                                                                                                          \\
                                                        &                                                                &                                                                                                                                                                                                                                                                                                                                                              &                                                                                                                                                                                                          \\
\multirow{3}{*}{Impact of significance level}           & \multirow{3}{*}{\autoref{sec:impact of significance level}}    & \multirow{3}{*}{\begin{tabular}[c]{@{}c@{}}\autoref{fig:impact of significance level}\\ \autoref{tab:overall audit accuracy based on Grubbs (sigma=0.001)}\\ \autoref{tab:overall audit accuracy based on Grubbs (sigma=0.0001)}\end{tabular}}                                                                                                               & The change value of TPR and TNR when the significance level   ($\sigma$) varies to 0.001 and 0.0001 compared to the default 0.01.                                                                        \\
                                                        &                                                                &                                                                                                                                                                                                                                                                                                                                                              &                                                                                                                                                                                                          \\
                                                        &                                                                &                                                                                                                                                                                                                                                                                                                                                              &                                                                                                                                                                                                          \\
\multirow{3}{*}{Impact of trajectory size}              & \multirow{3}{*}{\autoref{sec:impact_of_trajectory_size}}       & \multirow{3}{*}{\begin{tabular}[c]{@{}c@{}}\autoref{fig:impact_of_trajectory_size}\\ \autoref{tab:overall_audit_accuracy_based_on_Grubbs_0.25_trajsize}\\ \autoref{tab:overall_audit_accuracy_based_on_Grubbs_0.5_trajsize}\end{tabular}}                                                                                                                    & The change value of TPR and TNR when the trajectory size varies to 25\% and 50\% compared to the default 100\% (full length).                                                                            \\
                                                        &                                                                &                                                                                                                                                                                                                                                                                                                                                              &                                                                                                                                                                                                          \\
                                                        &                                                                &                                                                                                                                                                                                                                                                                                                                                              &                                                                                                                                                                                                          \\
\multirow{3}{*}{Real-world application}                 & \multirow{3}{*}{\autoref{sec:real-world application}}          & \multirow{3}{*}{\begin{tabular}[c]{@{}c@{}}\autoref{tab:audit result on halfcheetah}\\ \autoref{tab:details of the halfcheetah dataset}\\ \autoref{tab:details of the models trained on the halfcheetah dataset}\end{tabular}}                                                                                                                               & The TPR and TNR results on the Half Cheetah datasets, which are published by DeepMind and Google separately.                                                                                           \\
                                                        &                                                                &                                                                                                                                                                                                                                                                                                                                                              &                                                                                                                                                                                                          \\
                                                        &                                                                &                                                                                                                                                                                                                                                                                                                                                              &                                                                                                                                                                                                          \\
\multirow{4}{*}{Robustness: ensemble   architecture}    & \multirow{4}{*}{\autoref{sec:ensemble architecture}}           & \multirow{4}{*}{\begin{tabular}[c]{@{}c@{}}\autoref{tab:overall audit accuracy against model ensemble}\\ \autoref{fig:model ensemble k5-2 on lunarlander}\\ \autoref{fig:model ensemble k5-2 on ant}\\ \autoref{fig:model ensemble k5-2 on halfcheetah}\end{tabular}}                                                                                        & The TPR and TNR results of \sysname against model ensemble ($K=5$).                                                                                                                                      \\
                                                        &                                                                &                                                                                                                                                                                                                                                                                                                                                              &                                                                                                                                                                                                          \\
                                                        &                                                                &                                                                                                                                                                                                                                                                                                                                                              &                                                                                                                                                                                                          \\
                                                        &                                                                &                                                                                                                                                                                                                                                                                                                                                              &                                                                                                                                                                                                          \\
\multirow{7}{*}{Robustness: perturbing models   output} & \multirow{7}{*}{\autoref{sec:perturbing models output}}        & \multirow{7}{*}{\begin{tabular}[c]{@{}c@{}}\autoref{fig:robustness}\\ \autoref{fig:robustness 0.01 on lunarlander}\\ \autoref{fig:robustness 0.1 on lunarlander}\\ \autoref{fig:robustness 0.01 on bipedalwalker}\\ \autoref{fig:robustness 0.1 on bipedalwalker}\\ \autoref{fig:robustness 0.01 on ant}\\ \autoref{fig:robustness 0.1 on ant}\end{tabular}} & The TPR and TNR results of \sysname against models' action distortion.                                                                                                                                   \\
                                                        &                                                                &                                                                                                                                                                                                                                                                                                                                                              &                                                                                                                                                                                                          \\
                                                        &                                                                &                                                                                                                                                                                                                                                                                                                                                              &                                                                                                                                                                                                          \\
                                                        &                                                                &                                                                                                                                                                                                                                                                                                                                                              &                                                                                                                                                                                                          \\
                                                        &                                                                &                                                                                                                                                                                                                                                                                                                                                              &                                                                                                                                                                                                          \\
                                                        &                                                                &                                                                                                                                                                                                                                                                                                                                                              &                                                                                                                                                                                                          \\
                                                        &                                                                &                                                                                                                                                                                                                                                                                                                                                              &                                                                                                                                                                                                          \\
\bottomrule
\end{tabularx}
\end{table*}

\begin{table*}[t]
\caption{As a supplementary of~\cite{DCSJCCZ24}, we provide the TPR and TNR results of \sysname based on 3$\sigma$ principle. 
The mean and standard deviation of TPR and TNR in each row represent the audit results for one combination of task and model by four distance metrics. 
Bold indicates the highest sum of TPR and TNR, \ie, accuracy, in a row. 
}
\label{tab:overall audit accuracy based on 3sigma}
\centering
    \setlength{\tabcolsep}{0.8em}
    \renewcommand{\arraystretch}{1.1}
    \footnotesize
\begin{tabular}{cccccccccc} 
\toprule
\multirow{2}{*}{\begin{tabular}[c]{@{}c@{}}\textbf{Task}\\\textbf{ Name}\end{tabular}} & \multirow{2}{*}{\begin{tabular}[c]{@{}c@{}}\textbf{Offline}\\\textbf{ Model}\end{tabular}} & \multicolumn{2}{c}{\textbf{L1 Norm}}               & \multicolumn{2}{c}{\textbf{L2 Norm}}              & \multicolumn{2}{c}{\begin{tabular}[c]{@{}c@{}}\textbf{Cosine}\\\textbf{ Distance}\end{tabular}} & \multicolumn{2}{c}{\begin{tabular}[c]{@{}c@{}}\textbf{Wasserstein}\\\textbf{ Distance}\end{tabular}}  \\ 
\cline{3-10}
                                                                                       &                                                                                            & TPR                     & TNR                      & TPR                     & TNR                     & TPR                     & TNR                                                                   & TPR                     & TNR                                                                         \\ 
\hline
\multirow{4}{*}{\begin{tabular}[c]{@{}c@{}}Lunar\\ Lander\end{tabular}}                & BC                                                                                         & \textbf{96.53$\pm$1.36} & \textbf{100.00$\pm$0.00} & 95.47$\pm$2.81          & 100.00$\pm$0.00         & 95.73$\pm$2.58          & 100.00$\pm$0.00                                                       & 96.13$\pm$2.02          & 100.00$\pm$0.00                                                             \\
                                                                                       & BCQ                                                                                        & 96.13$\pm$3.01          & 100.00$\pm$0.00          & 94.80$\pm$3.18          & 100.00$\pm$0.00         & 94.67$\pm$2.92          & 100.00$\pm$0.00                                                       & \textbf{96.40$\pm$2.92} & \textbf{99.95$\pm$0.36}                                                     \\
                                                                                       & IQL                                                                                        & \textbf{97.20$\pm$3.24} & \textbf{99.97$\pm$0.28}  & 96.27$\pm$2.44          & 100.00$\pm$0.00         & 96.53$\pm$2.40          & 100.00$\pm$0.00                                                       & 96.53$\pm$4.14          & 98.90$\pm$3.33                                                              \\
                                                                                       & TD3PlusBC                                                                                  & \textbf{95.60$\pm$4.39} & \textbf{99.54$\pm$2.53}  & 92.80$\pm$5.07          & 99.91$\pm$0.47          & 93.33$\pm$5.40          & 99.93$\pm$0.40                                                        & 96.67$\pm$2.88          & 96.86$\pm$7.24                                                              \\ 
\cline{2-10}
\multirow{4}{*}{\begin{tabular}[c]{@{}c@{}}Bipedal\\ Walker\end{tabular}}              & BC                                                                                         & 96.56$\pm$4.27          & 100.00$\pm$0.00          & 95.78$\pm$4.58          & 100.00$\pm$0.00         & \textbf{98.33$\pm$2.50} & \textbf{100.00$\pm$0.00}                                              & 97.22$\pm$4.05          & 100.00$\pm$0.00                                                             \\
                                                                                       & BCQ                                                                                        & 96.67$\pm$4.20          & 100.00$\pm$0.00          & 94.78$\pm$7.43          & 100.00$\pm$0.00         & \textbf{98.67$\pm$1.63} & \textbf{100.00$\pm$0.00}                                              & 97.11$\pm$3.69          & 100.00$\pm$0.00                                                             \\
                                                                                       & IQL                                                                                        & 94.33$\pm$7.45          & 100.00$\pm$0.00          & 93.78$\pm$7.25          & 100.00$\pm$0.00         & \textbf{98.89$\pm$2.17} & \textbf{100.00$\pm$0.00}                                              & 94.00$\pm$9.45          & 100.00$\pm$0.00                                                             \\
                                                                                       & TD3PlusBC                                                                                  & \textbf{97.00$\pm$4.46} & \textbf{99.90$\pm$1.16}  & 94.11$\pm$8.63          & 97.80$\pm$12.09         & 95.33$\pm$6.66          & 97.78$\pm$12.19                                                       & 96.44$\pm$5.95          & 93.87$\pm$19.73                                                             \\ 
\cline{2-10}
\multirow{4}{*}{Ant}                                                                   & BC                                                                                         & 90.67$\pm$5.30          & 100.00$\pm$0.00          & 93.33$\pm$4.62          & 100.00$\pm$0.00         & 99.20$\pm$0.88          & 88.00$\pm$27.55                                                       & \textbf{95.20$\pm$2.99} & \textbf{99.99$\pm$0.07}                                                     \\
                                                                                       & BCQ                                                                                        & 90.40$\pm$8.68          & 99.96$\pm$0.42           & \textbf{94.13$\pm$3.83} & \textbf{99.94$\pm$0.56} & 98.00$\pm$2.00          & 88.47$\pm$26.83                                                       & 93.47$\pm$6.81          & 99.95$\pm$0.49                                                              \\
                                                                                       & IQL                                                                                        & 90.67$\pm$6.93          & 100.00$\pm$0.00          & 89.60$\pm$3.99          & 100.00$\pm$0.00         & 97.20$\pm$3.89          & 88.30$\pm$27.38                                                       & \textbf{91.20$\pm$9.03} & \textbf{100.00$\pm$0.00}                                                    \\
                                                                                       & TD3PlusBC                                                                                  & 95.62$\pm$5.19          & 99.74$\pm$1.79           & 94.12$\pm$5.03          & 99.35$\pm$2.58          & 99.08$\pm$1.54          & 88.52$\pm$26.25                                                       & \textbf{97.74$\pm$2.66} & \textbf{99.60$\pm$2.13}                                                     \\
\bottomrule
\end{tabular}
\end{table*}

\begin{figure*}[t]
\includegraphics[width=\hsize]{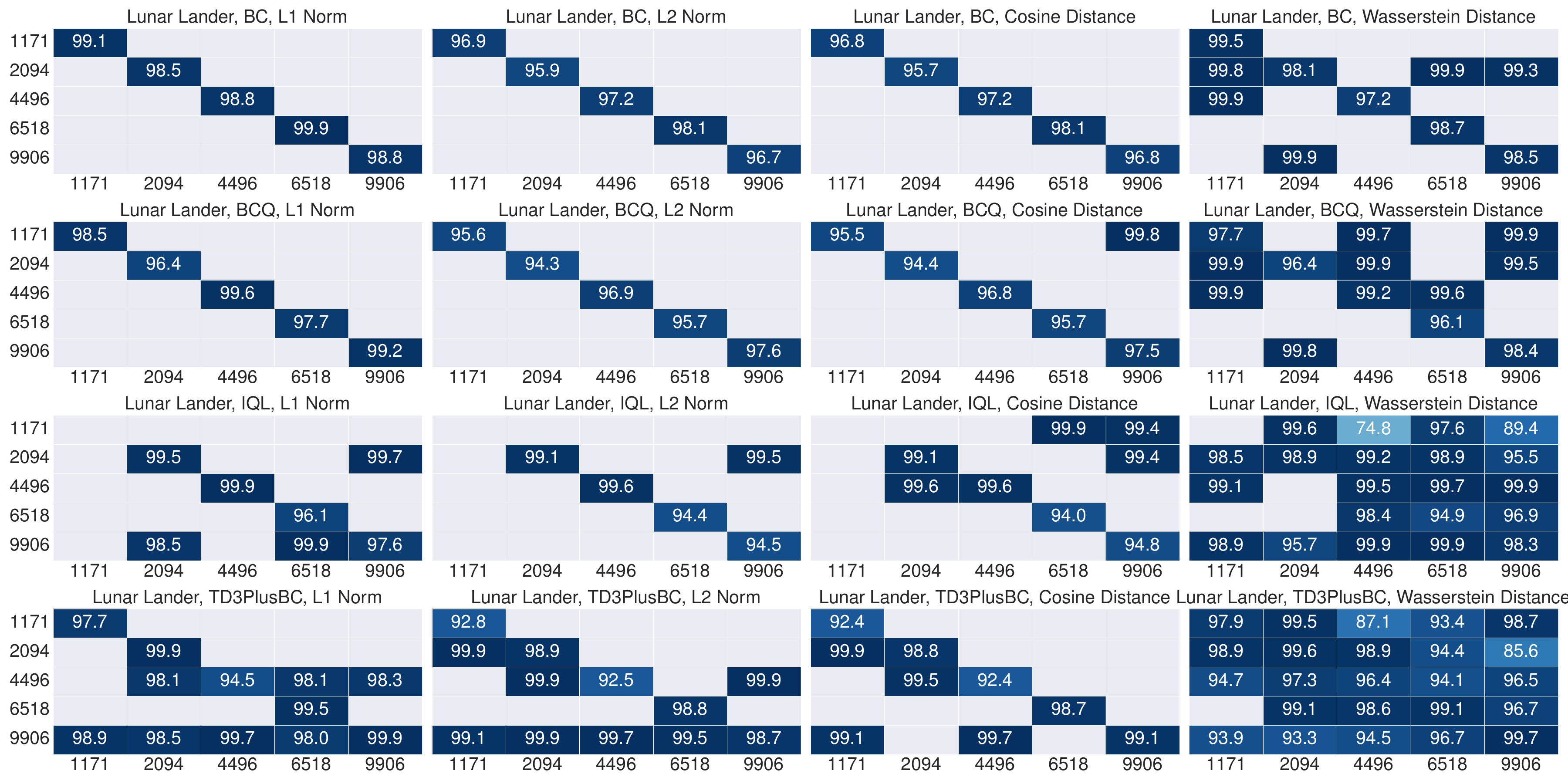}
\vspace{-0.6cm}
\caption{The audit accuracy between every two Lunar Lander datasets. 
The caption of each plot demonstrates the offline DRL model's type, the task, and the distance metric. 
The x labels are the names of datasets to be audited, \ie, the target datasets. 
The y labels are the names of datasets the suspect models learned, \ie, the actual datasets. 
Thus, the diagonal values show the audit accuracy when the actual dataset is the same as the target dataset, \ie, TPR, and the non-diagonal values are the TNR results. 
The positions without value mean 100\% accuracy. }
\label{fig:audit result on lunarlander}
\end{figure*}

\begin{figure*}[t]
\includegraphics[width=\hsize]{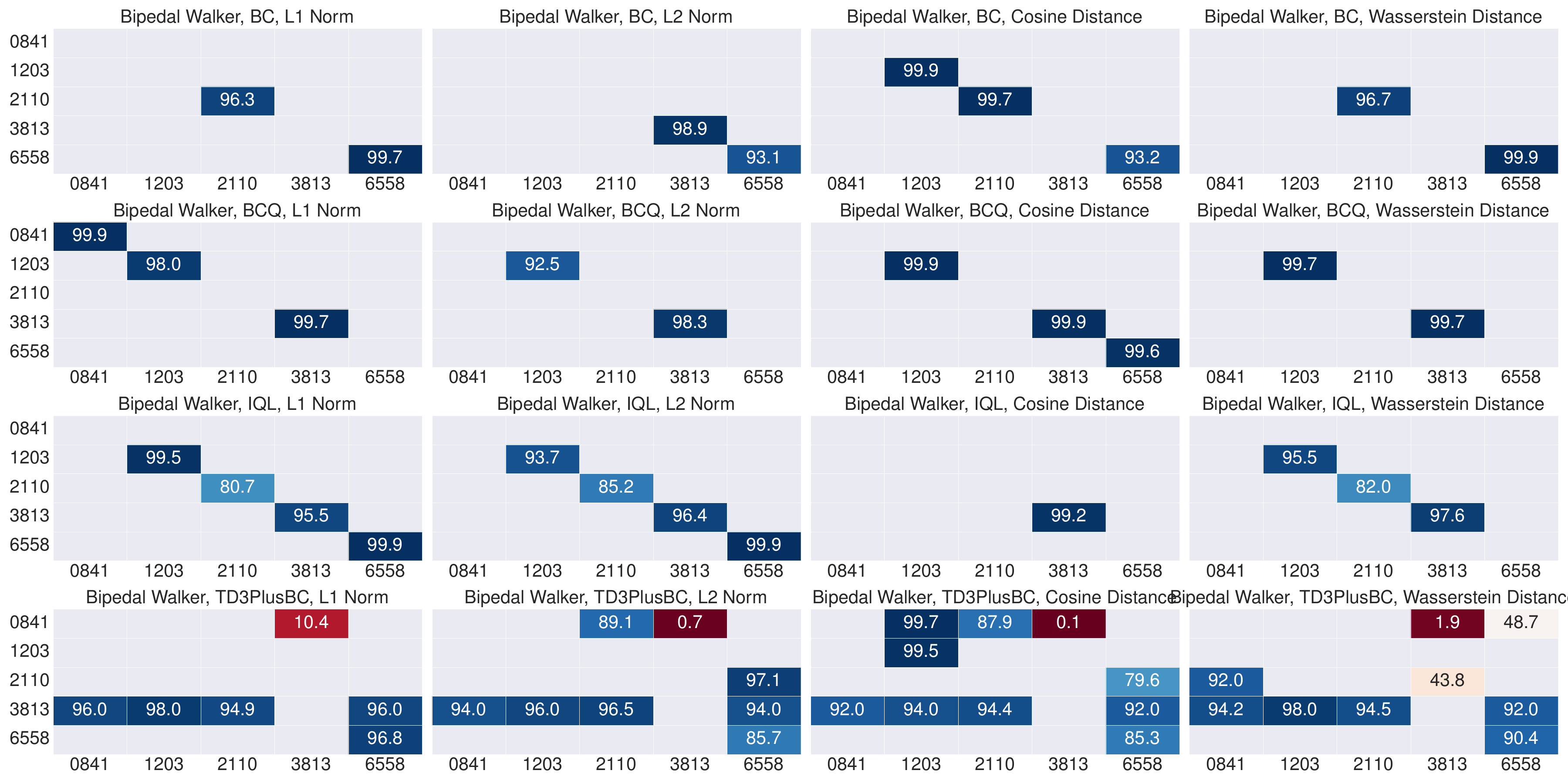}
\vspace{-0.2cm}
\caption{The audit accuracy between every two Bipedal Walker datasets. 
The caption of each plot demonstrates the offline DRL model's type, the task, and the distance metric. 
The x labels are the names of datasets to be audited, \ie, the target datasets. 
The y labels are the names of datasets the suspect models learned, \ie, the actual datasets. 
Thus, the diagonal values show the audit accuracy when the actual dataset is the same as the target dataset, \ie, TPR, and the non-diagonal values are the TNR results. 
The positions without value mean 100\% accuracy. 
}
\label{fig:audit result on bipedalwalker}
\end{figure*}

\begin{figure*}[t]
\includegraphics[width=\hsize]{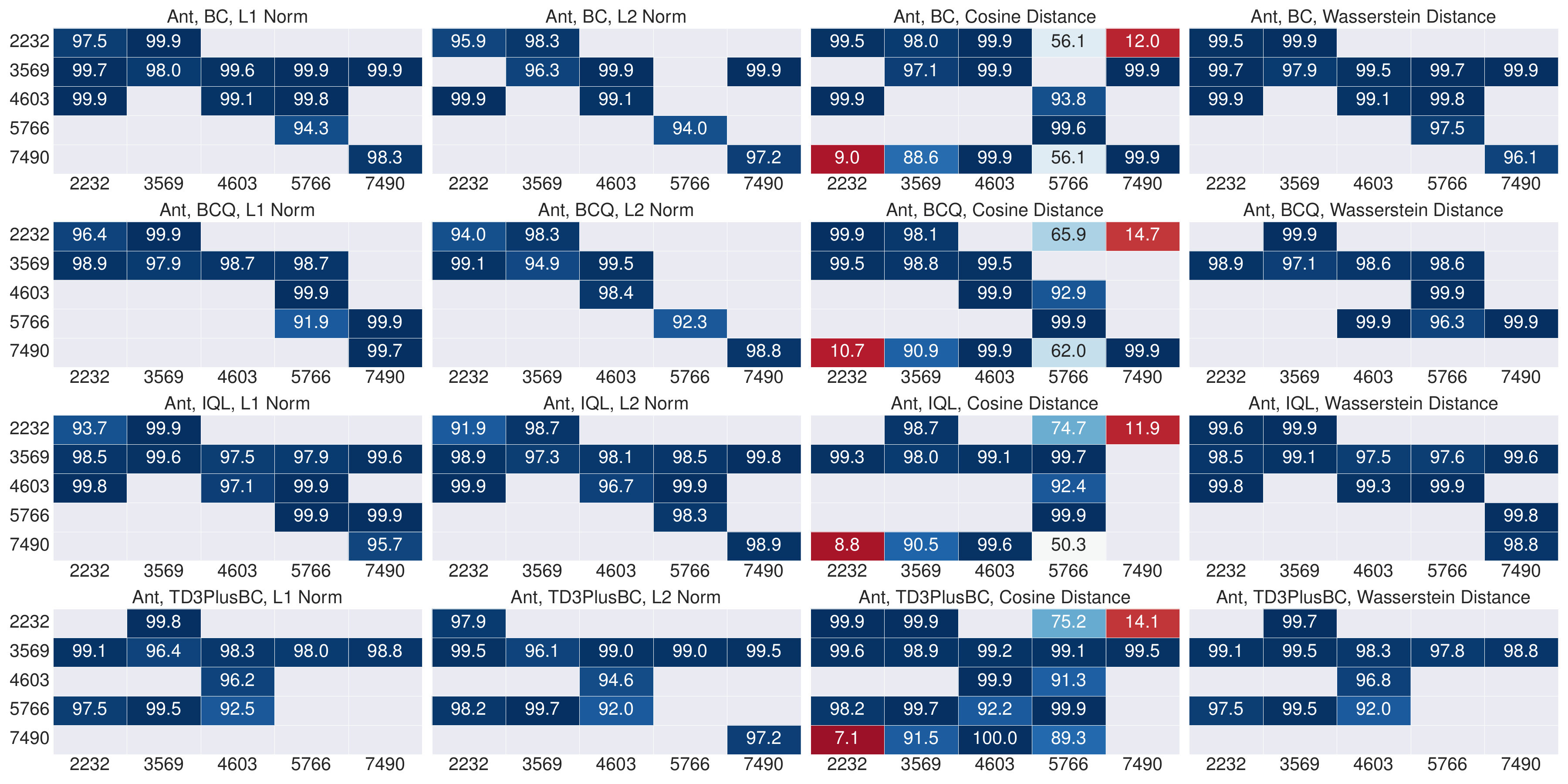}
\vspace{-0.6cm}
\caption{The audit accuracy between every two Ant datasets. 
The caption of each plot demonstrates the offline DRL model's type, the task, and the distance metric. 
The x labels are the names of datasets to be audited, \ie, the target datasets. 
The y labels are the names of datasets the suspect models learned, \ie, the actual datasets. 
Thus, the diagonal values show the audit accuracy when the actual dataset is the same as the target dataset, \ie, TPR, and the non-diagonal values are the TNR results. 
The positions without value mean 100\% accuracy. 
}
\label{fig:audit result on ant}
\end{figure*}

\begin{table*}[t]
\centering
\caption{The impact of shadow models’ amount. 
The TPR and TNR results of \sysname with 9 shadow models.}
\label{tab:overall audit accuracy based on Grubbs (9 Shadow Models)}
\setlength{\tabcolsep}{0.6em}
\renewcommand{\arraystretch}{1.1}
\footnotesize
\begin{tabular}{cccccccccc} 
\toprule
\multirow{2}{*}{\begin{tabular}[c]{@{}c@{}}\textbf{Task}\\\textbf{ Name}\end{tabular}} & \multirow{2}{*}{\begin{tabular}[c]{@{}c@{}}\textbf{Offline}\\\textbf{ Model}\end{tabular}} & \multicolumn{2}{c}{\textbf{L1 Norm}}                                & \multicolumn{2}{c}{\textbf{L2 Norm}}                                & \multicolumn{2}{c}{\begin{tabular}[c]{@{}c@{}}\textbf{Cosine}\\\textbf{ Distance}\end{tabular}} & \multicolumn{2}{c}{\begin{tabular}[c]{@{}c@{}}\textbf{Wasserstein}\\\textbf{ Distance}\end{tabular}}  \\ 
\cline{3-10}
                                                                                       &                                                                                            & TPR                              & TNR                              & TPR                              & TNR                              & TPR                             & TNR                                                           & TPR                             & TNR                                                                 \\ 
\hline
\multirow{4}{*}{\begin{tabular}[c]{@{}c@{}}Lunar\\ Lander\end{tabular}}                & BC                                                                                         & 97.09$\pm$1.09  & 100.00$\pm$0.00 & 94.97$\pm$1.31  & 100.00$\pm$0.00 & 95.09$\pm$1.41 & 100.00$\pm$0.00                              & 96.55$\pm$1.98 & 99.93$\pm$0.26                                     \\
                                                                                       & BCQ                                                                                        & 96.97$\pm$1.65  & 100.00$\pm$0.00 & 94.53$\pm$1.20  & 100.00$\pm$0.00 & 94.48$\pm$1.09 & 99.98$\pm$0.10                               & 97.03$\pm$1.51 & 99.78$\pm$0.38                                     \\
                                                                                       & IQL                                                                                        & 96.89$\pm$1.96  & 99.90$\pm$0.37  & 95.22$\pm$2.85  & 99.98$\pm$0.07  & 95.03$\pm$3.13 & 99.91$\pm$0.22                               & 96.82$\pm$2.14 & 96.85$\pm$6.45                                     \\
                                                                                       & TD3PlusBC                                                                                  & 97.24$\pm$2.17  & 99.32$\pm$0.84  & 93.77$\pm$3.64  & 99.82$\pm$0.45  & 93.81$\pm$3.71 & 99.78$\pm$0.49                               & 97.54$\pm$1.16 & 95.45$\pm$3.71                                     \\ 
\cline{2-10}
\multirow{4}{*}{\begin{tabular}[c]{@{}c@{}}Bipedal\\ Walker\end{tabular}}              & BC                                                                                         & 95.14$\pm$3.54  & 100.00$\pm$0.00 & 89.68$\pm$10.06 & 100.00$\pm$0.00 & 97.70$\pm$3.59 & 100.00$\pm$0.00                              & 94.80$\pm$3.69 & 100.00$\pm$0.00                                    \\
                                                                                       & BCQ                                                                                        & 93.90$\pm$5.98  & 100.00$\pm$0.00 & 95.47$\pm$3.37  & 100.00$\pm$0.00 & 98.69$\pm$0.93 & 100.00$\pm$0.00                              & 95.35$\pm$4.04 & 100.00$\pm$0.00                                    \\
                                                                                       & IQL                                                                                        & 88.55$\pm$10.61 & 100.00$\pm$0.00 & 87.79$\pm$8.56  & 100.00$\pm$0.00 & 98.80$\pm$1.27 & 100.00$\pm$0.00                              & 90.68$\pm$8.87 & 100.00$\pm$0.00                                    \\
                                                                                       & TD3PlusBC                                                                                  & 97.39$\pm$5.22  & 94.30$\pm$20.23 & 96.57$\pm$6.86  & 90.88$\pm$22.63 & 97.30$\pm$4.95 & 88.39$\pm$24.21                              & 96.08$\pm$7.85 & 84.40$\pm$30.65                                    \\ 
\cline{2-10}
\multirow{4}{*}{Ant}                                                                   & BC                                                                                         & 90.61$\pm$6.99  & 99.93$\pm$0.16  & 92.25$\pm$4.98  & 99.95$\pm$0.17  & 98.91$\pm$0.99 & 85.17$\pm$28.30                              & 96.23$\pm$3.90 & 99.92$\pm$0.16                                     \\
                                                                                       & BCQ                                                                                        & 92.65$\pm$3.46  & 99.78$\pm$0.50  & 90.00$\pm$5.47  & 99.88$\pm$0.27  & 98.02$\pm$1.01 & 85.88$\pm$28.10                              & 98.11$\pm$1.42 & 99.76$\pm$0.52                                     \\
                                                                                       & IQL                                                                                        & 97.05$\pm$1.06  & 99.58$\pm$0.92  & 94.44$\pm$2.40  & 99.63$\pm$0.72  & 99.12$\pm$0.43 & 85.16$\pm$28.62                              & 98.50$\pm$1.30 & 99.57$\pm$0.93                                     \\
                                                                                       & TD3PlusBC                                                                                  & 93.57$\pm$7.04  & 99.35$\pm$1.27  & 93.15$\pm$4.39  & 99.59$\pm$1.05  & 99.35$\pm$0.75 & 87.99$\pm$26.18                              & 97.86$\pm$1.32 & 99.30$\pm$1.36                                     \\
\bottomrule
\end{tabular}
\end{table*}

\begin{table*}[t]
\centering
\caption{The impact of shadow models’ amount. 
The TPR and TNR results of \sysname with 21 shadow models.}
\label{tab:overall audit accuracy based on Grubbs (21 Shadow Models)}
    \setlength{\tabcolsep}{0.6em}
    \renewcommand{\arraystretch}{1.1}
    \footnotesize
\begin{tabular}{cccccccccc} 
\toprule
\multirow{2}{*}{\begin{tabular}[c]{@{}c@{}}\textbf{Task}\\\textbf{ Name}\end{tabular}} & \multirow{2}{*}{\begin{tabular}[c]{@{}c@{}}\textbf{Offline}\\\textbf{ Model}\end{tabular}} & \multicolumn{2}{c}{\textbf{L1 Norm}}                                & \multicolumn{2}{c}{\textbf{L2 Norm}}                               & \multicolumn{2}{c}{\begin{tabular}[c]{@{}c@{}}\textbf{Cosine}\\\textbf{ Distance}\end{tabular}} & \multicolumn{2}{c}{\begin{tabular}[c]{@{}c@{}}\textbf{Wasserstein}\\\textbf{ Distance}\end{tabular}}  \\ 
\cline{3-10}
                                                                                       &                                                                                            & TPR                              & TNR                              & TPR                             & TNR                              & TPR                             & TNR                                                           & TPR                             & TNR                                                                 \\ 
\hline
\multirow{4}{*}{\begin{tabular}[c]{@{}c@{}}Lunar\\ Lander\end{tabular}}                & BC                                                                                         & 99.25$\pm$0.97  & 100.00$\pm$0.00 & 98.13$\pm$1.88 & 100.00$\pm$0.00 & 98.00$\pm$1.84 & 100.00$\pm$0.00                              & 99.11$\pm$0.93 & 99.96$\pm$0.10                                     \\
                                                                                       & BCQ                                                                                        & 99.56$\pm$0.32  & 100.00$\pm$0.00 & 98.40$\pm$0.68 & 100.00$\pm$0.00 & 98.27$\pm$0.47 & 99.99$\pm$0.04                               & 98.80$\pm$0.78 & 99.75$\pm$0.41                                     \\
                                                                                       & IQL                                                                                        & 97.95$\pm$2.45  & 99.96$\pm$0.17  & 97.11$\pm$3.12 & 99.98$\pm$0.09  & 96.85$\pm$3.40 & 99.92$\pm$0.19                               & 97.11$\pm$3.65 & 97.47$\pm$5.34                                     \\
                                                                                       & TD3PlusBC                                                                                  & 97.87$\pm$3.45  & 99.45$\pm$0.84  & 96.09$\pm$3.91 & 99.73$\pm$0.58  & 95.78$\pm$4.29 & 99.95$\pm$0.14                               & 98.00$\pm$2.60 & 96.27$\pm$3.43                                     \\ 
\cline{2-10}
\multirow{4}{*}{\begin{tabular}[c]{@{}c@{}}Bipedal\\ Walker\end{tabular}}              & BC                                                                                         & 97.07$\pm$3.60  & 100.00$\pm$0.00 & 97.24$\pm$4.49 & 100.00$\pm$0.00 & 97.69$\pm$4.51 & 100.00$\pm$0.00                              & 98.36$\pm$2.67 & 100.00$\pm$0.00                                    \\
                                                                                       & BCQ                                                                                        & 100.00$\pm$0.00 & 100.00$\pm$0.00 & 99.56$\pm$0.69 & 100.00$\pm$0.00 & 99.07$\pm$1.65 & 100.00$\pm$0.00                              & 99.96$\pm$0.09 & 100.00$\pm$0.00                                    \\
                                                                                       & IQL                                                                                        & 95.91$\pm$4.93  & 100.00$\pm$0.00 & 96.36$\pm$4.57 & 100.00$\pm$0.00 & 99.51$\pm$0.26 & 100.00$\pm$0.00                              & 96.44$\pm$4.54 & 100.00$\pm$0.00                                    \\
                                                                                       & TD3PlusBC                                                                                  & 99.91$\pm$0.18  & 95.05$\pm$19.14 & 99.87$\pm$0.27 & 93.96$\pm$20.97 & 99.82$\pm$0.36 & 92.79$\pm$21.27                              & 99.91$\pm$0.18 & 91.78$\pm$21.28                                    \\ 
\cline{2-10}
\multirow{4}{*}{Ant}                                                                   & BC                                                                                         & 98.13$\pm$1.55  & 99.91$\pm$0.18  & 97.73$\pm$1.19 & 99.86$\pm$0.41  & 99.73$\pm$0.53 & 86.82$\pm$26.97                              & 97.55$\pm$1.57 & 99.90$\pm$0.21                                     \\
                                                                                       & BCQ                                                                                        & 97.16$\pm$2.73  & 99.81$\pm$0.42  & 96.67$\pm$2.18 & 99.84$\pm$0.42  & 99.69$\pm$0.41 & 87.53$\pm$26.74                              & 98.58$\pm$1.69 & 99.80$\pm$0.43                                     \\
                                                                                       & IQL                                                                                        & 95.91$\pm$3.87  & 99.64$\pm$0.77  & 96.49$\pm$3.89 & 99.68$\pm$0.63  & 99.51$\pm$0.67 & 86.53$\pm$27.80                              & 97.65$\pm$2.19 & 99.64$\pm$0.78                                     \\
                                                                                       & TD3PlusBC                                                                                  & 99.44$\pm$0.60  & 99.23$\pm$1.53  & 98.27$\pm$1.03 & 99.36$\pm$1.64  & 99.76$\pm$0.33 & 88.42$\pm$25.93                              & 99.79$\pm$0.27 & 99.18$\pm$1.65                                     \\
\bottomrule
\end{tabular}
\end{table*}

\begin{table*}[t]
\centering
\caption{The impact of significance level. 
The TPR and TNR results of \sysname with $\sigma=0.001$.}
\label{tab:overall audit accuracy based on Grubbs (sigma=0.001)}
    \setlength{\tabcolsep}{0.6em}
    \renewcommand{\arraystretch}{1.1}
    \footnotesize
\begin{tabular}{cccccccccc}
\toprule
\multirow{2}{*}{\begin{tabular}[c]{@{}c@{}}\textbf{Task}\\\textbf{ Name}\end{tabular}} & \multirow{2}{*}{\begin{tabular}[c]{@{}c@{}}\textbf{Offline}\\\textbf{ Model}\end{tabular}} & \multicolumn{2}{c}{\textbf{L1 Norm}}                               & \multicolumn{2}{c}{\textbf{L2 Norm}}                               & \multicolumn{2}{c}{\begin{tabular}[c]{@{}c@{}}\textbf{Cosine}\\\textbf{ Distance}\end{tabular}} & \multicolumn{2}{c}{\begin{tabular}[c]{@{}c@{}}\textbf{Wasserstein}\\\textbf{ Distance}\end{tabular}}  \\ 
\cline{3-10}
                                                                                       &                                                                                            & TPR                             & TNR                              & TPR                             & TNR                              & TPR                             & TNR                                                           & TPR                              & TNR                                                                \\ 
\hline
\multirow{4}{*}{\begin{tabular}[c]{@{}c@{}}Lunar\\ Lander\end{tabular}}                & BC                                                                                         & 99.63$\pm$0.19 & 100.00$\pm$0.00 & 98.21$\pm$0.55 & 100.00$\pm$0.00 & 98.21$\pm$0.63 & 100.00$\pm$0.00                              & 99.31$\pm$0.38  & 99.84$\pm$0.38                                    \\
                                                                                       & BCQ                                                                                        & 99.15$\pm$0.67 & 100.00$\pm$0.00 & 97.60$\pm$1.13 & 100.00$\pm$0.00 & 97.63$\pm$1.04 & 99.97$\pm$0.13                               & 98.59$\pm$0.95  & 99.61$\pm$0.53                                    \\
                                                                                       & IQL                                                                                        & 99.31$\pm$0.90 & 99.83$\pm$0.40  & 98.56$\pm$1.51 & 99.96$\pm$0.16  & 98.51$\pm$1.59 & 99.79$\pm$0.51                               & 99.04$\pm$1.21  & 94.88$\pm$8.21                                    \\
                                                                                       & TD3PlusBC                                                                                  & 99.20$\pm$1.10 & 99.22$\pm$0.89  & 97.47$\pm$2.33 & 99.61$\pm$0.55  & 97.55$\pm$2.22 & 99.75$\pm$0.51                               & 99.49$\pm$0.56  & 92.45$\pm$5.32                                    \\ 
\cline{2-10}
\multirow{4}{*}{\begin{tabular}[c]{@{}c@{}}Bipedal\\ Walker\end{tabular}}              & BC                                                                                         & 99.97$\pm$0.05 & 100.00$\pm$0.00 & 98.67$\pm$2.67 & 100.00$\pm$0.00 & 98.64$\pm$2.66 & 100.00$\pm$0.00                              & 100.00$\pm$0.00 & 100.00$\pm$0.00                                   \\
                                                                                       & BCQ                                                                                        & 99.95$\pm$0.06 & 100.00$\pm$0.00 & 99.73$\pm$0.34 & 100.00$\pm$0.00 & 99.95$\pm$0.06 & 100.00$\pm$0.00                              & 99.97$\pm$0.05  & 100.00$\pm$0.00                                   \\
                                                                                       & IQL                                                                                        & 97.04$\pm$5.47 & 100.00$\pm$0.00 & 95.81$\pm$5.06 & 100.00$\pm$0.00 & 99.87$\pm$0.27 & 100.00$\pm$0.00                              & 97.68$\pm$4.51  & 100.00$\pm$0.00                                   \\
                                                                                       & TD3PlusBC                                                                                  & 99.92$\pm$0.16 & 89.69$\pm$22.99 & 97.17$\pm$5.65 & 85.59$\pm$27.05 & 97.20$\pm$5.60 & 82.52$\pm$30.48                              & 99.68$\pm$0.64  & 80.18$\pm$35.21                                   \\ 
\cline{2-10}
\multirow{4}{*}{Ant}                                                                   & BC                                                                                         & 99.52$\pm$0.50 & 99.86$\pm$0.25  & 98.48$\pm$0.80 & 99.88$\pm$0.40  & 99.55$\pm$0.66 & 80.58$\pm$33.24                              & 99.36$\pm$0.49  & 99.85$\pm$0.25                                    \\
                                                                                       & BCQ                                                                                        & 98.91$\pm$1.68 & 99.71$\pm$0.60  & 97.97$\pm$1.43 & 99.81$\pm$0.47  & 99.87$\pm$0.15 & 81.88$\pm$31.60                              & 99.52$\pm$0.64  & 99.69$\pm$0.63                                    \\
                                                                                       & IQL                                                                                        & 98.88$\pm$1.27 & 99.53$\pm$0.97  & 98.42$\pm$1.80 & 99.60$\pm$0.73  & 99.71$\pm$0.52 & 80.90$\pm$32.60                              & 99.92$\pm$0.06  & 99.49$\pm$1.05                                    \\
                                                                                       & TD3PlusBC                                                                                  & 99.62$\pm$0.46 & 98.92$\pm$2.08  & 98.73$\pm$0.97 & 99.24$\pm$1.91  & 99.79$\pm$0.36 & 84.36$\pm$28.07                              & 99.71$\pm$0.58  & 98.78$\pm$2.15                                    \\
\bottomrule
\end{tabular}
\end{table*}

\begin{table*}[t]
\centering
\caption{The impact of significance level. 
The TPR and TNR results of \sysname with $\sigma=0.0001$.}
\label{tab:overall audit accuracy based on Grubbs (sigma=0.0001)}
    \setlength{\tabcolsep}{0.6em}
    \renewcommand{\arraystretch}{1.1}
    \footnotesize
\begin{tabular}{cccccccccc} 
\toprule
\multirow{2}{*}{\begin{tabular}[c]{@{}c@{}}\textbf{Task}\\\textbf{ Name}\end{tabular}} & \multirow{2}{*}{\begin{tabular}[c]{@{}c@{}}\textbf{Offline}\\\textbf{ Model}\end{tabular}} & \multicolumn{2}{c}{\textbf{L1 Norm}}                                & \multicolumn{2}{c}{\textbf{L2 Norm}}                               & \multicolumn{2}{c}{\begin{tabular}[c]{@{}c@{}}\textbf{Cosine}\\\textbf{ Distance}\end{tabular}} & \multicolumn{2}{c}{\begin{tabular}[c]{@{}c@{}}\textbf{Wasserstein}\\\textbf{ Distance}\end{tabular}}  \\ 
\cline{3-10}
                                                                                       &                                                                                            & TPR                              & TNR                              & TPR                             & TNR                              & TPR                              & TNR                                                          & TPR                              & TNR                                                                \\ 
\hline
\multirow{4}{*}{\begin{tabular}[c]{@{}c@{}}Lunar\\ Lander\end{tabular}}                & BC                                                                                         & 99.87$\pm$0.12  & 100.00$\pm$0.00 & 98.93$\pm$0.34 & 100.00$\pm$0.00 & 99.04$\pm$0.35  & 100.00$\pm$0.00                             & 99.79$\pm$0.22  & 99.56$\pm$0.79                                    \\
                                                                                       & BCQ                                                                                        & 99.49$\pm$0.46  & 100.00$\pm$0.00 & 98.48$\pm$0.79 & 100.00$\pm$0.00 & 98.48$\pm$0.82  & 99.95$\pm$0.20                              & 99.33$\pm$0.54  & 98.88$\pm$1.54                                    \\
                                                                                       & IQL                                                                                        & 99.71$\pm$0.52  & 99.66$\pm$0.55  & 98.99$\pm$1.04 & 99.91$\pm$0.21  & 98.91$\pm$1.17  & 99.49$\pm$1.15                              & 99.52$\pm$0.70  & 91.46$\pm$10.53                                   \\
                                                                                       & TD3PlusBC                                                                                  & 99.55$\pm$0.56  & 98.80$\pm$1.27  & 98.48$\pm$1.48 & 98.96$\pm$1.24  & 98.56$\pm$1.49  & 99.38$\pm$0.87                              & 99.84$\pm$0.16  & 88.16$\pm$6.84                                    \\ 
\cline{2-10}
\multirow{4}{*}{\begin{tabular}[c]{@{}c@{}}Bipedal\\ Walker\end{tabular}}              & BC                                                                                         & 100.00$\pm$0.00 & 100.00$\pm$0.00 & 98.67$\pm$2.67 & 100.00$\pm$0.00 & 98.80$\pm$2.40  & 100.00$\pm$0.00                             & 100.00$\pm$0.00 & 100.00$\pm$0.00                                   \\
                                                                                       & BCQ                                                                                        & 100.00$\pm$0.00 & 100.00$\pm$0.00 & 99.81$\pm$0.26 & 100.00$\pm$0.00 & 100.00$\pm$0.00 & 100.00$\pm$0.00                             & 100.00$\pm$0.00 & 100.00$\pm$0.00                                   \\
                                                                                       & IQL                                                                                        & 98.53$\pm$2.87  & 100.00$\pm$0.00 & 96.27$\pm$4.74 & 100.00$\pm$0.00 & 99.87$\pm$0.27  & 100.00$\pm$0.00                             & 98.67$\pm$2.67  & 100.00$\pm$0.00                                   \\
                                                                                       & TD3PlusBC                                                                                  & 100.00$\pm$0.00 & 79.08$\pm$38.26 & 97.23$\pm$5.55 & 75.43$\pm$39.10 & 97.33$\pm$5.33  & 74.60$\pm$39.23                             & 99.97$\pm$0.05  & 73.98$\pm$38.98                                   \\ 
\cline{2-10}
\multirow{4}{*}{Ant}                                                                   & BC                                                                                         & 99.95$\pm$0.11  & 99.80$\pm$0.41  & 99.15$\pm$0.67 & 99.85$\pm$0.43  & 99.73$\pm$0.41  & 77.36$\pm$35.68                             & 99.73$\pm$0.22  & 99.79$\pm$0.42                                    \\
                                                                                       & BCQ                                                                                        & 99.33$\pm$1.21  & 99.56$\pm$0.76  & 98.75$\pm$1.11 & 99.80$\pm$0.48  & 99.92$\pm$0.11  & 78.87$\pm$33.94                             & 99.81$\pm$0.23  & 99.52$\pm$0.79                                    \\
                                                                                       & IQL                                                                                        & 99.73$\pm$0.29  & 99.35$\pm$1.25  & 98.91$\pm$1.45 & 99.56$\pm$0.80  & 99.89$\pm$0.21  & 77.78$\pm$34.93                             & 100.00$\pm$0.00 & 99.32$\pm$1.31                                    \\
                                                                                       & TD3PlusBC                                                                                  & 99.96$\pm$0.05  & 98.65$\pm$2.42  & 99.35$\pm$0.50 & 99.11$\pm$2.07  & 99.92$\pm$0.11  & 81.33$\pm$29.74                             & 99.90$\pm$0.19  & 98.06$\pm$3.18                                    \\
\bottomrule
\end{tabular}
\end{table*}

\begin{table*}[t]
\centering
\caption{The impact of trajectory size. 
The TPR and TNR results of \sysname with 25\% trajectory size.}
\label{tab:overall_audit_accuracy_based_on_Grubbs_0.25_trajsize}
\setlength{\tabcolsep}{0.6em}
\renewcommand{\arraystretch}{1.1}
\footnotesize
\begin{tabular}{cccccccccc} 
    \toprule
    \multirow{2}{*}{\begin{tabular}[c]{@{}c@{}}\textbf{Task}\\\textbf{ Name}\end{tabular}} & \multirow{2}{*}{\begin{tabular}[c]{@{}c@{}}\textbf{Offline}\\\textbf{ Model}\end{tabular}} & \multicolumn{2}{c}{\textbf{L1 Norm}}                               & \multicolumn{2}{c}{\textbf{L2 Norm}}                               & \multicolumn{2}{c}{\begin{tabular}[c]{@{}c@{}}\textbf{Cosine}\\\textbf{ Distance}\end{tabular}} & \multicolumn{2}{c}{\begin{tabular}[c]{@{}c@{}}\textbf{Wasserstein}\\\textbf{ Distance}\end{tabular}}  \\ 
    \cline{3-10}
                                                                                           &                                                                                            & TPR                             & TNR                              & TPR                             & TNR                              & TPR                             & TNR                                                           & TPR                             & TNR                                                                 \\ 
    \hline
    \multirow{4}{*}{\begin{tabular}[c]{@{}c@{}}Lunar\\ Lander\end{tabular}}                & BC                                                                                         & 98.13$\pm$1.05 & 99.53$\pm$1.30  & 96.27$\pm$2.00 & 99.64$\pm$1.09  & 96.29$\pm$1.97 & 99.13$\pm$2.41                               & 98.10$\pm$0.92 & 97.74$\pm$2.30                                     \\
                                                                                           & BCQ                                                                                        & 98.45$\pm$0.51 & 99.28$\pm$1.19  & 97.33$\pm$0.76 & 99.59$\pm$0.71  & 96.91$\pm$1.06 & 99.01$\pm$1.33                               & 98.56$\pm$1.10 & 94.02$\pm$4.44                                     \\
                                                                                           & IQL                                                                                        & 98.11$\pm$1.65 & 95.42$\pm$5.71  & 96.72$\pm$2.57 & 97.10$\pm$3.96  & 96.80$\pm$2.25 & 92.42$\pm$5.78                               & 98.19$\pm$1.55 & 84.64$\pm$9.09                                     \\
                                                                                           & TD3PlusBC                                                                                  & 98.00$\pm$2.42 & 95.44$\pm$4.57  & 96.43$\pm$3.13 & 96.86$\pm$3.24  & 95.95$\pm$2.56 & 92.95$\pm$3.92                               & 98.45$\pm$1.43 & 81.51$\pm$9.13                                     \\ 
    \cline{2-10}
    \multirow{4}{*}{\begin{tabular}[c]{@{}c@{}}Bipedal\\ Walker\end{tabular}}              & BC                                                                                         & 99.20$\pm$0.97 & 100.00$\pm$0.00 & 97.47$\pm$3.12 & 100.00$\pm$0.00 & 98.61$\pm$2.71 & 100.00$\pm$0.00                              & 99.36$\pm$0.90 & 100.00$\pm$0.00                                    \\
                                                                                           & BCQ                                                                                        & 98.59$\pm$2.63 & 100.00$\pm$0.00 & 97.68$\pm$2.95 & 100.00$\pm$0.00 & 99.68$\pm$0.27 & 100.00$\pm$0.00                              & 98.61$\pm$2.45 & 100.00$\pm$0.00                                    \\
                                                                                           & IQL                                                                                        & 96.80$\pm$5.37 & 100.00$\pm$0.00 & 95.73$\pm$5.42 & 100.00$\pm$0.00 & 99.41$\pm$0.45 & 100.00$\pm$0.00                              & 97.01$\pm$5.45 & 100.00$\pm$0.00                                    \\
                                                                                           & TD3PlusBC                                                                                  & 97.55$\pm$4.91 & 94.08$\pm$21.64 & 97.20$\pm$5.60 & 89.72$\pm$24.95 & 96.93$\pm$5.74 & 90.65$\pm$23.69                              & 97.41$\pm$5.17 & 84.06$\pm$33.72                                    \\ 
    \cline{2-10}
    \multirow{4}{*}{Ant}                                                                   & BC                                                                                         & 98.85$\pm$0.67 & 99.90$\pm$0.35  & 97.04$\pm$1.24 & 99.84$\pm$0.47  & 99.49$\pm$0.88 & 92.58$\pm$19.10                              & 98.96$\pm$0.81 & 99.90$\pm$0.35                                     \\
                                                                                           & BCQ                                                                                        & 98.11$\pm$1.40 & 99.85$\pm$0.34  & 97.36$\pm$1.61 & 99.78$\pm$0.46  & 99.49$\pm$0.76 & 92.64$\pm$19.34                              & 99.12$\pm$0.59 & 99.84$\pm$0.34                                     \\
                                                                                           & IQL                                                                                        & 98.45$\pm$1.00 & 99.90$\pm$0.30  & 96.45$\pm$1.47 & 99.85$\pm$0.40  & 99.68$\pm$0.51 & 92.56$\pm$20.01                              & 99.04$\pm$0.55 & 99.80$\pm$0.49                                     \\
                                                                                           & TD3PlusBC                                                                                  & 98.80$\pm$1.33 & 99.76$\pm$0.43  & 96.92$\pm$1.60 & 99.70$\pm$0.67  & 99.22$\pm$1.18 & 93.58$\pm$17.31                              & 99.28$\pm$1.07 & 99.74$\pm$0.45                                     \\
    \bottomrule
\end{tabular}
\end{table*}

\begin{table*}[t]
\centering
\caption{The impact of trajectory size. 
The TPR and TNR results of \sysname with 50\% trajectory size.} 
\label{tab:overall_audit_accuracy_based_on_Grubbs_0.5_trajsize}
\setlength{\tabcolsep}{0.6em}
\renewcommand{\arraystretch}{1.1}
\footnotesize
\begin{tabular}{cccccccccc} 
\toprule
\multirow{2}{*}{\begin{tabular}[c]{@{}c@{}}\textbf{Task}\\\textbf{ Name}\end{tabular}} & \multirow{2}{*}{\begin{tabular}[c]{@{}c@{}}\textbf{Offline}\\\textbf{ Model}\end{tabular}} & \multicolumn{2}{c}{\textbf{L1 Norm}}                               & \multicolumn{2}{c}{\textbf{L2 Norm}}                               & \multicolumn{2}{c}{\begin{tabular}[c]{@{}c@{}}\textbf{Cosine}\\\textbf{ Distance}\end{tabular}} & \multicolumn{2}{c}{\begin{tabular}[c]{@{}c@{}}\textbf{Wasserstein}\\\textbf{ Distance}\end{tabular}}  \\ 
\cline{3-10}
                                                                                        &                                                                                            & TPR                             & TNR                              & TPR                             & TNR                              & TPR                             & TNR                                                           & TPR                             & TNR                                                                 \\ 
\hline
\multirow{4}{*}{\begin{tabular}[c]{@{}c@{}}Lunar\\ Lander\end{tabular}}                & BC                                                                                         & 98.37$\pm$0.68 & 100.00$\pm$0.00 & 97.07$\pm$0.90 & 100.00$\pm$0.00 & 97.25$\pm$0.72 & 100.00$\pm$0.02                              & 98.58$\pm$0.50 & 98.50$\pm$1.91                                     \\
                                                                                        & BCQ                                                                                        & 98.16$\pm$0.55 & 99.96$\pm$0.19  & 96.11$\pm$0.83 & 99.95$\pm$0.20  & 96.40$\pm$0.84 & 99.58$\pm$0.80                               & 97.57$\pm$1.64 & 96.28$\pm$3.72                                     \\
                                                                                        & IQL                                                                                        & 98.03$\pm$2.25 & 98.93$\pm$1.35  & 96.80$\pm$2.84 & 99.29$\pm$1.06  & 97.25$\pm$2.18 & 96.38$\pm$2.48                               & 98.27$\pm$2.29 & 86.89$\pm$10.96                                    \\
                                                                                        & TD3PlusBC                                                                                  & 98.03$\pm$2.33 & 98.28$\pm$2.03  & 96.37$\pm$3.41 & 99.27$\pm$0.97  & 96.51$\pm$2.98 & 95.94$\pm$4.14                               & 98.24$\pm$1.79 & 84.30$\pm$8.08                                     \\ 
\cline{2-10}
\multirow{4}{*}{\begin{tabular}[c]{@{}c@{}}Bipedal\\ Walker\end{tabular}}              & BC                                                                                         & 99.44$\pm$0.75 & 100.00$\pm$0.00 & 97.81$\pm$2.83 & 100.00$\pm$0.00 & 98.67$\pm$2.67 & 100.00$\pm$0.00                              & 99.41$\pm$0.86 & 100.00$\pm$0.00                                    \\
                                                                                        & BCQ                                                                                        & 98.75$\pm$2.38 & 100.00$\pm$0.00 & 97.92$\pm$2.81 & 100.00$\pm$0.00 & 99.89$\pm$0.10 & 100.00$\pm$0.00                              & 99.55$\pm$0.72 & 100.00$\pm$0.00                                    \\
                                                                                        & IQL                                                                                        & 95.68$\pm$6.60 & 100.00$\pm$0.00 & 95.47$\pm$5.34 & 100.00$\pm$0.00 & 99.81$\pm$0.31 & 100.00$\pm$0.00                              & 96.40$\pm$6.11 & 100.00$\pm$0.00                                    \\
                                                                                        & TD3PlusBC                                                                                  & 98.35$\pm$3.31 & 94.29$\pm$21.47 & 97.20$\pm$5.60 & 91.75$\pm$22.51 & 96.96$\pm$6.02 & 90.88$\pm$23.07                              & 97.41$\pm$5.17 & 89.04$\pm$27.90                                    \\ 
\cline{2-10}
\multirow{4}{*}{Ant}                                                                   & BC                                                                                         & 98.21$\pm$0.98 & 99.92$\pm$0.24  & 97.04$\pm$1.33 & 99.85$\pm$0.43  & 99.49$\pm$0.83 & 88.52$\pm$25.25                              & 98.59$\pm$0.81 & 99.92$\pm$0.24                                     \\
                                                                                        & BCQ                                                                                        & 97.76$\pm$2.05 & 99.85$\pm$0.36  & 96.72$\pm$1.50 & 99.81$\pm$0.44  & 99.60$\pm$0.60 & 89.27$\pm$24.10                              & 98.88$\pm$1.30 & 99.84$\pm$0.36                                     \\
                                                                                        & IQL                                                                                        & 97.71$\pm$1.81 & 99.73$\pm$0.61  & 96.53$\pm$1.65 & 99.82$\pm$0.40  & 99.79$\pm$0.30 & 88.67$\pm$25.63                              & 98.99$\pm$0.65 & 99.70$\pm$0.66                                     \\
                                                                                        & TD3PlusBC                                                                                  & 98.52$\pm$1.81 & 99.61$\pm$0.77  & 96.99$\pm$1.57 & 99.74$\pm$0.64  & 99.82$\pm$0.25 & 90.62$\pm$24.30                              & 99.13$\pm$1.29 & 99.58$\pm$0.79                                     \\
\bottomrule
\end{tabular}
\end{table*}

\begin{figure*}[!t]
\includegraphics[width=\hsize]{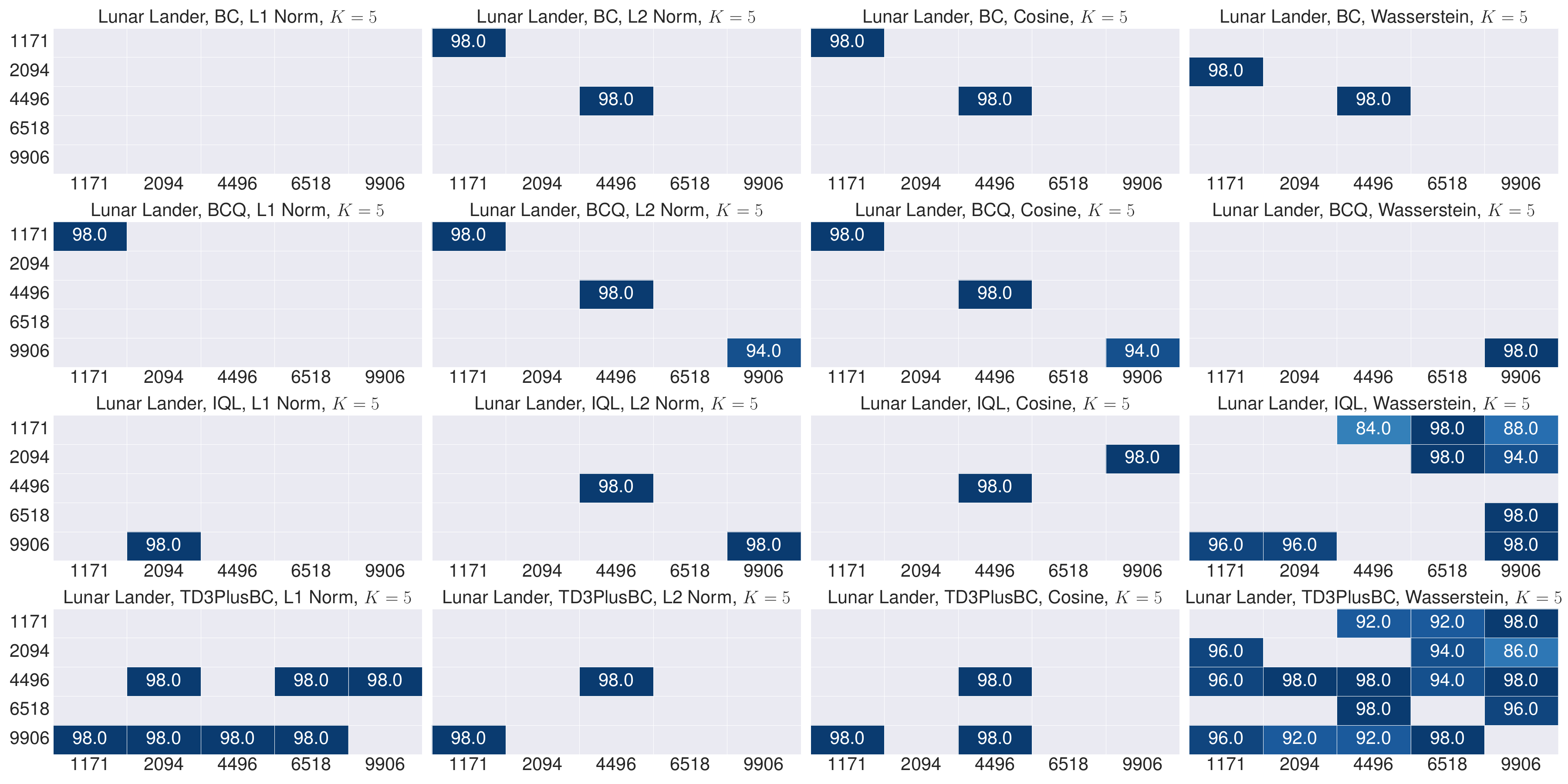}
\vspace{-0.2cm}
\caption{The audit accuracy against model ensemble for Lunar Lander.
The caption of each plot demonstrates the offline DRL model's type, the task, the distance metric, and the hyperparameter $K$ of the model ensemble.  
The x labels are the names of datasets to be audited, \ie, the target datasets. 
The y labels are the names of datasets the suspect models learned, \ie, the actual datasets. 
Thus, the diagonal values show the audit accuracy when the actual dataset is the same as the target dataset, \ie, TPR, and the non-diagonal values are the TNR results. 
The positions without value mean 100\% accuracy. 
}
\label{fig:model ensemble k5-2 on lunarlander}
\end{figure*}

\begin{figure*}[!t]
\includegraphics[width=\hsize]{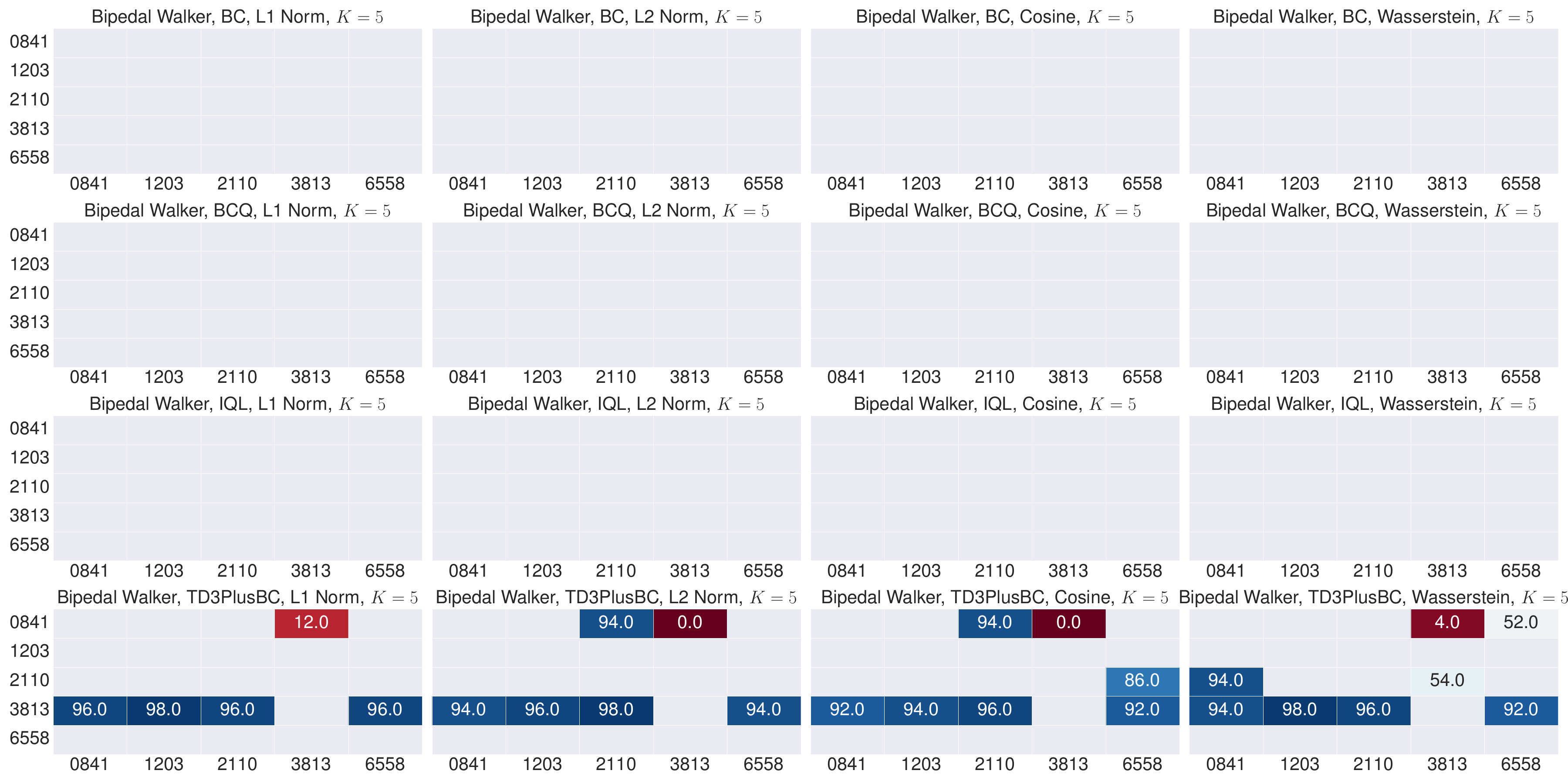}
\vspace{-0.2cm}
\caption{The audit accuracy against model ensemble for Bipedal Walker.
The caption of each plot demonstrates the offline DRL model's type, the task, the distance metric, and the hyperparameter $K$ of the model ensemble.  
The x labels are the names of datasets to be audited, \ie, the target datasets. 
The y labels are the names of datasets the suspect models learned, \ie, the actual datasets. 
Thus, the diagonal values show the audit accuracy when the actual dataset is the same as the target dataset, \ie, TPR, and the non-diagonal values are the TNR results. 
The positions without value mean 100\% accuracy. 
}
\label{fig:model ensemble k5-2 on bipedalwalker}
\end{figure*}

\begin{figure*}[!t]
\includegraphics[width=\hsize]{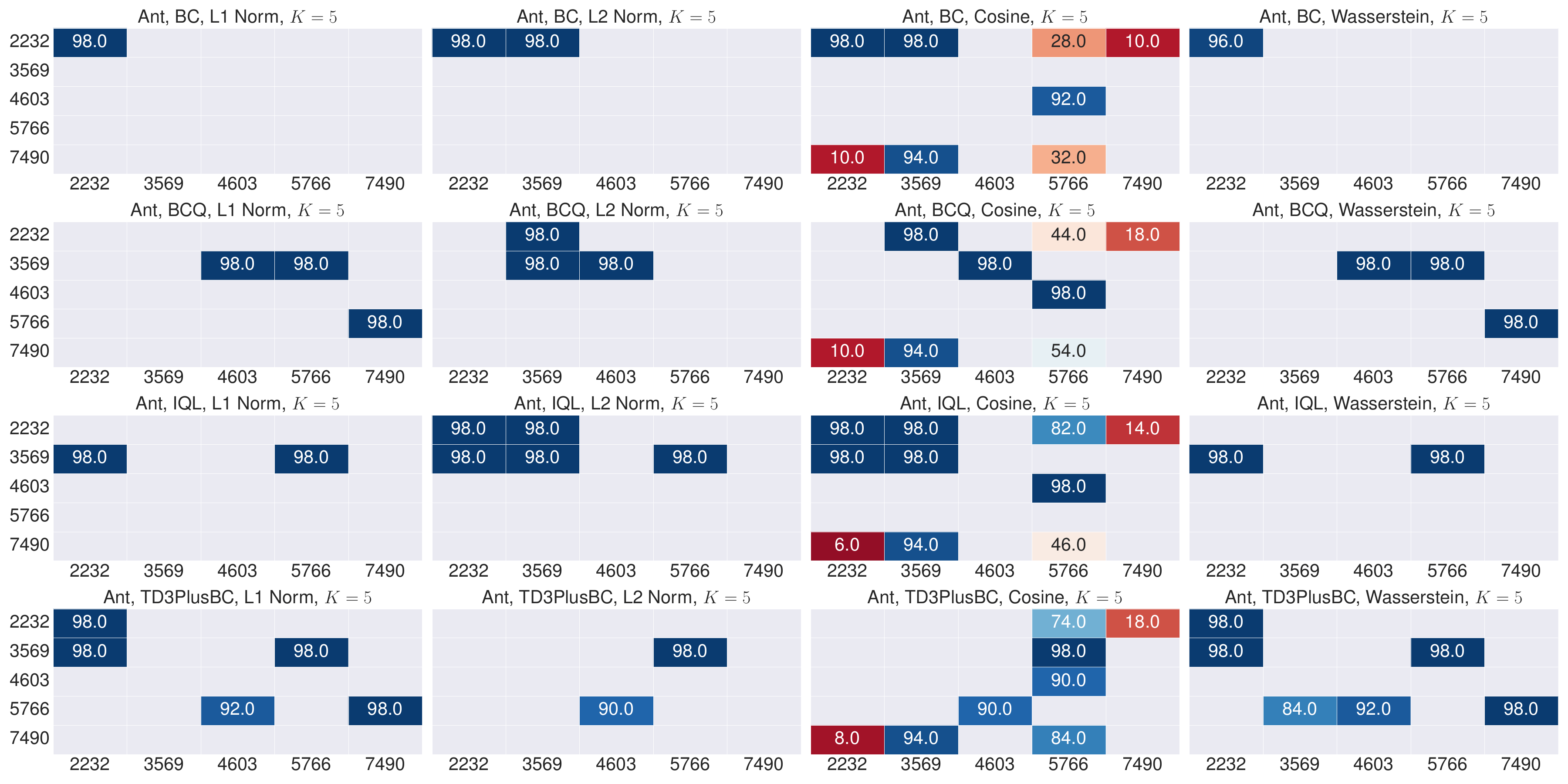}
\vspace{-0.2cm}
\caption{The audit accuracy against model ensemble for Ant.
The caption of each plot demonstrates the offline DRL model's type, the task, the distance metric, and the hyperparameter $K$ of the model ensemble.  
The x labels are the names of datasets to be audited, \ie, the target datasets. 
The y labels are the names of datasets the suspect models learned, \ie, the actual datasets. 
Thus, the diagonal values show the audit accuracy when the actual dataset is the same as the target dataset, \ie, TPR, and the non-diagonal values are the TNR results. 
The positions without value mean 100\% accuracy. 
}
\label{fig:model ensemble k5-2 on ant}
\end{figure*}

\begin{figure*}[!t]
\includegraphics[width=\hsize]{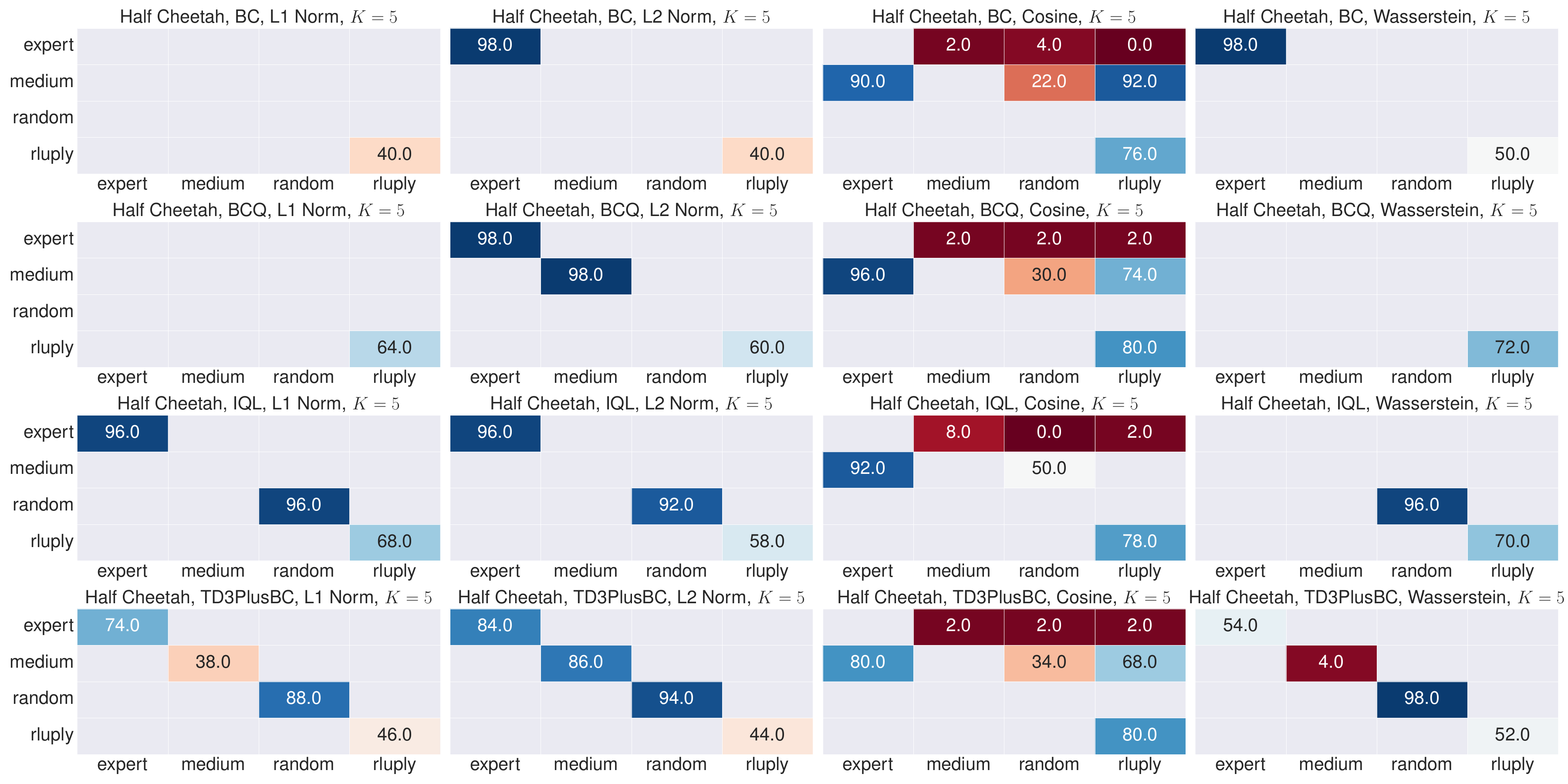}
\vspace{-0.2cm}
\caption{The audit accuracy against model ensemble for Half Cheetah.
The caption of each plot demonstrates the offline DRL model's type, the task, the distance metric, and the hyperparameter $K$ of the model ensemble.  
The x labels are the names of datasets to be audited, \ie, the target datasets. 
The y labels are the names of datasets the suspect models learned, \ie, the actual datasets. 
Thus, the diagonal values show the audit accuracy when the actual dataset is the same as the target dataset, \ie, TPR, and the non-diagonal values are the TNR results. 
The positions without value mean 100\% accuracy. 
}
\label{fig:model ensemble k5-2 on halfcheetah}
\end{figure*}

\begin{figure*}[!t]
\includegraphics[width=\hsize]{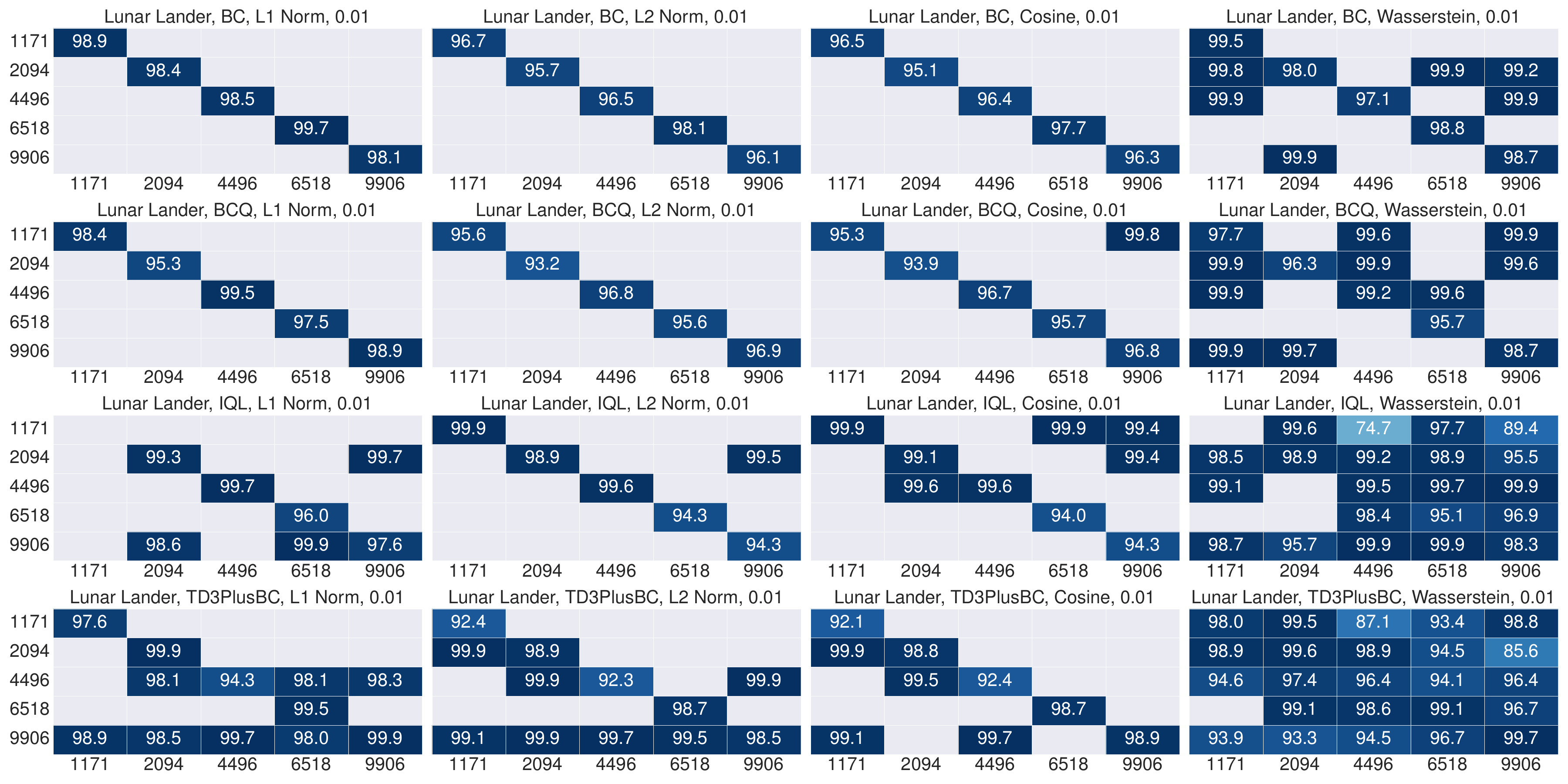}
\vspace{-0.2cm}
\caption{The audit accuracy with Gaussian noise ($\mu=0, \sigma=0.01$) on the suspect models' action for Lunar Lander.
The caption of each plot demonstrates the offline DRL model's type, the task, the distance metric, and the noise strength. 
The x labels are the names of datasets to be audited, \ie, the target datasets. 
The y labels are the names of datasets the suspect models learned, \ie, the actual datasets. 
Thus, the diagonal values show the audit accuracy when the actual dataset is the same as the target dataset, \ie, TPR, and the non-diagonal values are the TNR results. 
The positions without value mean 100\% accuracy. 
}
\label{fig:robustness 0.01 on lunarlander}
\end{figure*}

\begin{figure*}[!t]
\includegraphics[width=\hsize]{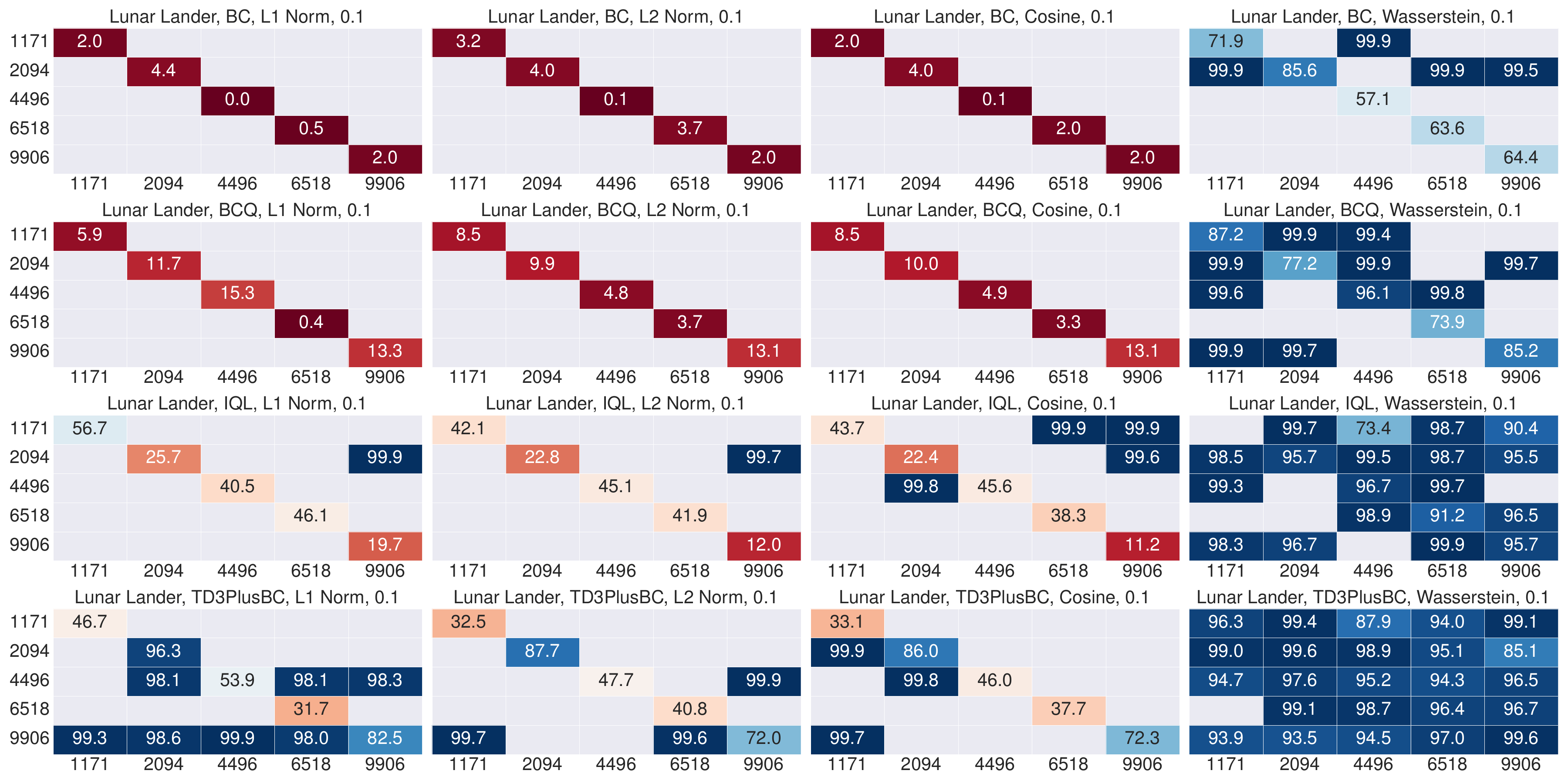}
\vspace{-0.2cm}
\caption{The audit accuracy with Gaussian noise ($\mu=0, \sigma=0.1$) on the suspect models' action for Lunar Lander.
The caption of each plot demonstrates the offline DRL model's type, the task, the distance metric, and the noise strength. 
The x labels are the names of datasets to be audited, \ie, the target datasets. 
The y labels are the names of datasets the suspect models learned, \ie, the actual datasets. 
Thus, the diagonal values show the audit accuracy when the actual dataset is the same as the target dataset, \ie, TPR, and the non-diagonal values are the TNR results. 
The positions without value mean 100\% accuracy. 
}
\label{fig:robustness 0.1 on lunarlander}
\end{figure*}

\begin{figure*}[!t]
\includegraphics[width=\hsize]{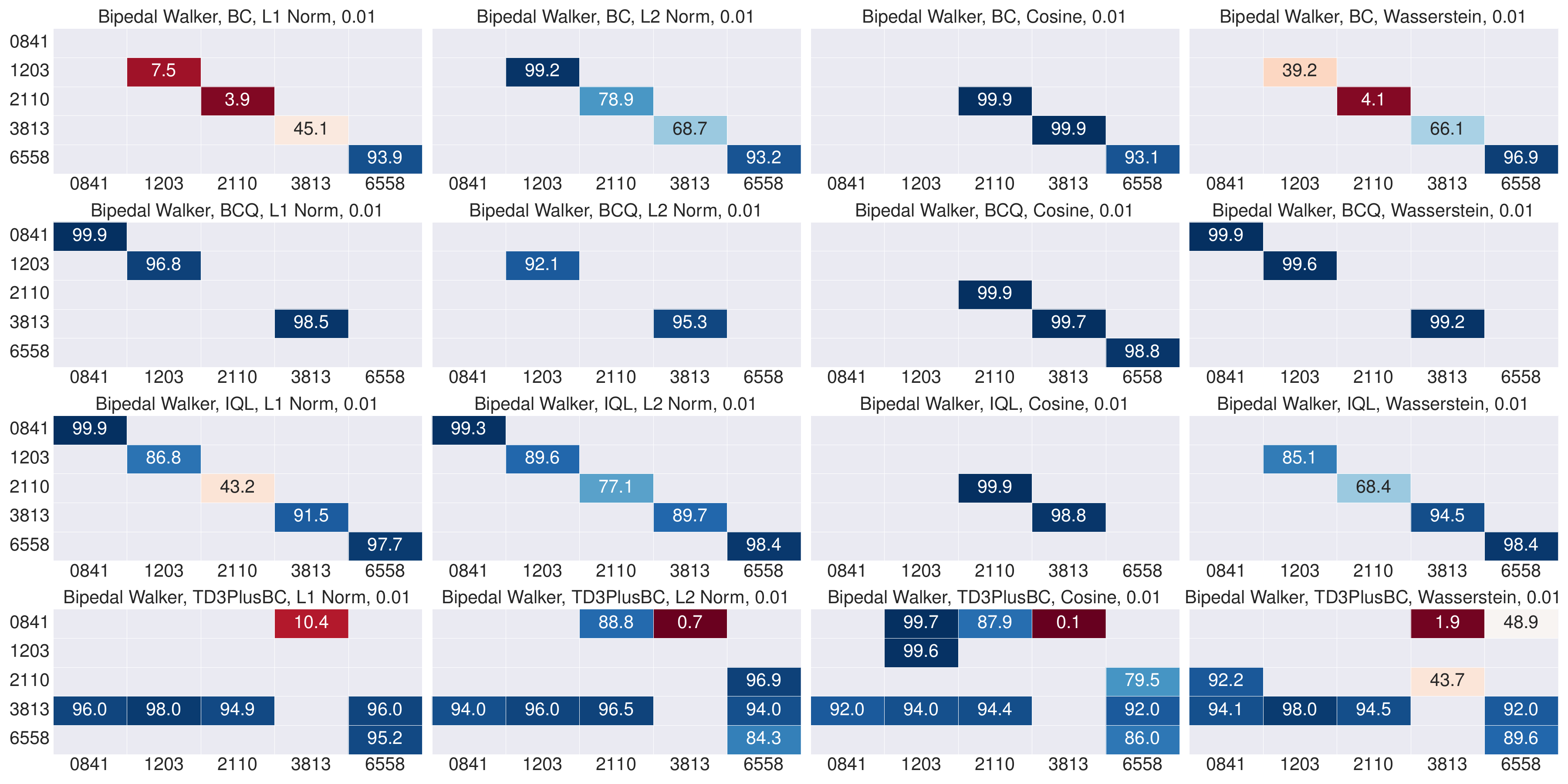}
\vspace{-0.2cm}
\caption{
The audit accuracy with Gaussian noise ($\mu=0, \sigma=0.01$) on the suspect models' action for Bipedal Walker.
The caption of each plot demonstrates the offline DRL model's type, the task, the distance metric, and the noise strength. 
The x labels are the names of datasets to be audited, \ie, the target datasets. 
The y labels are the names of datasets the suspect models learned, \ie, the actual datasets. 
Thus, the diagonal values show the audit accuracy when the actual dataset is the same as the target dataset, \ie, TPR, and the non-diagonal values are the TNR results. 
The positions without value mean 100\% accuracy. 
}
\label{fig:robustness 0.01 on bipedalwalker}
\end{figure*}

\begin{figure*}[!t]
\includegraphics[width=\hsize]{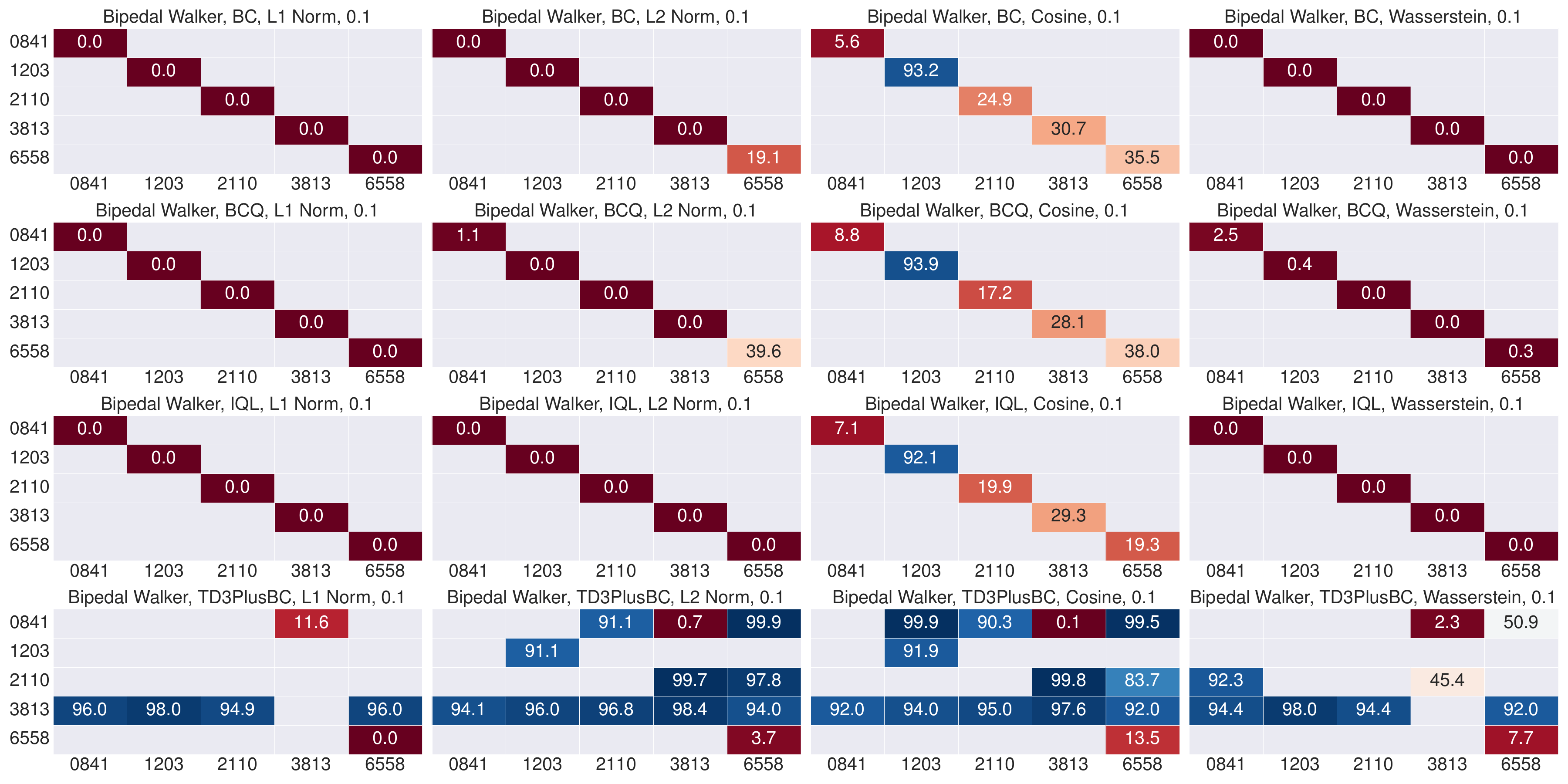}
\vspace{-0.2cm}
\caption{The audit accuracy with Gaussian noise ($\mu=0, \sigma=0.1$) on the suspect models' action for Bipedal Walker.
The caption of each plot demonstrates the offline DRL model's type, the task, the distance metric, and the noise strength. 
The x labels are the names of datasets to be audited, \ie, the target datasets. 
The y labels are the names of datasets the suspect models learned, \ie, the actual datasets. 
Thus, the diagonal values show the audit accuracy when the actual dataset is the same as the target dataset, \ie, TPR, and the non-diagonal values are the TNR results. 
The positions without value mean 100\% accuracy. }
\label{fig:robustness 0.1 on bipedalwalker}
\end{figure*}

\begin{figure*}[!t]
\includegraphics[width=\hsize]{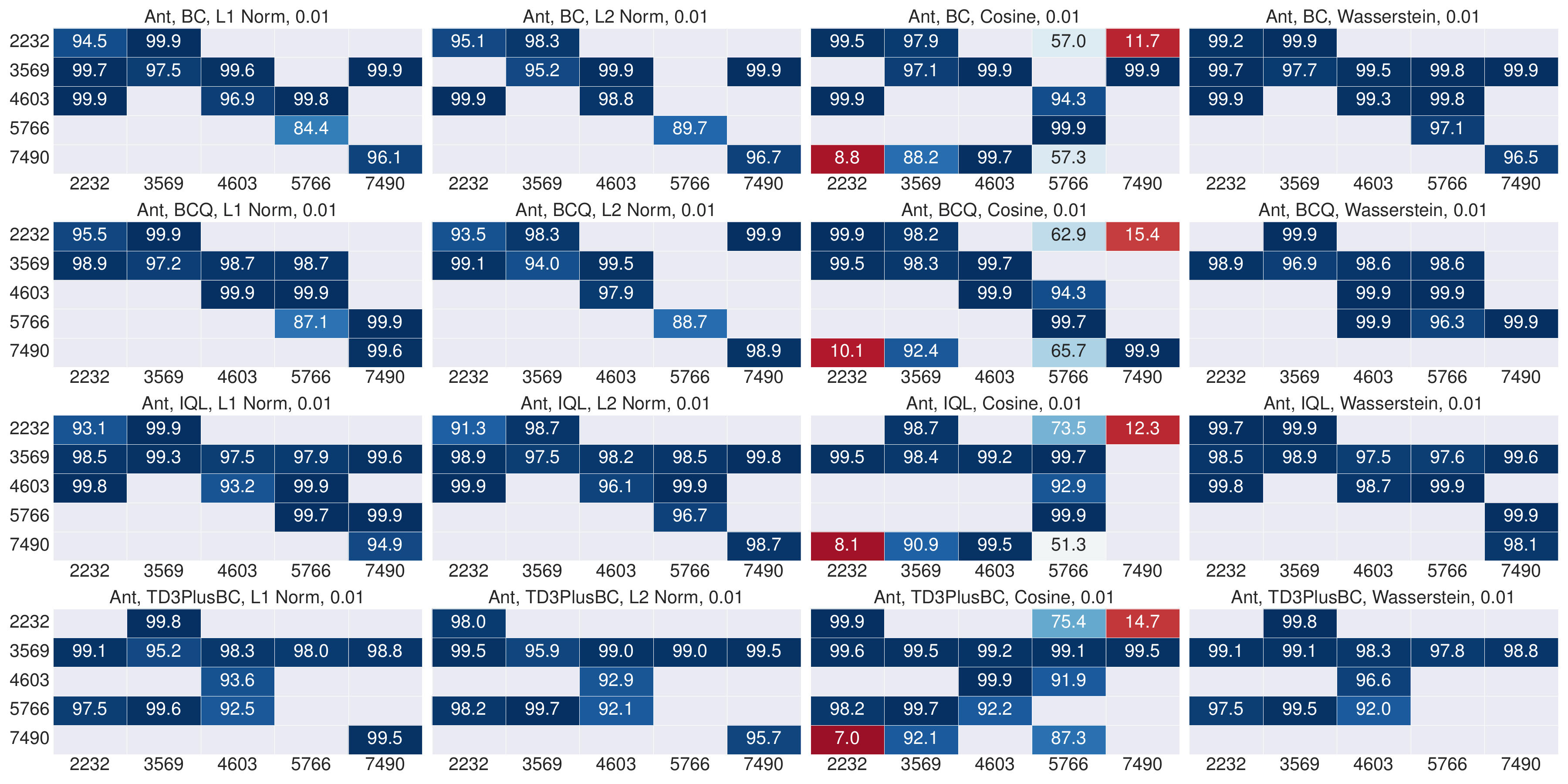}
\vspace{-0.2cm}
\caption{The audit accuracy with Gaussian noise ($\mu=0, \sigma=0.01$) on the suspect models' action for Ant.
The caption of each plot demonstrates the offline DRL model's type, the task, the distance metric, and the noise strength. 
The x labels are the names of datasets to be audited, \ie, the target datasets. 
The y labels are the names of datasets the suspect models learned, \ie, the actual datasets. 
Thus, the diagonal values show the audit accuracy when the actual dataset is the same as the target dataset, \ie, TPR, and the non-diagonal values are the TNR results. 
The positions without value mean 100\% accuracy. 
}
\label{fig:robustness 0.01 on ant}
\end{figure*}

\begin{figure*}[!t]
\includegraphics[width=\hsize]{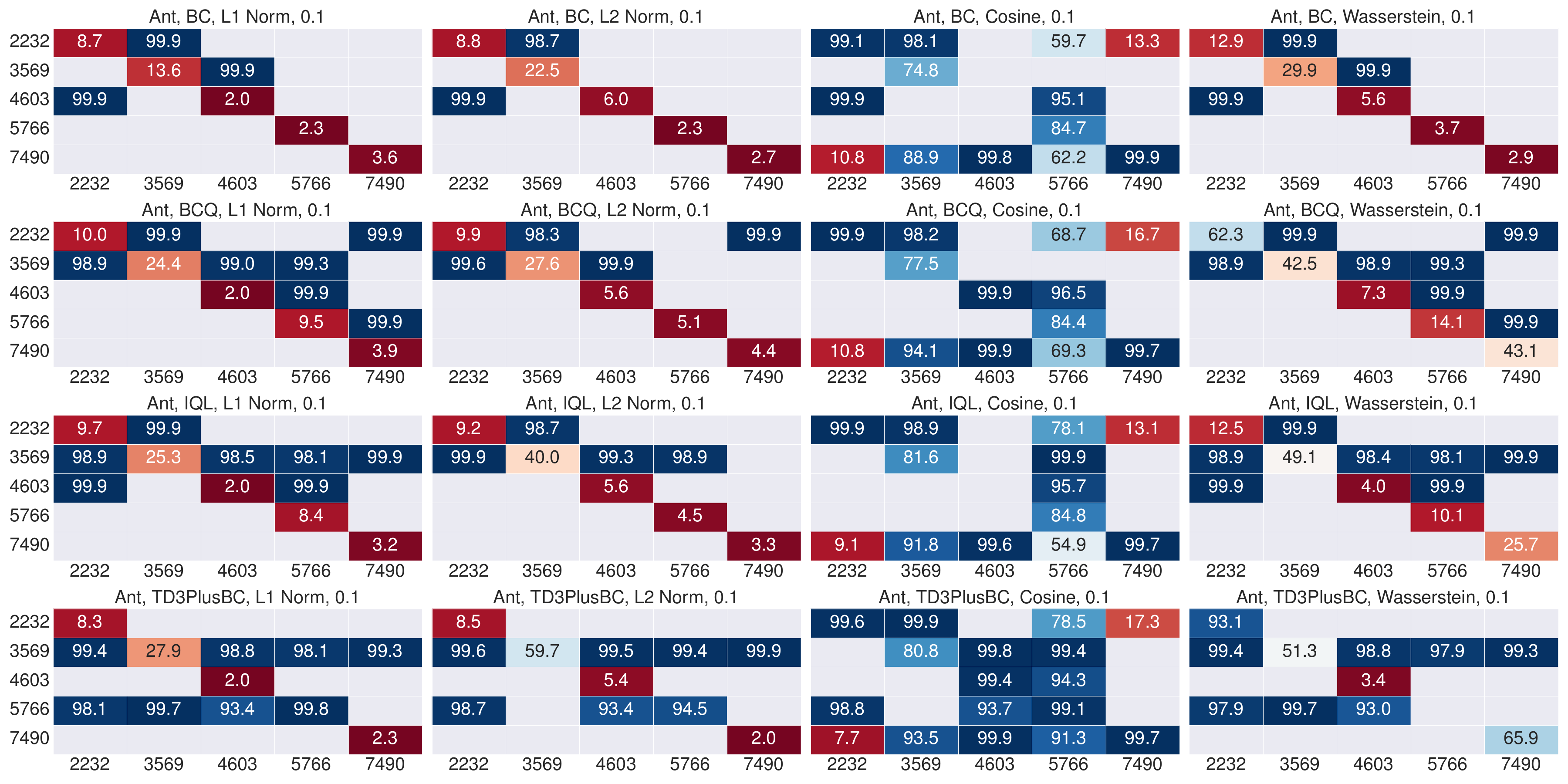}
\vspace{-0.2cm}
\caption{The audit accuracy with Gaussian noise ($\mu=0, \sigma=0.1$) on the suspect models' action for Ant.
The caption of each plot demonstrates the offline DRL model's type, the task, the distance metric, and the noise strength. 
The x labels are the names of datasets to be audited, \ie, the target datasets. 
The y labels are the names of datasets the suspect models learned, \ie, the actual datasets. 
Thus, the diagonal values show the audit accuracy when the actual dataset is the same as the target dataset, \ie, TPR, and the non-diagonal values are the TNR results. 
The positions without value mean 100\% accuracy. 
}
\label{fig:robustness 0.1 on ant}
\end{figure*}

\begin{table*}[t]
\footnotesize
\centering
\caption{The details of the online models for generating the offline datasets. The model performance shows the cumulative reward for 10 separate evaluations. }
\label{tab:the online models}
\setlength{\tabcolsep}{0.3em}
\renewcommand{\arraystretch}{1.1}
\begin{tabular}{ccccc} 
\toprule
\textbf{Task Name}              & \textbf{Online Model} & \textbf{Train Step}  & \textbf{Model Name} & \textbf{Model Performance}  \\ 
\hline
\multirow{5}{*}{Lunar Lander}   & \multirow{5}{*}{SAC}  & \multirow{5}{*}{1e6} & 1171                & 275.47$\pm$14.38            \\
                                &                       &                      & 2094                & 50.79$\pm$65.95             \\
                                &                       &                      & 4496                & 195.02$\pm$143.15           \\
                                &                       &                      & 6518                & 246.40$\pm$33.91            \\
                                &                       &                      & 9906                & 209.33$\pm$91.73            \\ 
\hline
\multirow{5}{*}{Bipedal Walker} & \multirow{5}{*}{PPO}  & \multirow{5}{*}{1e6} & 0841                & 285.55$\pm$60.84            \\
                                &                       &                      & 1203                & 286.94$\pm$53.46            \\
                                &                       &                      & 2110                & 283.58$\pm$47.35            \\
                                &                       &                      & 3813                & 235.88$\pm$103.83           \\
                                &                       &                      & 6558                & 285.16$\pm$65.92            \\ 
\hline
\multirow{5}{*}{Ant}            & \multirow{5}{*}{SAC}  & \multirow{5}{*}{2e6} & 2232                & 5377.70$\pm$1653.17         \\
                                &                       &                      & 3569                & 1924.58$\pm$1180.96         \\
                                &                       &                      & 4603                & 5531.45$\pm$844.10          \\
                                &                       &                      & 5766                & 3025.89$\pm$547.36          \\
                                &                       &                      & 7490                & 5897.37$\pm$477.34          \\
\bottomrule
\end{tabular}
\end{table*}

\begin{table*}[t]
\caption{The details of the offline DRL datasets}
\label{tab:the offline datasets}
\footnotesize
\setlength{\tabcolsep}{0.4em}
\renewcommand{\arraystretch}{1.1}
\centering
\begin{tabular}{ccccc} 
    \toprule
    \textbf{Task Name}              & \textbf{Number of Transitions} & \textbf{Dataset Name} & \textbf{Number of Trajectories} & \textbf{Length of trajectory}  \\ 
    \hline
    \multirow{5}{*}{Lunar Lander}   & \multirow{5}{*}{5e5}           & 1171                 & 2175                            & 229.83$\pm$83.51               \\
                                    &                                & 2094                 & 578                             & 864.19$\pm$231.88              \\
                                    &                                & 4496                 & 1252                            & 399.30$\pm$240.88              \\
                                    &                                & 6518                 & 1878                            & 266.13$\pm$99.65               \\
                                    &                                & 9906                 & 1566                            & 319.21$\pm$231.06              \\ 
    \hline
    \multirow{5}{*}{Bipedal Walker} & \multirow{5}{*}{1e6}           & 0841                 & 1019                            & 981.03$\pm$190.79              \\
                                    &                                & 1203                 & 1027                            & 973.07$\pm$118.42              \\
                                    &                                & 2110                 & 877                             & 1139.55$\pm$151.10             \\
                                    &                                & 3813                 & 887                             & 1126.63$\pm$379.05             \\
                                    &                                & 6558                 & 1041                            & 959.77$\pm$146.13              \\ 
    \hline
    \multirow{5}{*}{Ant}            & \multirow{5}{*}{2e6}           & 2232                 & 2093                            & 955.46$\pm$177.72              \\
                                    &                                & 3569                 & 3497                            & 571.66$\pm$375.40              \\
                                    &                                & 4603                 & 2096                            & 954.01$\pm$175.82              \\
                                    &                                & 5766                 & 2217                            & 901.84$\pm$236.93              \\
                                    &                                & 7490                 & 2103                            & 951.02$\pm$187.93              \\
    \bottomrule
\end{tabular}
\end{table*}

\begin{table*}
\centering
\caption{As a supplementary of~\cite{DCSJCCZ24}, we provide more details of the BC offline models. The model performance shows the cumulative reward for 10 separate evaluations. }
\label{tab:BC models}
\begin{tabular}{cccccccc} 
\toprule
\begin{tabular}[c]{@{}c@{}}\textbf{Offline}\\\textbf{Model}\end{tabular} & \begin{tabular}[c]{@{}c@{}}\textbf{Task}\\\textbf{Name}\end{tabular}     & \begin{tabular}[c]{@{}c@{}}\textbf{Dataset}\\\textbf{Name}\end{tabular} & \begin{tabular}[c]{@{}c@{}}\textbf{Model Performance}\\ ~\textbf{(No Defense)}\end{tabular} & \begin{tabular}[c]{@{}c@{}}\textbf{Model Performance}\\\textbf{(Trajectory Splitting)}\end{tabular} & \begin{tabular}[c]{@{}c@{}}\textbf{Model Performance}\\\textbf{(Model Ensemble)}\end{tabular} & \begin{tabular}[c]{@{}c@{}}\textbf{Model Performance}\\ ~\textbf{(Gauss. 0.01)}\end{tabular} & \begin{tabular}[c]{@{}c@{}}\textbf{Model Performance}\\ ~\textbf{(Gauss. 0.1)}\end{tabular}  \\ 
\hline
\multirow{15}{*}{BC}                                                     & \multirow{5}{*}{\begin{tabular}[c]{@{}c@{}}Lunar\\Lander\end{tabular}}   & 1171                                                                    & 272.06$\pm$5.14                                                                             & 269.17$\pm$11.28                                                                   & 266.20$\pm$13.81                                                                              & 270.84$\pm$6.38                                                                              & 269.01$\pm$11.12                                                                             \\
                                                                         &                                                                          & 2094                                                                    & 39.04$\pm$34.07                                                                             & 46.65$\pm$37.73                                                                    & 45.74$\pm$116.66                                                                              & 55.03$\pm$34.09                                                                              & 53.99$\pm$30.89                                                                              \\
                                                                         &                                                                          & 4496                                                                    & 173.62$\pm$58.93                                                                            & 183.34$\pm$47.00                                                                   & 189.41$\pm$102.32                                                                             & 161.95$\pm$54.27                                                                             & 177.23$\pm$35.35                                                                             \\
                                                                         &                                                                          & 6518                                                                    & 211.85$\pm$43.52                                                                            & 215.13$\pm$57.35                                                                   & 199.44$\pm$118.66                                                                             & 223.84$\pm$41.63                                                                             & 219.14$\pm$44.10                                                                             \\
                                                                         &                                                                          & 9906                                                                    & 225.45$\pm$32.91                                                                            & 213.70$\pm$40.65                                                                   & 234.34$\pm$67.54                                                                              & 215.19$\pm$35.72                                                                             & 215.76$\pm$33.81                                                                             \\ 
\cline{2-8}
                                                                         & \multirow{5}{*}{\begin{tabular}[c]{@{}c@{}}Bipedal\\Walker\end{tabular}} & 0841                                                                    & 264.92$\pm$29.11                                                                            & 277.42$\pm$18.00                                                                   & 241.77$\pm$117.88                                                                             & 257.99$\pm$26.73                                                                             & 268.83$\pm$25.55                                                                             \\
                                                                         &                                                                          & 1203                                                                    & 288.85$\pm$15.39                                                                            & 287.12$\pm$16.88                                                                   & 298.62$\pm$1.29                                                                               & 287.64$\pm$16.36                                                                             & 285.91$\pm$15.71                                                                             \\
                                                                         &                                                                          & 2110                                                                    & 276.80$\pm$26.03                                                                            & 277.15$\pm$24.63                                                                   & 265.78$\pm$98.38                                                                              & 283.05$\pm$20.17                                                                             & 286.19$\pm$14.48                                                                             \\
                                                                         &                                                                          & 3813                                                                    & 164.56$\pm$46.14                                                                            & 156.62$\pm$47.94                                                                   & 66.65$\pm$97.36                                                                               & 160.48$\pm$56.24                                                                             & 182.20$\pm$55.79                                                                             \\
                                                                         &                                                                          & 6558                                                                    & 281.02$\pm$48.65                                                                            & 277.24$\pm$54.87                                                                   & 308.01$\pm$0.87                                                                               & 284.69$\pm$22.77                                                                             & 268.39$\pm$39.12                                                                             \\ 
\cline{2-8}
                                                                         & \multirow{5}{*}{Ant}                                                     & 2232                                                                    & 5479.72$\pm$354.79                                                                          & 5427.47$\pm$609.23                                                                 & 5933.60$\pm$98.05                                                                             & 5324.99$\pm$441.27                                                                           & 4332.23$\pm$589.30                                                                           \\
                                                                         &                                                                          & 3569                                                                    & 1493.77$\pm$413.96                                                                          & 1523.73$\pm$473.18                                                                 & 1695.64$\pm$1255.83                                                                           & 1460.97$\pm$436.37                                                                           & 1412.06$\pm$391.99                                                                           \\
                                                                         &                                                                          & 4603                                                                    & 5424.74$\pm$422.83                                                                          & 5463.20$\pm$511.58                                                                 & 5269.72$\pm$1692.57                                                                           & 5470.37$\pm$473.25                                                                           & 4679.76$\pm$496.30                                                                           \\
                                                                         &                                                                          & 5766                                                                    & 2806.80$\pm$286.57                                                                          & 2863.00$\pm$291.50                                                                 & 2951.89$\pm$728.14                                                                            & 2899.11$\pm$313.43                                                                           & 2458.92$\pm$272.67                                                                           \\
                                                                         &                                                                          & 7490                                                                    & 5514.17$\pm$441.78                                                                          & 5410.28$\pm$467.33                                                                 & 5785.87$\pm$630.97                                                                            & 5451.01$\pm$430.96                                                                           & 4417.19$\pm$687.25                                                                           \\
\bottomrule
\end{tabular}
\end{table*}

\begin{table*}
    \centering
    \caption{As a supplementary of~\cite{DCSJCCZ24}, we provide more details of the BCQ offline models. The model performance shows the cumulative reward for 10 separate evaluations. }
    \label{tab:BCQ models}
    \begin{tabular}{cccccccc} 
    \toprule
    \begin{tabular}[c]{@{}c@{}}\textbf{Offline}\\\textbf{Model}\end{tabular} & \begin{tabular}[c]{@{}c@{}}\textbf{Task}\\\textbf{Name}\end{tabular}     & \begin{tabular}[c]{@{}c@{}}\textbf{Dataset}\\\textbf{Name}\end{tabular} & \begin{tabular}[c]{@{}c@{}}\textbf{Model Performance}\\ ~\textbf{(No Defense)}\end{tabular} & \begin{tabular}[c]{@{}c@{}}\textbf{Model Performance}\\\textbf{\textbf{(Trajectory Splitting)}}\end{tabular} & \begin{tabular}[c]{@{}c@{}}\textbf{Model Performance}\\\textbf{(\textcolor[rgb]{0.2,0.2,0.2}{Model Ensemble})}\end{tabular} & \begin{tabular}[c]{@{}c@{}}\textbf{Model Performance}\\ ~\textbf{(Gauss. 0.01)}\end{tabular} & \begin{tabular}[c]{@{}c@{}}\textbf{Model Performance}\\ ~\textbf{(Gauss. 0.1)}\end{tabular}  \\ 
    \hline
    \multirow{15}{*}{BCQ}                                                    & \multirow{5}{*}{\begin{tabular}[c]{@{}c@{}}Lunar\\Lander\end{tabular}}   & 1171                                                                    & 270.69$\pm$8.51                                                                             & 270.45$\pm$12.51                                                                            & 278.43$\pm$9.57                                                                                                             & 268.41$\pm$13.91                                                                             & 270.25$\pm$13.45                                                                             \\
                                                                             &                                                                          & 2094                                                                    & 52.67$\pm$26.38                                                                             & 64.70$\pm$22.08                                                                             & 30.79$\pm$81.96                                                                                                             & 57.80$\pm$30.64                                                                              & 55.73$\pm$28.36                                                                              \\
                                                                             &                                                                          & 4496                                                                    & 166.13$\pm$57.37                                                                            & 195.16$\pm$37.67                                                                            & 88.06$\pm$182.38                                                                                                            & 188.89$\pm$38.33                                                                             & 191.19$\pm$46.30                                                                             \\
                                                                             &                                                                          & 6518                                                                    & 234.99$\pm$30.93                                                                            & 227.41$\pm$36.30                                                                            & 233.08$\pm$45.90                                                                                                            & 236.13$\pm$33.25                                                                             & 235.19$\pm$25.16                                                                             \\
                                                                             &                                                                          & 9906                                                                    & 243.74$\pm$24.43                                                                            & 236.93$\pm$23.49                                                                            & 236.40$\pm$41.93                                                                                                            & 237.51$\pm$31.45                                                                             & 233.42$\pm$34.97                                                                             \\ 
    \cline{2-8}
                                                                             & \multirow{5}{*}{\begin{tabular}[c]{@{}c@{}}Bipedal\\Walker\end{tabular}} & 0841                                                                    & 228.05$\pm$43.06                                                                            & 235.75$\pm$39.17                                                                            & 229.03$\pm$117.30                                                                                                           & 247.17$\pm$37.65                                                                             & 249.94$\pm$28.21                                                                             \\
                                                                             &                                                                          & 1203                                                                    & 269.87$\pm$28.93                                                                            & 276.35$\pm$23.98                                                                            & 243.15$\pm$112.91                                                                                                           & 276.78$\pm$21.02                                                                             & 281.59$\pm$17.69                                                                             \\
                                                                             &                                                                          & 2110                                                                    & 281.34$\pm$18.56                                                                            & 282.23$\pm$20.88                                                                            & 264.16$\pm$97.68                                                                                                            & 270.97$\pm$24.22                                                                             & 270.78$\pm$26.49                                                                             \\
                                                                             &                                                                          & 3813                                                                    & 166.87$\pm$55.81                                                                            & 181.39$\pm$45.03                                                                            & 131.04$\pm$165.52                                                                                                           & 177.90$\pm$52.03                                                                             & 185.86$\pm$45.97                                                                             \\
                                                                             &                                                                          & 6558                                                                    & 271.52$\pm$57.34                                                                            & 271.09$\pm$75.34                                                                            & 306.09$\pm$4.03                                                                                                             & 275.55$\pm$30.70                                                                             & 262.56$\pm$43.95                                                                             \\ 
    \cline{2-8}
                                                                             & \multirow{5}{*}{Ant}                                                     & 2232                                                                    & 3844.45$\pm$875.84                                                                          & 3651.94$\pm$943.58                                                                          & 4295.42$\pm$2225.70                                                                                                         & 3587.01$\pm$816.81                                                                           & 2514.55$\pm$772.19                                                                           \\
                                                                             &                                                                          & 3569                                                                    & 1032.55$\pm$327.13                                                                          & 951.30$\pm$312.96                                                                           & 435.72$\pm$420.34                                                                                                           & 1013.29$\pm$283.76                                                                           & 942.25$\pm$266.46                                                                            \\
                                                                             &                                                                          & 4603                                                                    & 4554.06$\pm$676.32                                                                          & 4562.26$\pm$828.88                                                                          & 3980.17$\pm$2203.92                                                                                                         & 4480.04$\pm$639.38                                                                           & 3412.76$\pm$804.53                                                                           \\
                                                                             &                                                                          & 5766                                                                    & 2583.27$\pm$268.12                                                                          & 2502.33$\pm$323.71                                                                          & 2603.11$\pm$1075.03                                                                                                         & 2640.93$\pm$323.35                                                                           & 2031.98$\pm$293.49                                                                           \\
                                                                             &                                                                          & 7490                                                                    & 3653.48$\pm$1108.85                                                                         & 3755.22$\pm$1159.16                                                                         & 4012.54$\pm$2267.61                                                                                                         & 3552.11$\pm$1115.43                                                                          & 2432.82$\pm$892.02                                                                           \\
    \bottomrule
\end{tabular}
\end{table*}

\begin{table*}
    \centering
    \caption{As a supplementary of~\cite{DCSJCCZ24}, we provide more details of the IQL offline models. The model performance shows the cumulative reward for 10 separate evaluations. }
    \label{tab:IQL models}
    \begin{tabular}{cccccccc} 
    \toprule
    \begin{tabular}[c]{@{}c@{}}\textbf{Offline }\\\textbf{Model}\end{tabular} & \begin{tabular}[c]{@{}c@{}}\textbf{Task }\\\textbf{Name}\end{tabular}    & \begin{tabular}[c]{@{}c@{}}\textbf{Dataset }\\\textbf{Name}\end{tabular} & \begin{tabular}[c]{@{}c@{}}\textbf{Model Performance}\\ ~\textbf{(No Defense)}\end{tabular} & \begin{tabular}[c]{@{}c@{}}\textbf{Model Performance}\\\textbf{(Trajectory Splitting)}\end{tabular} & \begin{tabular}[c]{@{}c@{}}\textbf{Model Performance}\\\textbf{(Model Ensemble)}\end{tabular} & \begin{tabular}[c]{@{}c@{}}\textbf{Model Performance}\\ ~\textbf{(Gauss. 0.01)}\end{tabular} & \begin{tabular}[c]{@{}c@{}}\textbf{Model Performance}\\ ~\textbf{(Gauss. 0.1)}\end{tabular}  \\ 
    \hline
    \multirow{15}{*}{IQL}                                                     & \multirow{5}{*}{\begin{tabular}[c]{@{}c@{}}Lunar\\Lander\end{tabular}}   & 1171                                                                     & 268.55$\pm$10.81                                                                            & 271.91$\pm$5.10                                            & 275.03$\pm$21.12                                                                              & 266.55$\pm$13.58                                                                             & 265.35$\pm$14.22                                                                             \\
                                                                              &                                                                          & 2094                                                                     & 57.92$\pm$31.14                                                                             & 39.17$\pm$27.55                                            & 47.20$\pm$86.05                                                                               & 49.96$\pm$28.60                                                                              & 46.82$\pm$29.83                                                                              \\
                                                                              &                                                                          & 4496                                                                     & 181.48$\pm$40.09                                                                            & 190.44$\pm$51.16                                           & 138.19$\pm$219.58                                                                             & 194.79$\pm$43.22                                                                             & 181.89$\pm$38.97                                                                             \\
                                                                              &                                                                          & 6518                                                                     & 226.85$\pm$43.37                                                                            & 240.15$\pm$27.03                                           & 237.64$\pm$32.00                                                                              & 218.41$\pm$45.31                                                                             & 245.00$\pm$20.38                                                                             \\
                                                                              &                                                                          & 9906                                                                     & 237.33$\pm$29.23                                                                            & 238.88$\pm$20.10                                           & 221.57$\pm$104.60                                                                             & 245.07$\pm$20.08                                                                             & 231.25$\pm$25.67                                                                             \\ 
    \cline{2-8}
                                                                              & \multirow{5}{*}{\begin{tabular}[c]{@{}c@{}}Bipedal\\Walker\end{tabular}} & 0841                                                                     & 261.23$\pm$33.91                                                                            & 254.33$\pm$34.94                                           & 272.50$\pm$54.24                                                                              & 254.18$\pm$35.26                                                                             & 264.38$\pm$34.52                                                                             \\
                                                                              &                                                                          & 1203                                                                     & 284.63$\pm$17.38                                                                            & 291.47$\pm$14.52                                           & 271.86$\pm$54.77                                                                              & 285.79$\pm$17.42                                                                             & 285.38$\pm$14.02                                                                             \\
                                                                              &                                                                          & 2110                                                                     & 285.24$\pm$28.87                                                                            & 288.77$\pm$21.97                                           & 299.45$\pm$4.43                                                                               & 287.20$\pm$19.29                                                                             & 281.40$\pm$23.24                                                                             \\
                                                                              &                                                                          & 3813                                                                     & 169.20$\pm$38.88                                                                            & 155.45$\pm$57.20                                           & 172.28$\pm$123.79                                                                             & 172.41$\pm$43.21                                                                             & 163.51$\pm$55.19                                                                             \\
                                                                              &                                                                          & 6558                                                                     & 279.97$\pm$33.62                                                                            & 285.89$\pm$23.26                                           & 159.33$\pm$182.92                                                                             & 284.75$\pm$21.41                                                                             & 268.64$\pm$40.07                                                                             \\ 
    \cline{2-8}
                                                                              & \multirow{5}{*}{Ant}                                                     & 2232                                                                     & 4577.36$\pm$865.63                                                                          & 4437.55$\pm$766.02                                         & 4968.65$\pm$1337.85                                                                           & 4678.37$\pm$804.33                                                                           & 3420.01$\pm$912.08                                                                           \\
                                                                              &                                                                          & 3569                                                                     & 1406.45$\pm$447.39                                                                          & 1415.31$\pm$336.28                                         & 1563.86$\pm$1225.73                                                                           & 1421.81$\pm$459.03                                                                           & 1239.03$\pm$328.98                                                                           \\
                                                                              &                                                                          & 4603                                                                     & 5248.48$\pm$477.42                                                                          & 5148.72$\pm$476.71                                         & 5822.61$\pm$164.84                                                                            & 5232.36$\pm$536.94                                                                           & 4135.40$\pm$708.39                                                                           \\
                                                                              &                                                                          & 5766                                                                     & 2846.64$\pm$295.47                                                                          & 2779.50$\pm$233.82                                         & 2680.32$\pm$1019.01                                                                           & 2879.16$\pm$262.72                                                                           & 2338.67$\pm$263.68                                                                           \\
                                                                              &                                                                          & 7490                                                                     & 4814.81$\pm$556.16                                                                          & 4715.59$\pm$628.54                                         & 3367.15$\pm$2159.49                                                                           & 4877.90$\pm$707.65                                                                           & 3461.92$\pm$694.92                                                                           \\
    \bottomrule
\end{tabular}
\end{table*}
    
\begin{table*}
    \centering
    \caption{As a supplementary of~\cite{DCSJCCZ24}, we provide more details of the TD3PlusBC offline models. 
    The model performance shows the cumulative reward for 10 separate evaluations. }
    \label{tab:TD3PlusBC models}
    \begin{tabular}{cccccccc} 
    \toprule
    \begin{tabular}[c]{@{}c@{}}\textbf{Offline}\\\textbf{Model}\end{tabular} & \begin{tabular}[c]{@{}c@{}}\textbf{Task}\\\textbf{Name}\end{tabular}     & \begin{tabular}[c]{@{}c@{}}\textbf{Dataset}\\\textbf{Name}\end{tabular} & \begin{tabular}[c]{@{}c@{}}\textbf{Model Performance}\\ ~\textbf{(No Defense)}\end{tabular} & \begin{tabular}[c]{@{}c@{}}\textbf{Model Performance}\\\textbf{\textbf{(Trajectory Splitting)}}\end{tabular} & \begin{tabular}[c]{@{}c@{}}\textbf{Model Performance}\\\textbf{\textbf{(Model Ensemble)}}\end{tabular} & \begin{tabular}[c]{@{}c@{}}\textbf{Model Performance}\\ ~\textbf{(Gauss. 0.01)}\end{tabular} & \begin{tabular}[c]{@{}c@{}}\textbf{Model Performance}\\ ~\textbf{(Gauss. 0.1)}\end{tabular}  \\ 
    \hline
    \multirow{15}{*}{TD3PlusBC}                                              & \multirow{5}{*}{\begin{tabular}[c]{@{}c@{}}Lunar\\Lander\end{tabular}}   & 1171                                                                    & 263.65$\pm$21.47                                                                            & 266.03$\pm$13.80                                                                            & 263.34$\pm$16.97                                                                                       & 267.15$\pm$12.96                                                                             & 265.12$\pm$10.36                                                                             \\
                                                                             &                                                                          & 2094                                                                    & 99.72$\pm$47.21                                                                             & 95.78$\pm$34.82                                                                             & 71.96$\pm$111.26                                                                                       & 100.67$\pm$34.95                                                                             & 90.74$\pm$37.66                                                                              \\
                                                                             &                                                                          & 4496                                                                    & 201.58$\pm$42.49                                                                            & 207.92$\pm$33.77                                                                            & 159.76$\pm$135.46                                                                                      & 207.07$\pm$28.13                                                                             & 194.69$\pm$42.59                                                                             \\
                                                                             &                                                                          & 6518                                                                    & 242.58$\pm$21.54                                                                            & 229.34$\pm$29.64                                                                            & 248.09$\pm$30.91                                                                                       & 238.51$\pm$21.01                                                                             & 243.96$\pm$16.41                                                                             \\
                                                                             &                                                                          & 9906                                                                    & 235.98$\pm$25.48                                                                            & 241.78$\pm$21.68                                                                            & 206.93$\pm$113.68                                                                                      & 230.41$\pm$34.05                                                                             & 229.55$\pm$36.81                                                                             \\ 
    \cline{2-8}
                                                                             & \multirow{5}{*}{\begin{tabular}[c]{@{}c@{}}Bipedal\\Walker\end{tabular}} & 0841                                                                    & -102.88$\pm$56.03                                                                           & -100.63$\pm$59.89                                                                           & -108.40$\pm$0.22                                                                                       & -101.93$\pm$54.70                                                                            & -97.05$\pm$73.10                                                                             \\
                                                                             &                                                                          & 1203                                                                    & -86.65$\pm$22.94                                                                            & -87.17$\pm$22.47                                                                            & -95.59$\pm$17.13                                                                                       & -87.64$\pm$22.35                                                                             & -86.95$\pm$25.50                                                                             \\
                                                                             &                                                                          & 2110                                                                    & -101.94$\pm$22.86                                                                           & -100.43$\pm$26.01                                                                           & -80.32$\pm$14.02                                                                                       & -101.02$\pm$23.63                                                                            & -98.78$\pm$26.75                                                                             \\
                                                                             &                                                                          & 3813                                                                    & -114.96$\pm$14.97                                                                           & -115.04$\pm$14.37                                                                           & -126.18$\pm$2.36                                                                                       & -113.97$\pm$12.93                                                                            & -119.21$\pm$10.67                                                                            \\
                                                                             &                                                                          & 6558                                                                    & 154.70$\pm$148.81                                                                           & 138.26$\pm$165.87                                                                           & 303.03$\pm$2.30                                                                                        & 165.64$\pm$136.38                                                                            & 168.47$\pm$68.50                                                                             \\ 
    \cline{2-8}
                                                                             & \multirow{5}{*}{Ant}                                                     & 2232                                                                    & 259.94$\pm$116.75                                                                           & 216.71$\pm$118.76                                                                           & 258.28$\pm$297.07                                                                                      & 243.30$\pm$121.39                                                                            & 222.48$\pm$139.45                                                                            \\
                                                                             &                                                                          & 3569                                                                    & 549.14$\pm$192.30                                                                           & 563.13$\pm$156.03                                                                           & 566.88$\pm$655.52                                                                                      & 579.86$\pm$213.98                                                                            & 495.19$\pm$160.81                                                                            \\
                                                                             &                                                                          & 4603                                                                    & 374.17$\pm$199.78                                                                           & 370.58$\pm$217.99                                                                           & 151.13$\pm$112.93                                                                                      & 372.37$\pm$194.00                                                                            & 367.64$\pm$279.00                                                                            \\
                                                                             &                                                                          & 5766                                                                    & 396.13$\pm$130.69                                                                           & 368.56$\pm$115.13                                                                           & 369.73$\pm$275.57                                                                                      & 334.74$\pm$172.22                                                                            & 361.40$\pm$117.68                                                                            \\
                                                                             &                                                                          & 7490                                                                    & 314.48$\pm$222.06                                                                           & 326.59$\pm$153.02                                                                           & 689.21$\pm$637.77                                                                                      & 365.24$\pm$212.09                                                                            & 275.81$\pm$130.30                                                                            \\
    \bottomrule
    \end{tabular}
\end{table*}

\begin{table*}[t]
\footnotesize
\caption{The details of the HalfCheetah dataset}
\label{tab:details of the halfcheetah dataset}
\centering
\begin{tabular}{ccccc}
\toprule
\textbf{Task   Name}          & \textbf{Number of Transitions} & \textbf{Dataset Name} & \textbf{Number of Trajectories} & \textbf{Length of trajectory} \\ \hline
\multirow{4}{*}{Half Cheetah} & 1e6                            & D4RL   Expert        & 1001                            & 998.00  $\pm$0.06             \\
                              & 1e6                            & D4RL   Medium        & 1001                            & 997.90  $\pm$3.13             \\
                              & 1e6                            & D4RL   Random        & 1001                            & 998.00$\pm$0.00               \\
                              & 3.003e5                        & RL Unplugged         & 300                             & 1001.00$\pm$0.00              \\ \bottomrule
\end{tabular}
\end{table*}

\begin{table*}
    \centering
    \caption{As a supplementary of~\cite{DCSJCCZ24}, we provide more details of the models trained on the HalfCheetah dataset. The model performance shows the cumulative reward for 10 separate evaluations. }
    \label{tab:details of the models trained on the halfcheetah dataset}
    \begin{tabular}{cccccc} 
    \toprule
    \textbf{Task Name}             & \textbf{Offline Model}     & \textbf{Dataset Name} & \begin{tabular}[c]{@{}c@{}}\textbf{Model Performance}\\\textbf{(No Defense)}\end{tabular} &\begin{tabular}[c]{@{}c@{}}\textbf{Model Performance}\\\textbf{(Trajectory Splitting)}\end{tabular} & \begin{tabular}[c]{@{}c@{}}\textbf{\textbf{Model Performance}}\\\textbf{\textbf{(Model Ensemble)}}\end{tabular}  \\ 
    \hline
    \multirow{16}{*}{Half Cheetah} & \multirow{4}{*}{BC}        & D4RL   Expert         & 12620.94$\pm$307.84                                                                       & 12624.61$\pm$333.32            & 12868.22$\pm$180.39                                                                                              \\
                                   &                            & D4RL   Medium         & 4223.77$\pm$134.67                                                                        & 4265.82$\pm$96.17              & 4293.35$\pm$75.67                                                                                                \\
                                   &                            & D4RL   Random         & -0.33$\pm$0.24                                                                            & -0.33$\pm$0.22                 & -0.37$\pm$0.62                                                                                                   \\
                                   &                            & RL Unplugged          & -427.50$\pm$113.42                                                                        & -431.01$\pm$110.15             & -427.06$\pm$56.30                                                                                                \\ 
    \cline{2-6}
                                   & \multirow{4}{*}{BCQ}       & D4RL   Expert         & 10974.19$\pm$842.10                                                                       & 10735.35$\pm$1345.57           & 12334.59$\pm$539.99                                                                                              \\
                                   &                            & D4RL   Medium         & 4765.24$\pm$98.75                                                                         & 4746.03$\pm$108.99             & 4512.03$\pm$99.46                                                                                                \\
                                   &                            & D4RL   Random         & -1.13$\pm$0.43                                                                            & -1.15$\pm$0.54                 & -0.54$\pm$0.78                                                                                                   \\
                                   &                            & RL Unplugged          & -421.91$\pm$212.36                                                                        & -419.59$\pm$219.29             & -378.28$\pm$64.55                                                                                                \\ 
    \cline{2-6}
                                   & \multirow{4}{*}{IQL}       & D4RL   Expert         & 10163.20$\pm$1106.70                                                                      & 9920.53$\pm$879.89             & 11268.02$\pm$2640.57                                                                                             \\
                                   &                            & D4RL   Medium         & 4808.11$\pm$46.99                                                                         & 4800.87$\pm$59.75              & 4671.25$\pm$99.09                                                                                                \\
                                   &                            & D4RL   Random         & 1649.55$\pm$518.47                                                                        & 1644.31$\pm$551.32             & 1822.31$\pm$31.63                                                                                                \\
                                   &                            & RL Unplugged          & -378.74$\pm$151.65                                                                        & -367.87$\pm$156.81             & -311.62$\pm$16.31                                                                                                \\ 
    \cline{2-6}
                                   & \multirow{4}{*}{TD3PlusBC} & D4RL   Expert         & 12712.69$\pm$383.33                                                                       & 12752.25$\pm$274.38            & 11468.00$\pm$872.43                                                                                              \\
                                   &                            & D4RL   Medium         & 4969.74$\pm$56.31                                                                         & 4964.48$\pm$57.44              & 4871.85$\pm$82.15                                                                                                \\
                                   &                            & D4RL   Random         & 1046.23$\pm$226.61                                                                        & 1050.03$\pm$214.80             & 1128.32$\pm$3.15                                                                                                 \\
                                   &                            & RL Unplugged          & -181.50$\pm$205.29                                                                        & -175.49$\pm$225.09             & -385.80$\pm$54.32                                                                                                \\
    \bottomrule
    \end{tabular}
\end{table*}

\begin{table*}[ht]
\centering
\caption{The TPR and TNR results on the Half Cheetah task. 
The mean and standard deviation of TPR and TNR in each row represent the audit results for one combination of task and model by four distance metrics. 
Bold indicates the highest sum of TPR and TNR, \ie, accuracy, in a row. 
Each pair of TPR and TNR is derived from the diagonal
and non-diagonal values of the corresponding heatmap in \autoref{fig:audit result on halfcheetah}.
}
\label{tab:audit result on halfcheetah}
    \setlength{\tabcolsep}{0.6em}
    \renewcommand{\arraystretch}{1.1}
    \footnotesize
\begin{tabular}{cccccccccc} 
\toprule
\multirow{2}{*}{\begin{tabular}[c]{@{}c@{}}\textbf{Task}\\\textbf{ Name}\end{tabular}} & \multirow{2}{*}{\begin{tabular}[c]{@{}c@{}}\textbf{Offline}\\\textbf{ Model}\end{tabular}} & \multicolumn{2}{c}{\textbf{L1 Norm}}                               & \multicolumn{2}{c}{\textbf{L2 Norm}}                               & \multicolumn{2}{c}{\begin{tabular}[c]{@{}c@{}}\textbf{Cosine}\\\textbf{ Distance}\end{tabular}} & \multicolumn{2}{c}{\begin{tabular}[c]{@{}c@{}}\textbf{Wasserstein}\\\textbf{ Distance}\end{tabular}}  \\ 
\cline{3-10}
                                                                                       &                                                                                            & TPR                             & TNR                              & TPR                             & TNR                              & TPR                             & TNR                                                           & TPR                                      & TNR                                                        \\ 
\hline
\multirow{4}{*}{\begin{tabular}[c]{@{}c@{}}Half\\Cheetah\end{tabular}}                 & BC                                                                                         & 96.07$\pm$3.15 & 100.00$\pm$0.00 & 96.07$\pm$2.34 & 100.00$\pm$0.00 & 99.80$\pm$0.35 & 68.62$\pm$42.47                              & \textbf{98.47$\pm$1.13} & \textbf{100.00$\pm$0.00}                  \\
                                                                                       & BCQ                                                                                        & 95.37$\pm$0.55 & 100.00$\pm$0.00 & 95.83$\pm$1.20 & 100.00$\pm$0.00 & 99.57$\pm$0.47 & 70.14$\pm$41.14                              & \textbf{97.47$\pm$1.35} & \textbf{100.00$\pm$0.00}                  \\
                                                                                       & IQL                                                                                        & 95.47$\pm$0.77 & 100.00$\pm$0.00 & 95.68$\pm$1.02 & 100.00$\pm$0.00 & 99.78$\pm$0.23 & 71.38$\pm$41.05                              & \textbf{97.12$\pm$2.70} & \textbf{100.00$\pm$0.00}                  \\
                                                                                       & TD3PlusBC                                                                                  & 95.00$\pm$2.87 & 100.00$\pm$0.00 & 95.50$\pm$1.99 & 100.00$\pm$0.00 & 99.87$\pm$0.16 & 70.57$\pm$40.85                              & \textbf{98.27$\pm$1.09} & \textbf{100.00$\pm$0.00}                  \\
\bottomrule
\end{tabular}
\end{table*}

\begin{figure*}[t]
\includegraphics[width=\hsize]{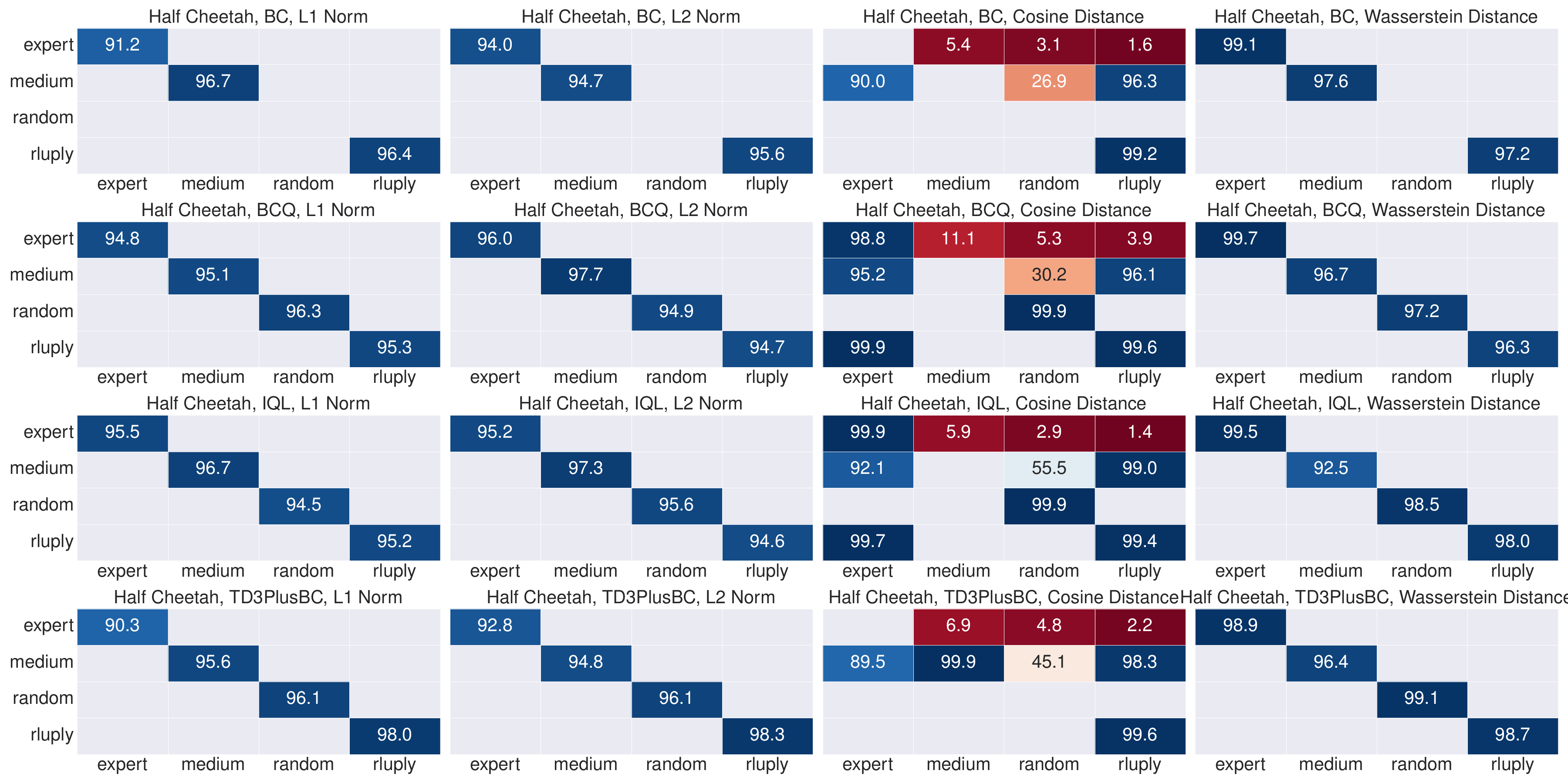}
\vspace{-0.2cm}
\caption{The audit accuracy between every two Half Cheetah datasets. 
The caption of each plot demonstrates the offline DRL model's type, the task, and the distance metric. 
The x labels are the names of datasets to be audited, \ie, the target datasets. 
The y labels are the names of datasets the suspect models learned, \ie, the actual datasets. 
Thus, the diagonal values show the audit accuracy when the actual dataset is the same as the target dataset, \ie, TPR, and the non-diagonal values are the TNR results. 
The positions without value mean 100\% accuracy. 
}

\label{fig:audit result on halfcheetah}
\end{figure*}

\begin{table*}
\centering
\caption{The TPR and TNR results of \sysname when splitting each trajectory into shorter ones ($S=5$). The mean and standard deviation of TPR and TNR in each row represent the audit results for one combination of task and model by four distance metrics. Each pair of TPR and TNR is derived from the diagonal and non-diagonal values of the corresponding heatmap in \autoref{fig:traj-splitting5-lunarlander} (Lunar Lander), \autoref{fig:traj-splitting5-bipedalwalker} (Bipedal Walker), \autoref{fig:traj-splitting5-ant} (Ant), and \autoref{fig:traj-splitting5-halfcheetah} (Half Cheetah). }
\label{tab:overall-audit-accuracy-trajectory-splitting}
\begin{tabular}{cccccccccc} 
\toprule
\multirow{2}{*}{\begin{tabular}[c]{@{}c@{}}\textbf{Task}\\\textbf{ Name}\end{tabular}} & \multirow{2}{*}{\begin{tabular}[c]{@{}c@{}}\textbf{Offline}\\\textbf{ Model}\end{tabular}} & \multicolumn{2}{c}{\textbf{L1 Norm}}                                                                       & \multicolumn{2}{c}{\textbf{L2 Norm}}                                                                       & \multicolumn{2}{c}{\begin{tabular}[c]{@{}c@{}}\textbf{Cosine}\\\textbf{ Distance}\end{tabular}}            & \multicolumn{2}{c}{\begin{tabular}[c]{@{}c@{}}\textbf{Wasserstein}\\\textbf{ Distance}\end{tabular}}        \\ 
\cline{3-10}
                                                                                       &                                                                                            & TPR                                                 & TNR                                                  & TPR                                                 & TNR                                                  & TPR                                                 & TNR                                                  & TPR                                                 & TNR                                                   \\ 
\hline
\multirow{4}{*}{\begin{tabular}[c]{@{}c@{}}Lunar\\ Lander\end{tabular}}                & BC                                                                                         & 99.01$\pm$0.46                     & 100.00$\pm$0.00                     & 96.96$\pm$0.73                     & 100.00$\pm$0.00                     & 96.96$\pm$0.73                     & 100.00$\pm$0.00                     & 98.43$\pm$0.73                     & 99.94$\pm$0.18                       \\
                                                                                       & BCQ                                                                                        & 98.29$\pm$1.10                     & 100.00$\pm$0.00                     & 95.87$\pm$1.12                     & 100.00$\pm$0.00                     & 95.87$\pm$1.06                     & 99.99$\pm$0.03                      & 97.60$\pm$1.14                     & 99.91$\pm$0.15                       \\
                                                                                       & IQL                                                                                        & 98.59$\pm$1.55                     & 99.91$\pm$0.31                      & 97.52$\pm$2.51                     & 99.97$\pm$0.12                      & 97.49$\pm$2.56                     & 99.92$\pm$0.19                      & 98.32$\pm$1.79                     & 97.10$\pm$5.66                       \\
                                                                                       & TD3PlusBC                                                                                  & 98.29$\pm$2.04                     & 99.48$\pm$0.79                      & 96.35$\pm$3.01                     & 99.89$\pm$0.22                      & 96.27$\pm$3.16                     & 99.91$\pm$0.23                      & 98.53$\pm$1.25                     & 95.59$\pm$3.77                       \\ 
\cline{2-10}
\multirow{4}{*}{\begin{tabular}[c]{@{}c@{}}Bipedal\\ Walker\end{tabular}}              & BC                                                                                         & 99.65$\pm$0.57                     & 100.00$\pm$0.00                     & 98.45$\pm$2.71                     & 100.00$\pm$0.00                     & 98.64$\pm$2.66                     & 100.00$\pm$0.00                     & 99.79$\pm$0.43                     & 100.00$\pm$0.00                      \\
                                                                                       & BCQ                                                                                        & 99.55$\pm$0.71                     & 100.00$\pm$0.00                     & 98.19$\pm$2.84                     & 100.00$\pm$0.00                     & 99.68$\pm$0.45                     & 100.00$\pm$0.00                     & 99.89$\pm$0.10                     & 100.00$\pm$0.00                      \\
                                                                                       & IQL                                                                                        & 95.17$\pm$7.39                     & 100.00$\pm$0.00                     & 95.01$\pm$5.49                     & 100.00$\pm$0.00                     & 99.81$\pm$0.31                     & 100.00$\pm$0.00                     & 95.33$\pm$7.01                     & 100.00$\pm$0.00                      \\
                                                                                       & TD3PlusBC                                                                                  & 99.39$\pm$1.23                     & 94.77$\pm$19.42                     & 97.15$\pm$5.71                     & 93.37$\pm$21.46                     & 96.93$\pm$6.00                     & 91.98$\pm$21.75                     & 98.13$\pm$3.73                     & 88.23$\pm$25.40                      \\ 
\cline{2-10}
\multirow{4}{*}{Ant}                                                                   & BC                                                                                         & 98.03$\pm$1.38                     & 99.93$\pm$0.12                      & 96.77$\pm$1.49                     & 99.90$\pm$0.36                      & 99.39$\pm$0.91                     & 86.07$\pm$27.76                     & 98.05$\pm$1.43                     & 99.91$\pm$0.15                       \\
                                                                                       & BCQ                                                                                        & 97.47$\pm$2.93                     & 99.80$\pm$0.44                      & 95.89$\pm$2.32                     & 99.84$\pm$0.41                      & 99.65$\pm$0.63                     & 86.86$\pm$27.33                     & 98.83$\pm$1.55                     & 99.79$\pm$0.47                       \\
                                                                                       & IQL                                                                                        & 97.68$\pm$2.08                     & 99.65$\pm$0.73                      & 96.77$\pm$2.50                     & 99.69$\pm$0.59                      & 99.63$\pm$0.62                     & 85.74$\pm$28.43                     & 99.31$\pm$0.49                     & 99.63$\pm$0.78                       \\
                                                                                       & TD3PlusBC                                                                                  & 98.71$\pm$1.63                     & 99.18$\pm$1.71                      & 97.20$\pm$1.79                     & 99.35$\pm$1.72                      & 99.81$\pm$0.31                     & 88.35$\pm$25.99                     & 99.22$\pm$1.31                     & 99.14$\pm$1.81                       \\ 
\cline{2-10}
\multirow{4}{*}{\begin{tabular}[c]{@{}c@{}}Half\\Cheetah\end{tabular}}                 & BC                                                                                         & \multicolumn{1}{l}{98.50$\pm$1.50} & \multicolumn{1}{l}{100.00$\pm$0.00} & \multicolumn{1}{l}{96.87$\pm$2.14} & \multicolumn{1}{l}{100.00$\pm$0.00} & \multicolumn{1}{l}{99.74$\pm$0.27} & \multicolumn{1}{l}{69.56$\pm$41.70} & \multicolumn{1}{l}{98.60$\pm$0.92} & \multicolumn{1}{l}{100.00$\pm$0.00}  \\
                                                                                       & BCQ                                                                                        & \multicolumn{1}{l}{96.83$\pm$1.54} & \multicolumn{1}{l}{100.00$\pm$0.00} & \multicolumn{1}{l}{96.27$\pm$1.36} & \multicolumn{1}{l}{100.00$\pm$0.00} & \multicolumn{1}{l}{99.93$\pm$0.12} & \multicolumn{1}{l}{68.58$\pm$41.88} & \multicolumn{1}{l}{97.43$\pm$1.38} & \multicolumn{1}{l}{100.00$\pm$0.00}  \\
                                                                                       & IQL                                                                                        & \multicolumn{1}{l}{97.00$\pm$2.00} & \multicolumn{1}{l}{100.00$\pm$0.00} & \multicolumn{1}{l}{96.25$\pm$1.13} & \multicolumn{1}{l}{100.00$\pm$0.00} & \multicolumn{1}{l}{99.56$\pm$0.20} & \multicolumn{1}{l}{72.63$\pm$40.22} & \multicolumn{1}{l}{97.06$\pm$2.73} & \multicolumn{1}{l}{100.00$\pm$0.00}  \\
                                                                                       & TD3PlusBC                                                                                  & \multicolumn{1}{l}{97.53$\pm$1.30} & \multicolumn{1}{l}{100.00$\pm$0.00} & \multicolumn{1}{l}{96.53$\pm$1.20} & \multicolumn{1}{l}{100.00$\pm$0.00} & \multicolumn{1}{l}{99.56$\pm$0.61} & \multicolumn{1}{l}{71.21$\pm$40.66} & \multicolumn{1}{l}{98.37$\pm$1.09} & \multicolumn{1}{l}{100.00$\pm$0.00}  \\
\bottomrule
\end{tabular}
\end{table*}

\begin{figure*}[!t]
\includegraphics[width=\hsize]{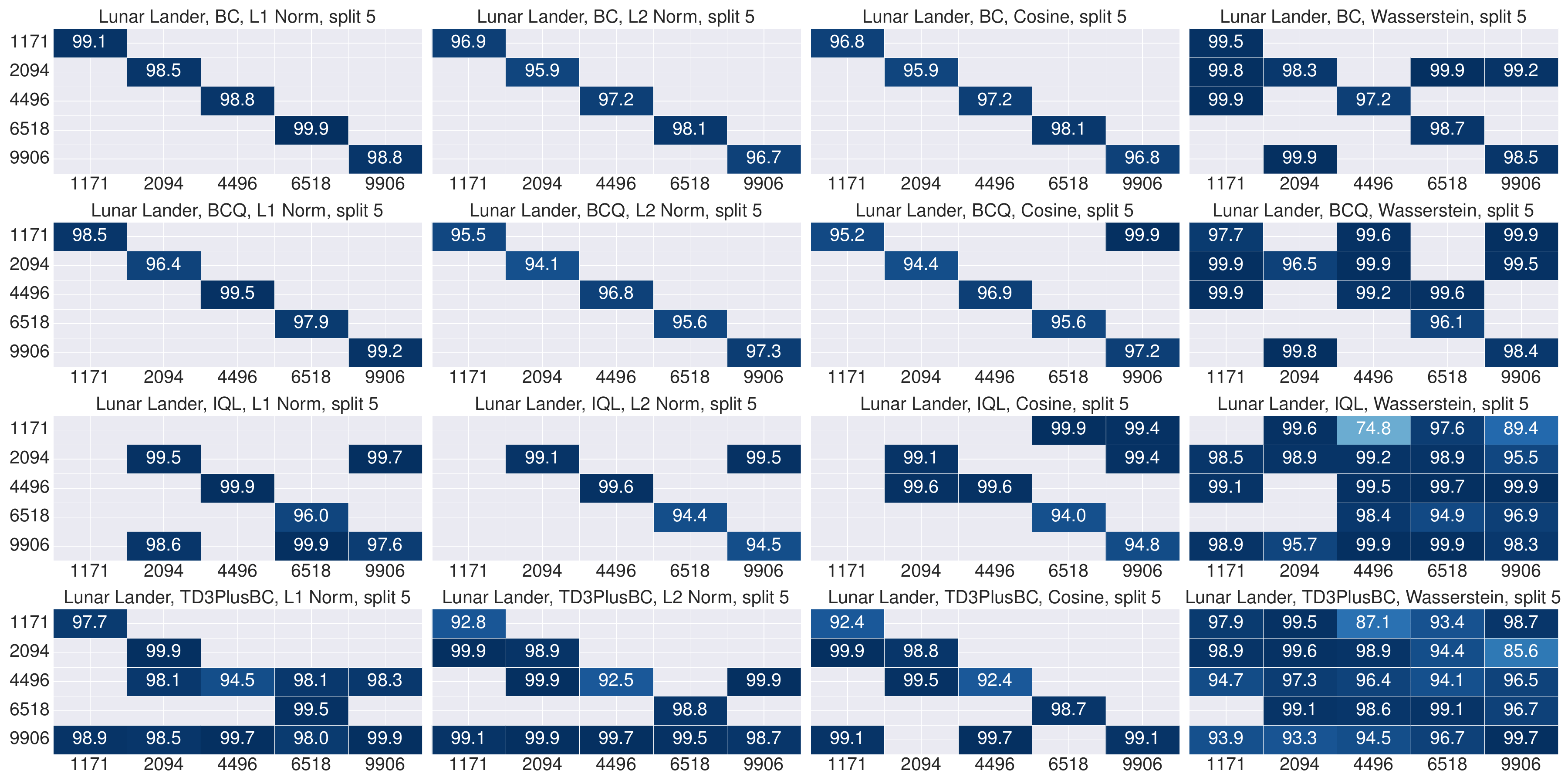}
\vspace{-0.2cm}
\caption{The audit accuracy of \sysname on Lunar Lander when splitting each trajectory into shorter ones ($S=5$).
The caption of each plot demonstrates the offline DRL model's type, the task, the distance metric, and the hyperparameter $S$ of trajectory splitting.  
The x labels are the names of datasets to be audited, \ie, the target datasets. 
The y labels are the names of datasets the suspect models learned, \ie, the actual datasets. 
Thus, the diagonal values show the audit accuracy when the actual dataset is the same as the target dataset, \ie, TPR, and the non-diagonal values are the TNR results. 
The positions without value mean 100\% accuracy. 
}
\label{fig:traj-splitting5-lunarlander}
\end{figure*}

\begin{figure*}[!t]
\includegraphics[width=\hsize]{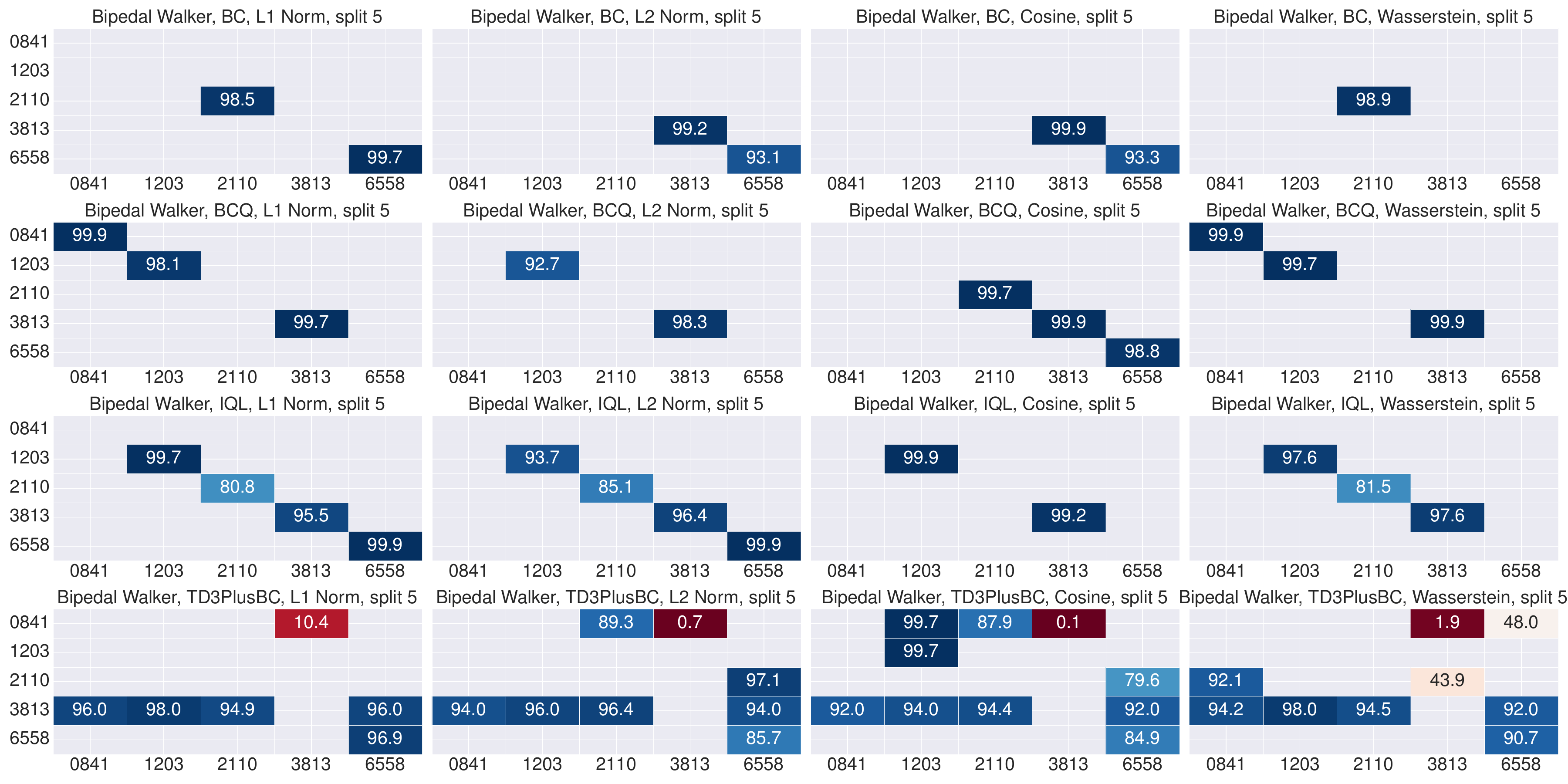}
\vspace{-0.2cm}
\caption{The audit accuracy of \sysname on Bipedal Walker when splitting each trajectory into shorter ones ($S=5$).
The caption of each plot demonstrates the offline DRL model's type, the task, the distance metric, and the hyperparameter $S$ of trajectory splitting.  
The x labels are the names of datasets to be audited, \ie, the target datasets. 
The y labels are the names of datasets the suspect models learned, \ie, the actual datasets. 
Thus, the diagonal values show the audit accuracy when the actual dataset is the same as the target dataset, \ie, TPR, and the non-diagonal values are the TNR results. 
The positions without value mean 100\% accuracy. 
}
\label{fig:traj-splitting5-bipedalwalker}
\end{figure*}

\begin{figure*}[!t]
\includegraphics[width=\hsize]{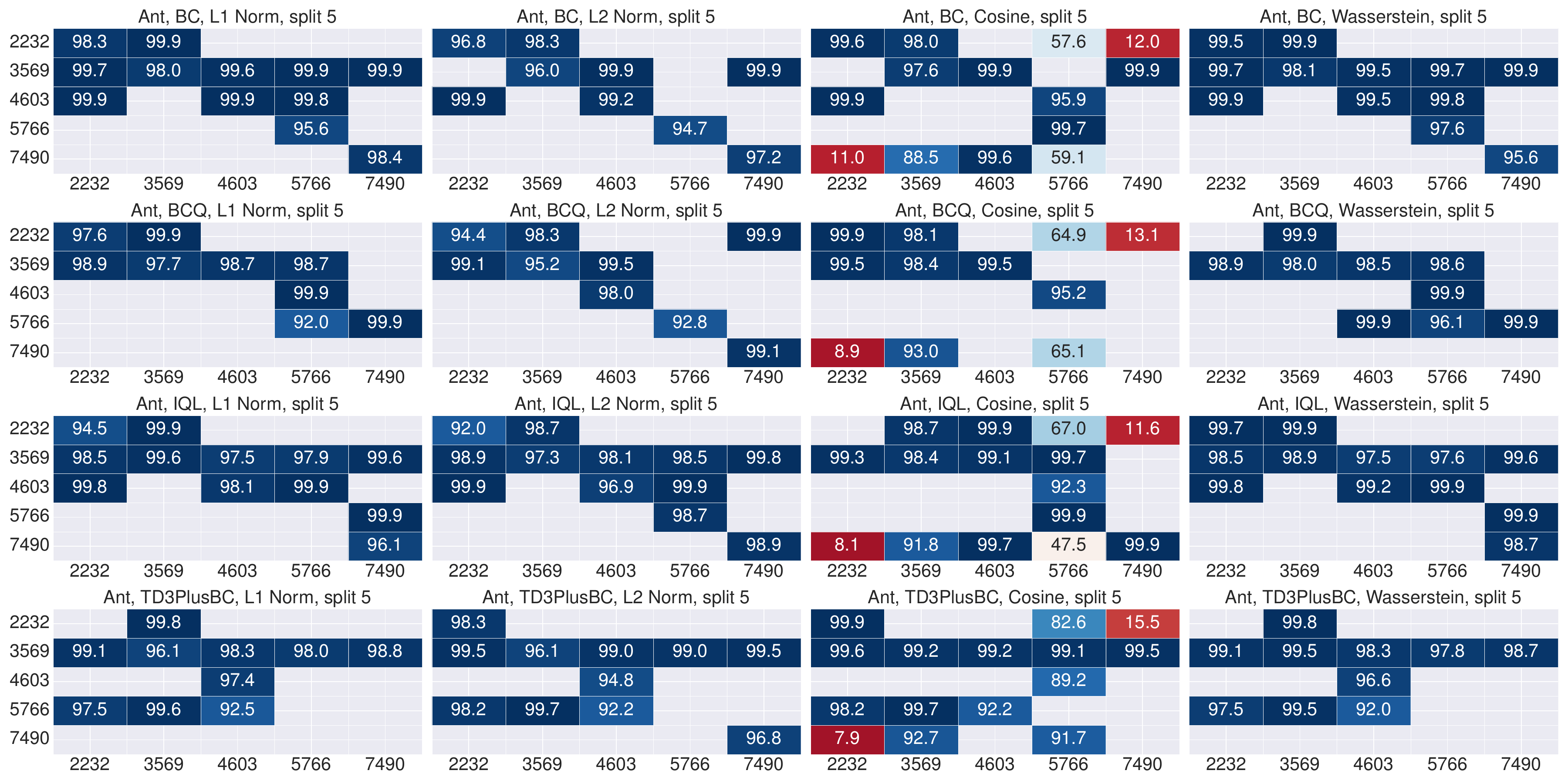}
\vspace{-0.2cm}
\caption{The audit accuracy of \sysname on Ant when splitting each trajectory into shorter ones ($S=5$).
The caption of each plot demonstrates the offline DRL model's type, the task, the distance metric, and the hyperparameter $S$ of trajectory splitting.  
The x labels are the names of datasets to be audited, \ie, the target datasets. 
The y labels are the names of datasets the suspect models learned, \ie, the actual datasets. 
Thus, the diagonal values show the audit accuracy when the actual dataset is the same as the target dataset, \ie, TPR, and the non-diagonal values are the TNR results. 
The positions without value mean 100\% accuracy. 
}
\label{fig:traj-splitting5-ant}
\end{figure*}

\begin{figure*}[!t]
\includegraphics[width=\hsize]{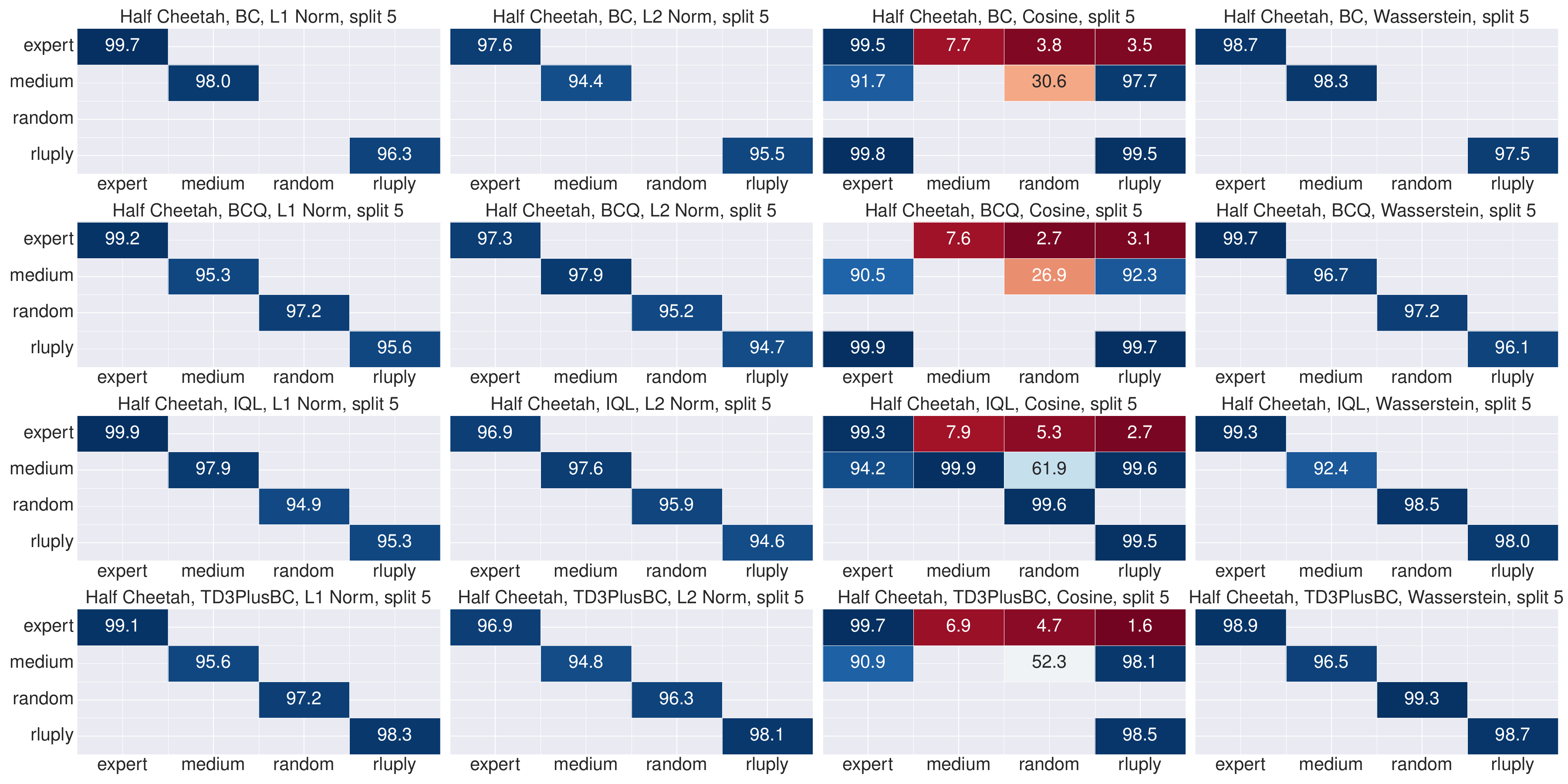}
\vspace{-0.2cm}
\caption{The audit accuracy of \sysname on Half Cheetah when splitting each trajectory into shorter ones ($S=5$).
The caption of each plot demonstrates the offline DRL model's type, the task, the distance metric, and the hyperparameter $S$ of trajectory splitting. 
The x labels are the names of datasets to be audited, \ie, the target datasets. 
The y labels are the names of datasets the suspect models learned, \ie, the actual datasets. 
Thus, the diagonal values show the audit accuracy when the actual dataset is the same as the target dataset, \ie, TPR, and the non-diagonal values are the TNR results. 
The positions without value mean 100\% accuracy. 
}
\label{fig:traj-splitting5-halfcheetah}
\end{figure*}

\end{document}